\documentclass[usegraphicx]{mn2e}

\newcommand{\colsw}{col}

\usepackage{amssymb}
\usepackage{mathptmx}

\newcommand{\Mpc}{\rm\thinspace Mpc}
\newcommand{\kpc}{\rm\thinspace kpc}

\newcommand{\km}{\rm\thinspace km}

\newcommand{\cm}{\rm\thinspace cm}
\newcommand{\yr}{\rm\thinspace yr}

\newcommand{\s}{\rm\thinspace s}

\newcommand{\K}{\rm\thinspace K}
\newcommand{\Kpcmc}{\hbox{$\K\cm^{-3}\,$}}
\newcommand{\Msun}{\hbox{$\rm\thinspace M_{\odot}$}}

\newcommand{\Msunpyr}{\hbox{$\Msun\yr^{-1}\,$}}
\newcommand{\keV}{\rm\thinspace keV}

\newcommand{\erg}{\rm\thinspace erg}

\newcommand{\ergps}{\hbox{$\erg\s^{-1}\,$}}

\newcommand{\keVcmsq}{\hbox{$\keV\cm^{2}\,$}}
\newcommand{\kmps}{\hbox{$\km\s^{-1}\,$}}

\newcommand{\kmpspMpc}{\hbox{$\kmps\Mpc^{-1}$}}
\newcommand{\Zsun}{\hbox{$\thinspace \mathrm{Z}_{\odot}$}}

\newcommand{\psqcm}{\hbox{$\cm^{-2}\,$}}
\newcommand{\pcmsq}{\hbox{$\cm^{-2}\,$}}

\begin{document}

\title{Mapping small-scale temperature and abundance structures in
the core of the Perseus cluster}
\author[J.S. Sanders et al]
{J.S. Sanders$^1$\thanks{E-mail: jss@ast.cam.ac.uk},
  A.C. Fabian$^1$, S.W. Allen$^1$ and R.W. Schmidt$^2$\\
  $^1$ Institute of Astronomy, Madingley Road, Cambridge CB3 0HA\\
  $^2$ Institut f\"ur Physik, Universit\"at Potsdam, Am neuen Palais
  10, D-14469 Potsdam, Germany.}

\maketitle

\begin{abstract}
  We report further results from a 191~ks \emph{Chandra} observation
  of the core of the Perseus cluster, Abell 426. The emission-weighted
  temperature and abundance structure is mapped detail. There are
  temperature variations down to $\sim 1 \kpc$ in the brightest
  regions.  Globally, the strongest X-ray surface brightness features
  appear to be caused by temperature changes. Density and temperature
  changes conspire to give approximate azimuthal balance in pressure
  showing that the gas is in hydrostatic equilibrium.  Si, S, Ar, Ca,
  Fe and Ni abundance profiles rise inward from about 100~kpc, peaking
  at about 30--40~kpc. Most of these abundances drop inwards of the
  peak, but Ne shows a central peak, all of which may be explained by
  resonance scattering. There is no evidence for a widespread
  additional cooler temperature component in the cluster with a
  temperature greater than a factor of two from the local temperature.
  There is however evidence for a widespread hard component which may
  be nonthermal. The temperature and abundance of gas in the cluster
  is observed to be correlated in a manner similar to that found
  between clusters.
\end{abstract}

\begin{keywords}
  X-rays: galaxies --- galaxies: clusters: individual: Perseus ---
  intergalactic medium --- cooling flows
\end{keywords}

\section{Introduction}
The Perseus cluster, Abell 426, is the brightest cluster of galaxies
in the sky in the X-ray band.  The most prominent features in the core
of the cluster are the cavities in the X-ray emitting gas (B\"ohringer
et al 1993), associated with the bubbles blown by the central source
3C~84 (Pedlar et al 1990). The coolest X-ray gas in the cluster lies
on the rims of these lobes (Fabian et al 2000, 2003a). A swirl of cool
gas occupies the inner 70 kpc radius of the cluster (Churazov et al
2000; Fabian et al 2000; Schmidt, Fabian \& Sanders 2002; Churazov et
al 2003a; Fabian et al 2003a), winding around the core in an
anti-clockwise direction.  The X-ray emission contains a number of
quasi-periodic fluctuations in emission, which are interpreted as
sound waves propagating in the intracluster medium (ICM), driven by
the active nucleus (Fabian et al 2003a). In addition there appears to
be a weak shock to the NE of the core, about 10 kpc from the bubble
rim. The abundance in the cluster rises towards the centre, peaking
around $0.7\Zsun$ about 40-50 kpc from the core, and dropping back
down in the centre (Schmidt et al 2002; Churazov et al 2003a).

We previously observed the Perseus cluster with \emph{Chandra} (Fabian
et al 2000; Schmidt et al 2002) for about~25 ks. The initial results
from a much longer 191.2~ks \emph{Chandra} observation were reported
in Fabian et al (2003a), showing evidence for the presence of shocks
and ripples in the core of the cluster. Here we present further
results from this observation, in particular the temperature and
abundance structures in this object, and tests for the presence of
multiphase gas. As this is the deepest \emph{Chandra} observation of a
bright nearby diffuse object, our results are affected by systematic
uncertainties which are not visible in shorter observations.  We do
not discuss the high-velocity system (HVS) in depth here (see Gillmon,
Sanders \& Fabian 2003), nor the relationship between the H$\alpha$
nebulosity and the X-ray emission (see Fabian et al 2003b).

The Perseus cluster is at a redshift of 0.0183. We assume that $H_0 =
70 \kmpspMpc$; therefore 1 kpc corresponds to about 2.7 arcsec.

\section{Data preparation}
\label{sect:prep}
The two datasets were combined into a single dataset of exposure
191.2~ks as in Fabian et al (2003a). This is a valid since the two
datasets were taken at essentially the same roll angle. We made a
number of checks whilst analysing the data to check that there are no
systematic uncertainties introduced by the merging procedure. The
calibration products created using the merged dataset were identical
in the quality of the fits to those generated from the individual
datasets.  No flaring was visible in a lightcurve made from events
away from the centre of the cluster.

The merged events file was corrected for time-dependent gain shift
using the Summer 2003 version of the \textsc{corr\_tgain} utility
(Vikhlinin 2003) and the corrgain2002-05-01.fits correction file. The
data were then reprocessed using the acisD2000-08-12gainN0003.fits
gain file, recalculating the PI values. PHA and position randomisation
was enabled.

Unless mentioned otherwise, data presented here were grouped to
include a minimum of 20 counts per spectral bin. Response files and
ancillary-response files were generated using the \textsc{mkwarf} and
\textsc{mkrmf} \textsc{ciao} tools, weighting the areas of the CCD in
response by using the number of counts between 0.5 and 7 keV. The FEF
file used for response generation was
acisD2000-01-29fef\_phaN0003.fits. The ancillary-response matrices
were corrected for the degradation in the low energy response
(Marshall et al 2003) using the \textsc{corrarf} routine (Vikhlinin
2002) to apply the \textsc{acisabs} absorption correction (Chartas \&
Getman 2002).  Abundances were measured assuming the abundance ratios
of Anders \& Grevesse (1989).  \textsc{xspec} version 11.2.0 (Arnaud
1996) was used to fit the spectra. The \textsc{mekal} (Mewe,
Gronenschild \& van den Oord 1985; Liedahl, Osterheld \& Goldstein
1995) spectral model used here is the default one available in that
version of \textsc{xspec}. The \textsc{apec} (Smith et al 2001)
spectral model we used was version 1.3.1.

\section{Systematic uncertainties}
This paper is based on the longest set of \emph{Chandra} observations
of a nearby diffuse source. The total number of good events from the
observation is $1.9 \times 10^7$. With this unprecedented dataset we
can create many high signal-to-noise spectra. Owing to the hard work
of the calibration team, models fit the data with residuals of less
than 10 per~cent over the whole usable spectral range. However, the
high quality of the data means that the tiny fractional residuals
yield a poor quality of fit (reduced-$\chi^2 \sim 2-3$).

It is therefore difficult to know whether small features in a spectral
model are fitting systematic residuals or real features in the
data, when the spectrum has high signal-to-noise. Results in
the paper must be taken with this caveat, although we attempt to
ensure that random errors exceed systematic errors. We hope that all
the results will be confirmed as the calibration improves. Relative
trends will be more reliable than individual results, although a
systematic residual may vary as a function of the underlying emission
or position on the detector.

\begin{figure}
  \includegraphics[angle=-90,width=\columnwidth]{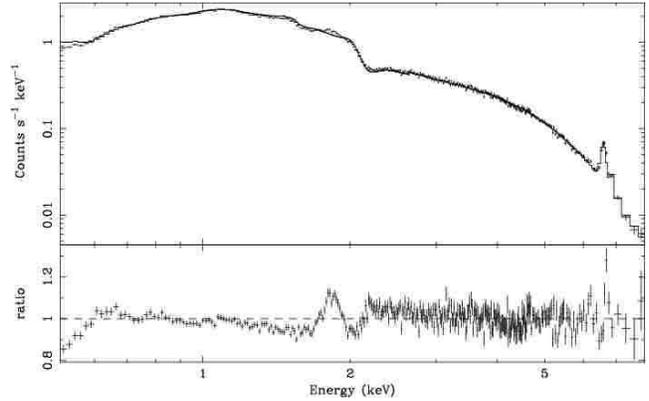}
  \caption{Fit of \textsc{mekal} model between 1 and 7 keV to a
    spectrum from the outer region of the ACIS-S3 CCD. The data were
    rebinned for display to have a minimum significance of $20 \sigma$
    per point, rebinning up to 20 channels.}
  \label{fig:bigedge_resid}
\end{figure}

The systematic uncertainties may be illustrated with the spectrum in
Fig.~\ref{fig:bigedge_resid}. The plot shows a \textsc{mekal} model
fit with absorption to the spectrum from an approximately triangular
region, to the South and South East of the ACIS-S3 CCD, around
$200-300$ arcsec from the core of the cluster. The temperature,
absorption, and abundance were allowed to vary in the fit which
extended between 1 and 7 keV. The abundance ratios were fixed to solar
values.

At energies less than 0.6 keV the model overestimates the flux from
the object. At 0.4 keV the model predicts twice as much flux as is
observed. Extra absorption (at least assuming solar abundance for the
absorber) does not improve the fit. We experimented using a different
correction for the low energy degradation,
\textsc{contamarf}\footnote{http://space.mit.edu/CXC/analysis/ACIS\_Contam/script.html}.
The results were very similar to using \textsc{acisabs}, except the
column density increased everywhere by about $2 \times 10^{20}
\psqcm$.

In addition to the problems at low energy, there is a large residual
around 2 keV, where the effective area of the mirror abruptly changes.
This residual can be removed by adding a Gaussian near the edge.
\textsc{mekal} or \textsc{apec} models cannot fit this residual if
individual abundances are allowed to vary. There may also be another
calibration problem starting near 1 keV, and increasing to 2 keV. It
is unclear how large this effect is, as changing Ni and Mg abundances
can allow the models to fit the data up to 1.8 keV, depending on the
quality of the data. The residual can be fitted with a wide Gaussian,
but the width of the Gaussian depends on the model fitted.

\begin{figure}
  \includegraphics[angle=-90,width=0.99\columnwidth]{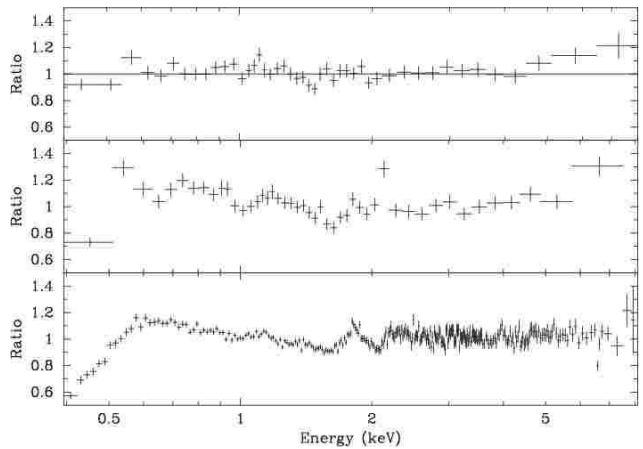}
  \caption{Ratios of models to data from identical regions
    for three different \emph{Chandra} datasets. Top dataset was taken
    on 1999-11-28 (published in Fabian et al 2000, Schmidt et al
    2002). Middle dataset was taken on 2001-01-29 (published in
    Schmidt et al 2002). Bottom dataset is our new merged dataset. The
    points were rebinned for display to have a minimum significance of
    $20 \sigma$ per point, rebinning up to 20 channels.}
  \label{fig:olddata_resid}
\end{figure}

To investigate whether the residuals are new problems, or whether they
exist in older data, we extracted spectra from the same region from
this dataset and those used by Fabian et al (2000) and Schmidt et al
(2002).  The region used was a $\sim 200$ arcsec box to the South of
the nucleus.  There are three \emph{Chandra} datasets: the one
published in Fabian et al (2000) and Schmidt et al (2002) [OBS-ID
503], another just published in Schmidt et al (2002) [OBS-ID 1513],
and the merged dataset we present here. In
Fig.~\ref{fig:olddata_resid} we show the ratios of the data to a best
fit \textsc{mekal} model (fitted to each dataset between 0.7 and 6
keV).  The datasets were reprocessed with the latest gain files, major
background flares were removed from the 503 dataset, and blank sky
background spectra were used (processing these in the same way as the
original dataset).  Ancillary responses were corrected with
\textsc{corrarf}.

The plot shows that the ratios to the model we see from this dataset
are consistent with those from the 1513 dataset, especially around 2
keV and below 0.6 keV. The 1513 dataset shows a hard energy excess,
probably as we did not make a careful enough removal of background
flares for this discussion. The ratios are compatible with the 503
dataset except in the lowest energy bin.

Our main analysis technique is to split the data up into a number of
independent spatial regions for separate spectral analysis (following
Sanders \& Fabian 2002; Schmidt et al 2002). Depending on the average
quality of each spectrum, and the model we fit, we vary the procedure
as described below.

If the number of counts in each spectrum is small (less than $\sim
9\times10^4$) then we fit the whole energy range 0.6 to 8 keV. The
data below 0.6 keV are unusable due to the drop-off in flux. The
residuals around 1.8 keV are no worse than in previous observations of
this object, and probably in observations of other objects with
\emph{Chandra}.

If the spectra we are fitting contain many counts ($\gtrsim 9 \times
10^4$), then we exclude the region from 1.3 to 2.3 keV.  This energy
range was excluded from bins with $S/N \gtrsim 300$.  It is a rather
severe treatment of the data, but a correction of the response in this
region requires knowledge of which residuals are real and which are
systematic. Excluding this band means the majority of the spectra when
fitted produce a reduced-$\chi^2$ less than 1.2, even for large
spatial regions. It is important to now that much of this paper
concerns the spatial distribution of temperatures, abundances etc.
Provided that the systematic uncertainties have little spatial
variation, the features and trends in our resulting maps will be real,
although there may be small effects in the absolute values.

\section{X-ray image}
\begin{figure}
  \includegraphics[width=\columnwidth]{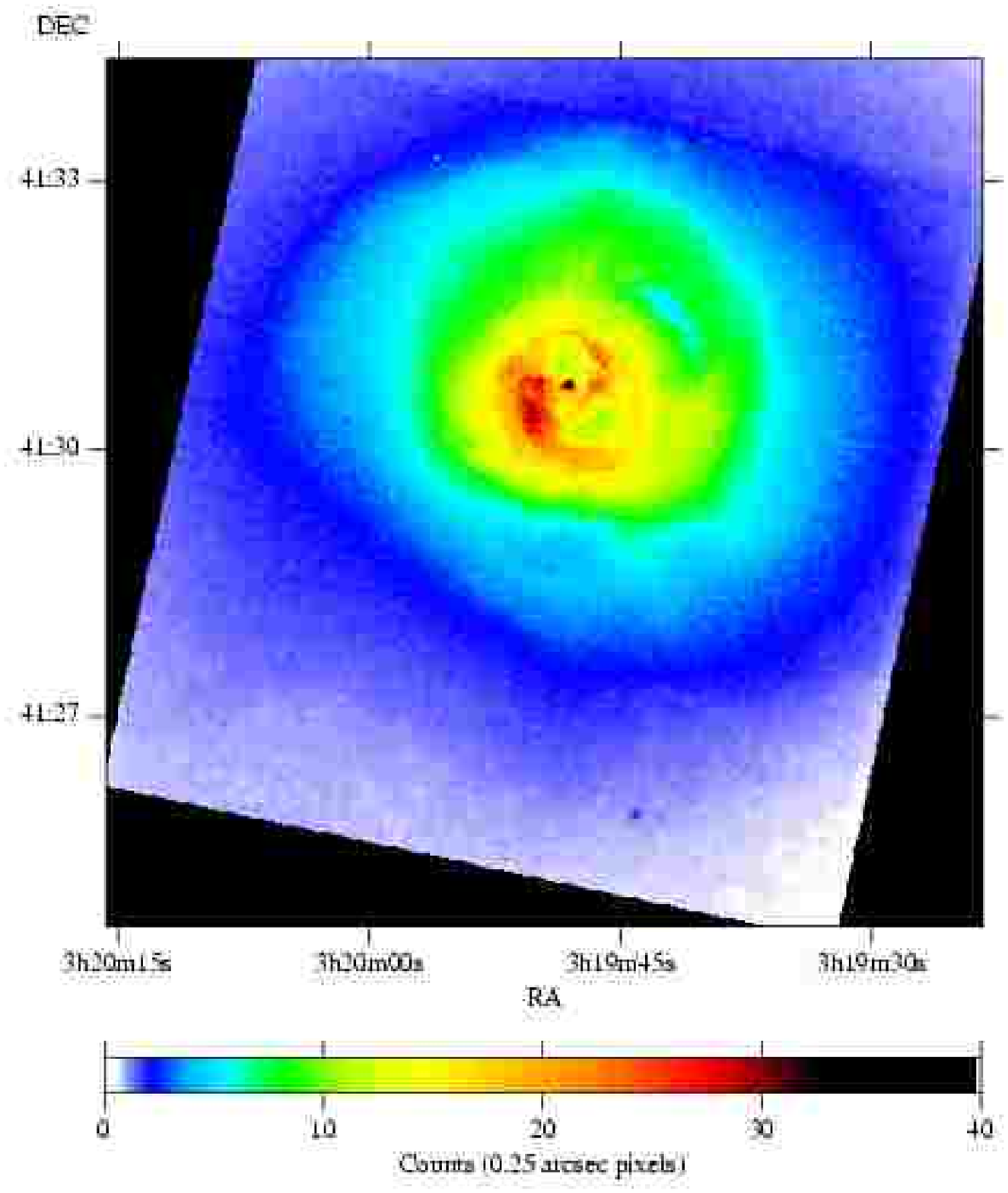}
  \includegraphics[width=\columnwidth]{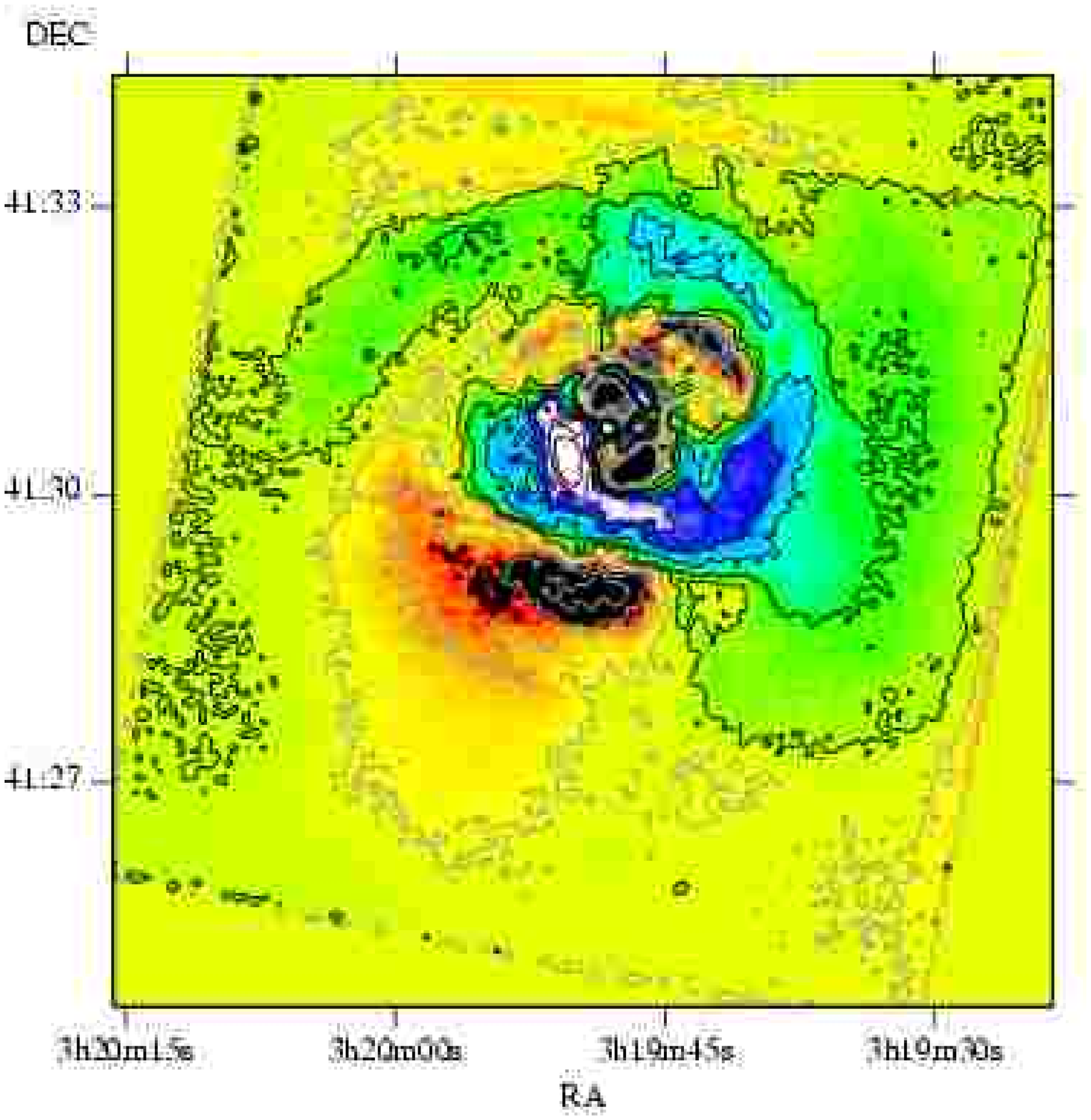}
  \caption{(Top) Full-band image of the cluster. Image was binned to have
    $S/N \sim 17$ in each bin, then smoothed with the \textsc{natgrid}
    natural-neighbour interpolation library. (Bottom) `Deprojected
    image' of the core of the cluster. Areas with excess counts are
    shown in white (with 7 black contours, spaced with square-root
    scaling between 1.2 and 35 excess counts per 0.49 arcsec pixel),
    and areas with a deficit of counts are black (with 5 white
    contours, spaced linearly between -10 and -0.5 counts per pixel).
    The image was smoothed with a 4-pixel width Gaussian.}
  \label{fig:image}
\end{figure}

We show in Fig.~\ref{fig:image}~(top) a full-band image of the cluster
(detailed images of the central region of the image can be seen in
Fabian et al 2003a).  The data were extracted using 0.25 arcsec bin
(0.5 pixel bins), corrected with an exposure map, and rebinned using
the bin-accretion algorithm of Cappellari \& Copin (2003) with a
signal to noise ratio ($S/N$) of $\sim 17$ in each bin (about 290
counts). The image was then smoothed using the \textsc{natgrid}
natural-neighbour interpolation
library\footnote{http://ngwww.ucar.edu/ngdoc/ng/ngmath/natgrid/nnhome.html}.
To highlight the azimuthal variation of the surface brightness, we
show a `deprojected image' of the same region in
Fig.~\ref{fig:image}~(bottom), created by subtracting from each pixel
the projected contribution from shells outside that pixel.

\section{Projected temperature structure}
\label{sect:temperature}
\begin{figure}
  \includegraphics[width=\columnwidth]{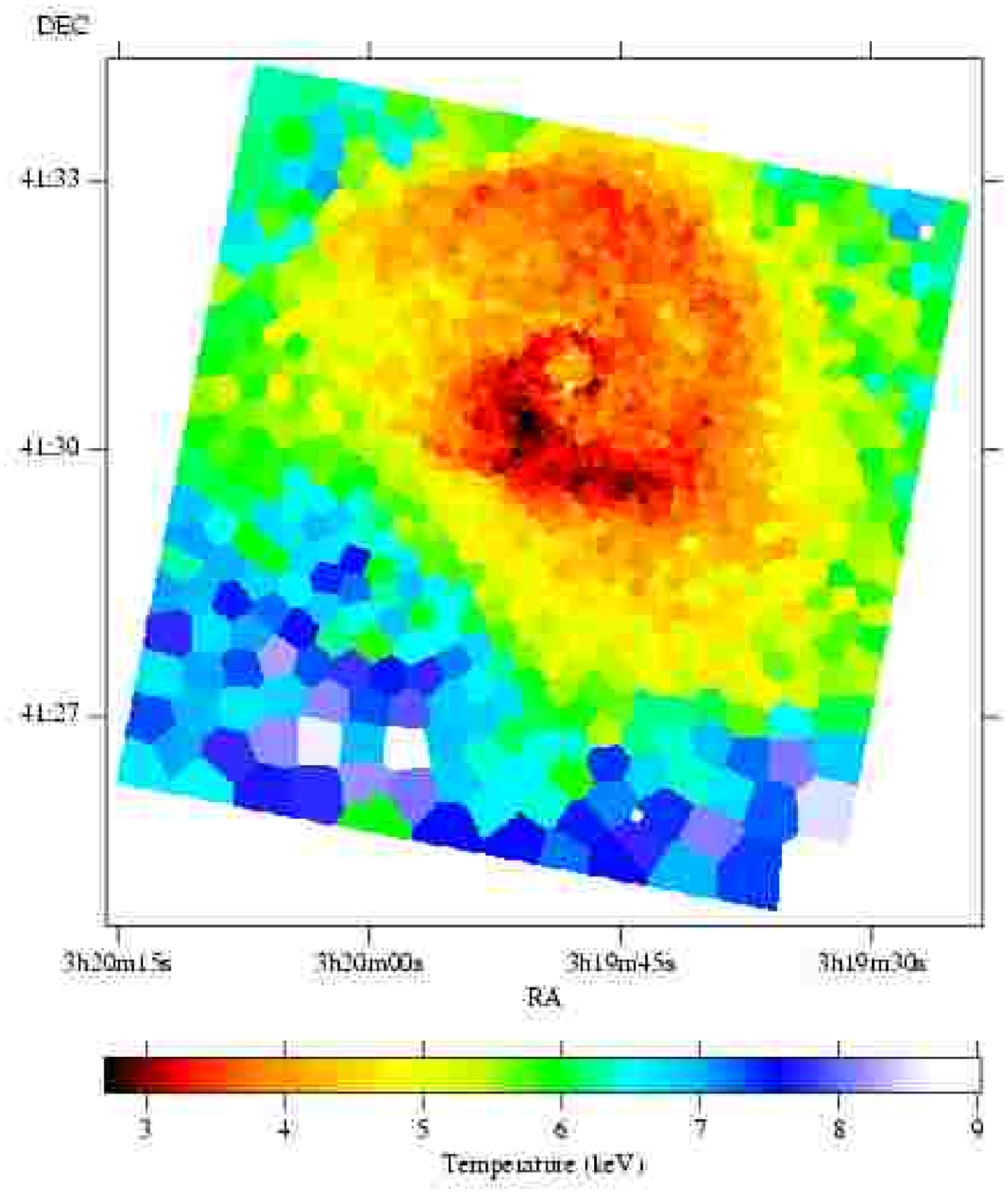}
  \includegraphics[width=\columnwidth]{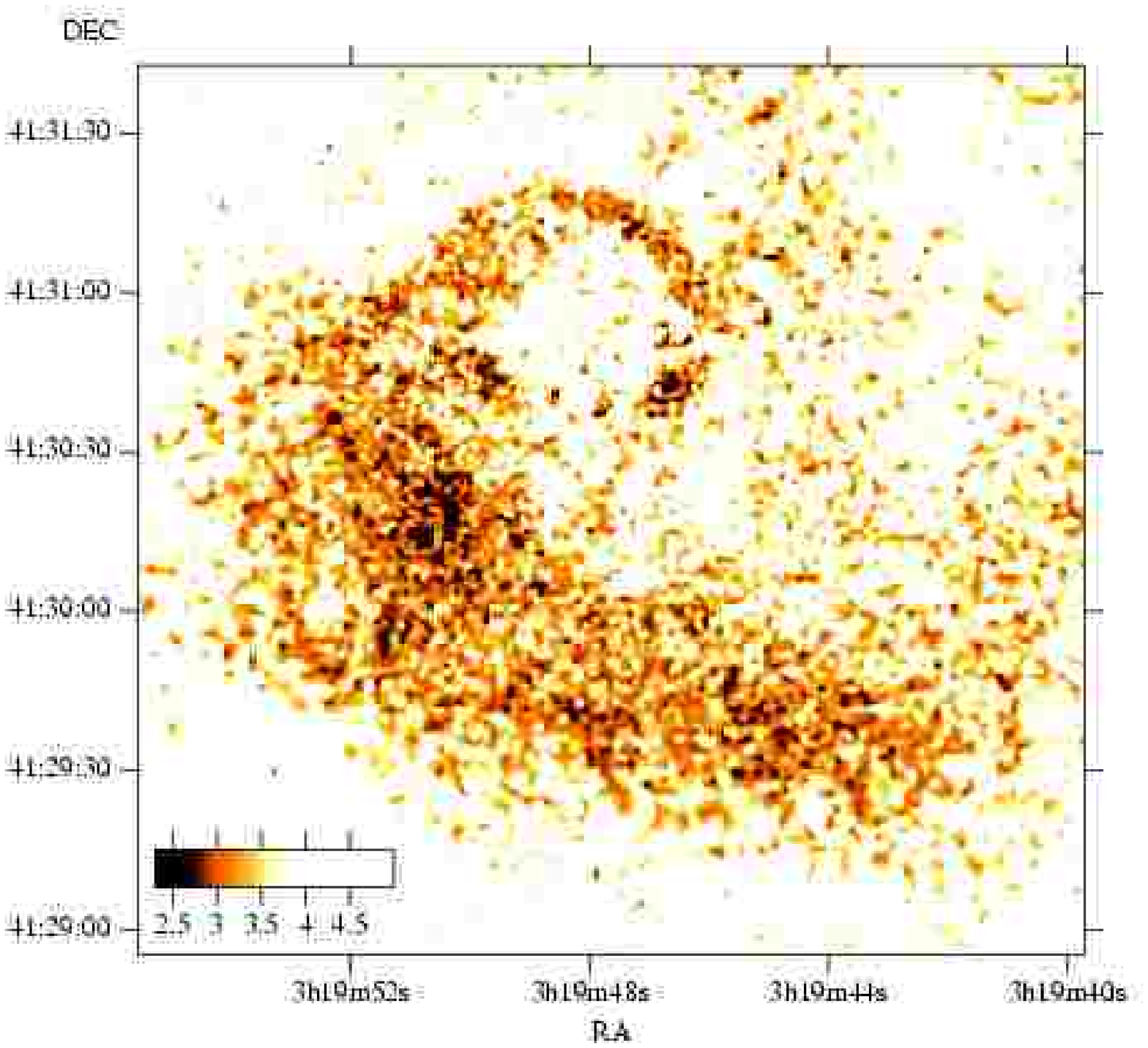}
  \caption{(Top) Projected emission-weighted temperature map of the core of
    the cluster, generated from fitting spectra from the $S/N \sim
    100$ map. The uncertainties increase from 3-4 per~cent in the
    centre, to 8-9 per~cent in the hottest regions. (Bottom) Detail
    from the central region of the cluster, created from the $S/N \sim
    19$ map, and smoothed with a Gaussian of 0.49 arcsec.}
  \label{fig:T_map}
\end{figure}

Fig.~\ref{fig:T_map} shows a detailed image of the emission-weighted
projected temperature structure of the cluster (see also Fig.~6 in
Fabian et al 2003a for a smoothed version).

The data were binned spatially into 1422 regions using the
bin-accretion algorithm of Cappellari \& Copin (2003), with $S/N \sim
100$ ($\sim 10,000$ counts between 0.5 and 7.5 keV). The generated
spectra were fit with a single-temperature \textsc{mekal} model
absorbed by a \textsc{phabs} model (Balucinska-Church \& McCammon
1992), between 0.6 and 8 keV, with variable absorption, abundance and
temperature in each bin, fixing the redshift to the appropriate value.
The results are shown plotted in Fig.~\ref{fig:T_map}~(top). Obvious
point sources were excluded (See Table \ref{tab:ptsources}), shown as
white circles in the temperature map (these sources have been cut out
from all the maps showing the results of spectral fitting in this
paper).

\begin{table}
  \centering
  \begin{tabular}{lll}
    RA (J2000) & Dec (J2000) & Radius (arcmin) \\ \hline
    03:19:26.79 & +41:32:25.7 & 0.082  \\
    03:19:33.76 & +41:29:56.2 & 0.0246 \\
    03:19:37.31 & +41:29:59.1 & 0.0246 \\
    03:19:43.97 & +41:33:04.6 & 0.0328 \\
    03:19:44.10 & +41:25:53.7 & 0.0738 \\
    03:19:48.09 & +41:31:01.7 & 0.0246 \\
    03:19:48.17 & +41:30:42.5 & 0.041  \\
    03:19:56.11 & +41:33:15.5 & 0.0328 \\
    03:20:01.45 & +41:31:27.2 & 0.0328 \\ \hline
  \end{tabular}
  \caption{Excluded point sources, sorted by RA.}
  \label{tab:ptsources}
\end{table}

We created a version of the map with higher spatial resolution, the
central region shown in Fig.~\ref{fig:T_map}~(bottom). The
temperatures in this map have large uncertainties, but it shows the
morphology well.  To create the high resolution map, the data were
again spatially binned with $S/N \sim 19$ ($\sim 360$ counts per
spectrum between 0.5 and 7.5 keV), producing 37,811 spatial regions,
from which spectra were extracted. Each spectrum was fit with the same
model, but using a fixed abundance and absorption with the appropriate
values obtained from the $S/N \sim 100$ map, only allowing the
temperature and normalisation to vary (This is similar to the
procedure used by Schmidt et al 2002). The spectra were fit using $C$
statistics (Cash 1979) rather than $\chi^2$ statistics, a fit
statistic more appropriate in the low signal-to-noise regime.  We fit
the spectra between 0.6 and 5 keV. Instead of calculating individual
responses and backgrounds for each spectrum, we used the appropriate
ones corresponding to the nearest $S/N \sim 100$ bin.

There is a considerable level of apparent structure in the temperature
distribution. For instance there are number of cooler clumps in the
region of bright emission to the south-east and south of the innermost
southern radio lobe. If we fit the spectrum of the largest cool clump
at (03:19:50.8, +41:30:13) then we obtain a temperature of $2.4 \pm
0.2$ keV.  There are also some hot features in the outer part of the
cluster, for example there is a $\sim 20$ arcsec radius region near
(03:20:03.4, +41:26:59) consisting of a clump of hot bins.  The best
fitting temperature of that region is $9.0 \pm 0.55$ keV. An
additional clump near (03:19:37.9, +41:25:37) has a best fitting
temperature of $10.3^{+1.5}_{-1.4}$ keV.  Comparing the temperature
structure against the locations of galaxies, there appear to be a
number of galaxies which appear to lie along the low temperature swirl
around the core. Many of the smallest scale (few arcsec) structures
seen in the detailed temperature map (Fig.~\ref{fig:T_map}~[bottom])
are likely to be noise, but coherent structures seen in the X-ray
surface brightness maps from the brightest part of the cluster
suggests there are likely to be temperature variations on this scale.
Examination of maps with intermediate resolution reveals structures
cooler than their surroundings by $\sim 0.6$~keV on scales of $\sim
2.5$ arcsec (1~kpc).

\begin{figure*}
  \begin{tabular}{ll}
  \includegraphics[width=0.82\columnwidth]{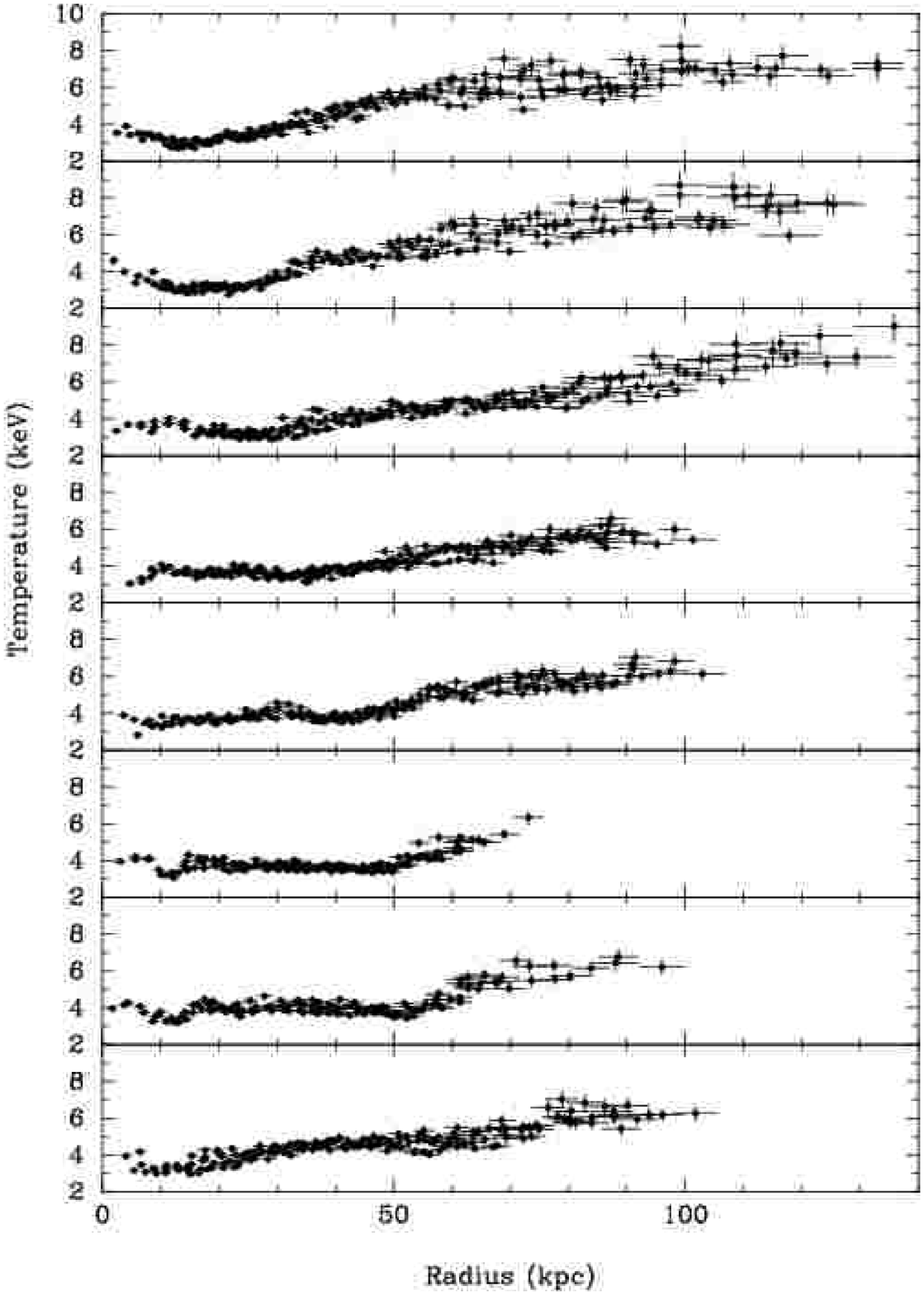}
  &
  \includegraphics[width=0.82\columnwidth]{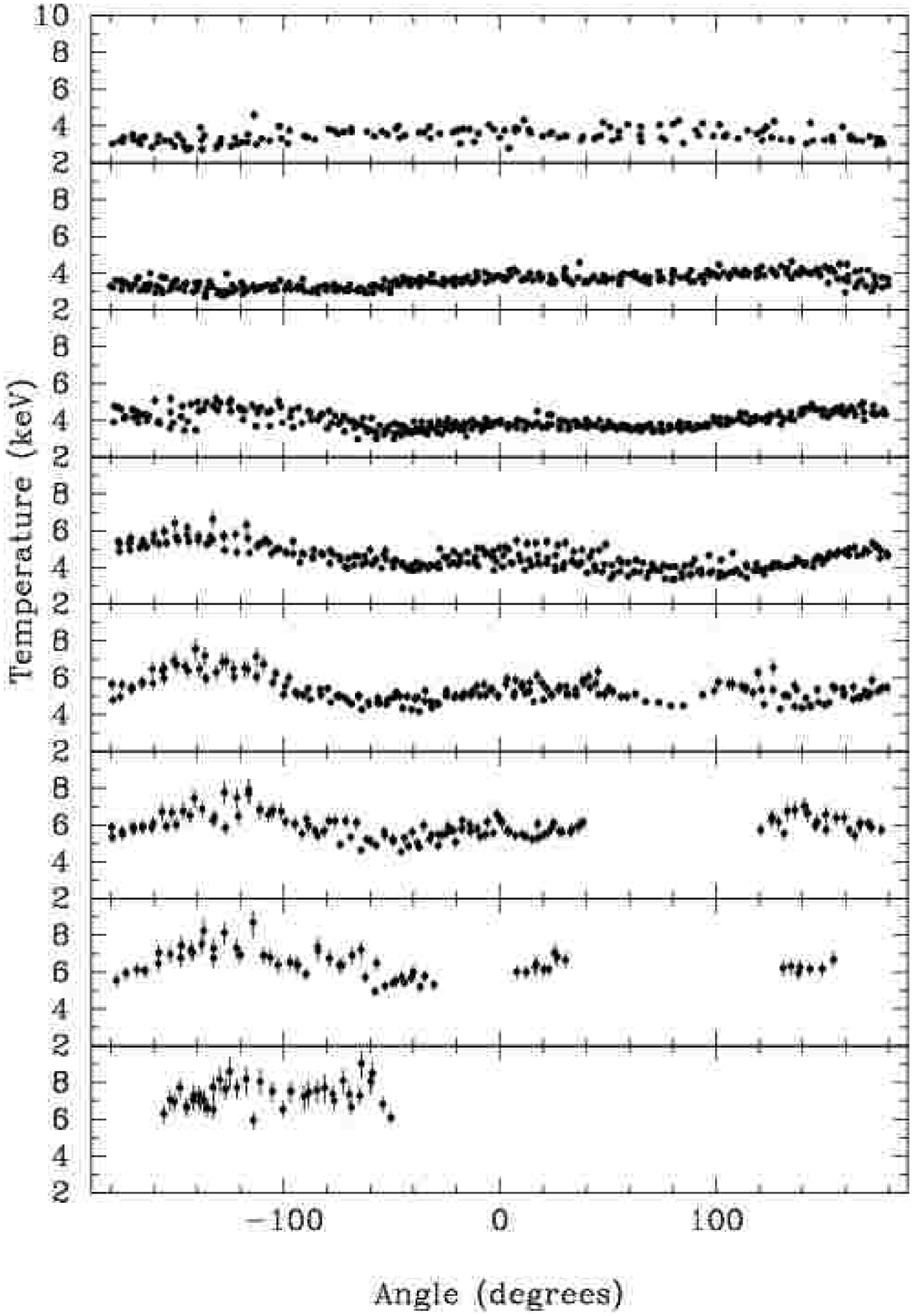}
  \end{tabular}
  \caption{Left: temperature of each spatial bin in the $S/N\sim 100$
    temperature map as a function of radius (kpc), divided up into
    eight annular sectors at multiples of $45^\circ$. The panels
    progress in a anti-clockwise direction, with the first panel
    spanning between East and South-East. The horizontal error bars
    indicated the RMS radius of the pixels in each bin, the vertical
    error bars indicate the $1\sigma$ uncertainty on the temperature.
    Right: temperature of each spatial bin as a function of angle
    (measured anti-clockwise with $0^\circ$ as the West), divided into
    eight radial annuli. The top panel shows the first annulus (inner
    15 kpc), moving out by 15 kpc for each subsequent panel. The final
    panel also shows all remaining points.}
  \label{fig:T_100_plots}
\end{figure*}

In Fig.~\ref{fig:T_100_plots} we show the radial and angular plots of
the temperatures obtained from the larger scale $S/N \sim 100$ fits.
The plot highlights the cool rims around the regions associated with
the radio lobes, the fairly flat, cool region of emission to the south
of the core, and the temperature increasing in the outer parts. Also
the wave-like ripples in the temperature structure caused by the cool
swirl are visible as a function of angle.

\subsection{Absorption structure}
\begin{figure}
  \includegraphics[width=\columnwidth]{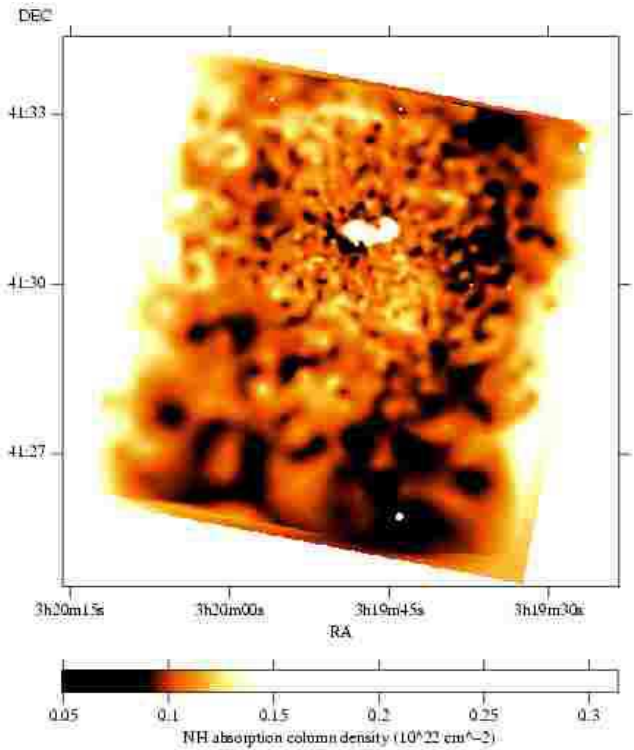}
  \caption{Map of $N_H$ (galactic plus intrinsic plus
    instrumentational), derived using bins with $S/N \sim 100$. The
    data were smoothed with \textsc{natgrid}.}
  \label{fig:NH_map}
\end{figure}

Looking for the signature of absorption in the dataset is a difficult
process. The first problem is the calibration problems below 0.6 keV,
the second the variation of the degradation of the low energy response
with position. It is thought this degradation is caused by build-up of
material on the filter in front of the detector (Marshall et al 2003),
and is more prevalent in those regions where the filter is cooler.  We
show the variation of the best-fitting hydrogen absorption (assuming
solar absorber) as a function of position in Fig.~\ref{fig:NH_map},
binning spectra spatially with $S/N \sim 100$. This map does not
include any correction for the spatial dependence of the QE
degradation, but does include \textsc{corrarf} correction for the
average effect. This map was produced by the same set of spectral fits
as the temperature map plotted in Fig.~\ref{fig:T_map}~(top), and
smoothed with \textsc{natgrid}.

The Galactic column density towards the Perseus cluster is uncertain
as it lies at low galactic latitude. The value obtained from Dickey \&
Lockman (1990) is $N_H \sim 1.5 \times 10^{21} \psqcm$. Most of the
values we measure are below that. The values in our absorption map are
$\sim 2 \times 10^{20} \pcmsq$ greater systematically if we use the
\textsc{contamarf} correction rather than \textsc{corrarf}.

The east-west absorption feature near the centre of the cluster is
associated with the high-velocity system (HVS), which we discuss
separately in Gillmon et al (2003).  Along the east and west edges of
the CCD are regions of high absorption, which may be where there is
build-up of material on the detector.  The level of absorption appears
to decrease from the E-W edges to minima about $1/4$ and $3/4$ of the
distance along the edge, but then there is an increase of absorption
again towards the centre. There are enhancements especially to the
edges of the cool swirling region at the centre, in particular near
(03:19:53.8 +41:32:50) and (03:19:43.7 +41:29:26). In addition there
is an enhancement below the position of the southern outermost radio
lobe. It appears that where gas is cooler we see more absorption.

It is very difficult to say for certain that the features we see are
real. The absorption map produced seems fairly independent of spectral
model and the signal to noise of the spatial binning; variable
abundance and variable abundance cooling flow models produce similar
maps. The effect may however be an artifact of the QE degradation or
another calibration problem, but it is interesting to note that the
features in the absorption map correspond to features in the cluster,
although this would be the case if the spectral model is incorrect. We
stress that the QE degradation is position dependent.

\subsection{Velocity structure}
\label{sect:redshift}
Dupke \& Bregman (2001) pioneered the use of \emph{ASCA} GIS spectra
to measure velocities in the Perseus cluster. The region of the
cluster we can probe with this observation is much smaller, so we
cannot compare our results with theirs.

Our initial attempts to measure the velocity structure were thwarted
by a systematic caused by the response matrix energy binning. The
redshift distribution was discretised into values separated by the
response matrix energy bin size. To avoid this problem we used a width
of 0.5~eV in the response matrices used in this section, rather than
the usual value of 10~eV. In addition early versions of the correction
of the time dependence of the gain did not completely correct the
problem, due to a bug in \textsc{mkrmf}. In this section we use the
November 2003 release of \textsc{corr\_tgain} rather than the Summer
2003 release.

We extracted spectra from bins with $S/N \sim 500$ (about $2.5\times
10^5$ counts). The spectra were fitted using a \textsc{mekal} model
between 3 and 8~keV, with the temperature, abundance, normalisation
and redshift free parameters. We used $C$~statistics on unbinned
spectra. This spectral region was used as the Fe-K lines are the best
indicators of velocity structure, as the centroid of the Fe-L complex
is more difficult to determine and its spectral region is contaminated
by emission from other elements.

\begin{figure}
  \includegraphics[angle=-90,width=\columnwidth]{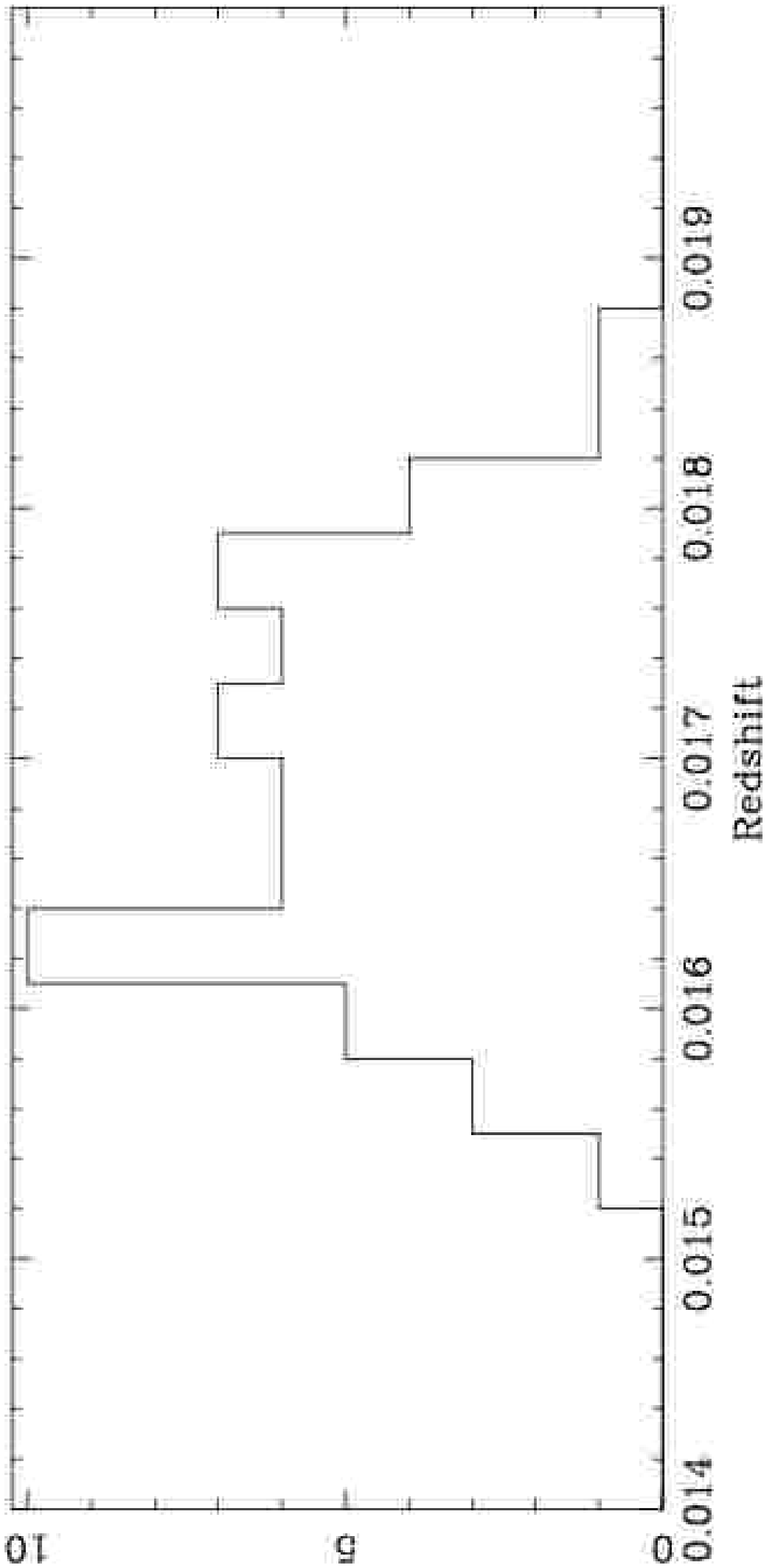}
  \\[3mm]
  \includegraphics[angle=-90,width=\columnwidth]{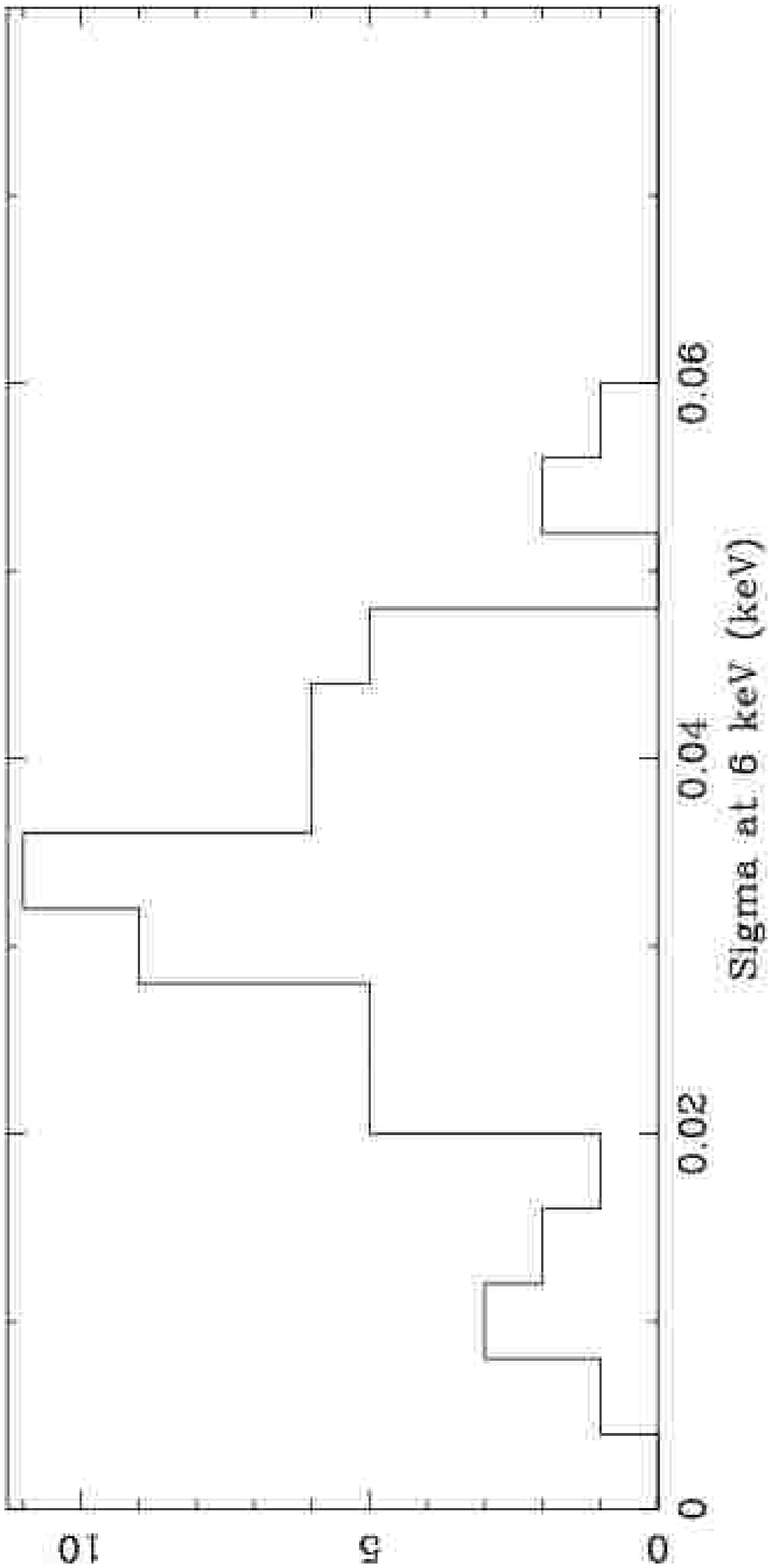}
  \\[3mm]
  \caption{(Top) Distribution of the best-fitting redshifts in each
    bin. The mean 1$\sigma$ error on each value is $6.6 \times
    10^{-4}$. (Bottom) Distribution of the \textsc{gsmooth} smoothing
    $\sigma$, measured at 6 keV. The mean error on each value is $\sim
    0.01\keV$.}
  \label{fig:redshift}
\end{figure}

\begin{figure}
  \includegraphics[width=\columnwidth]{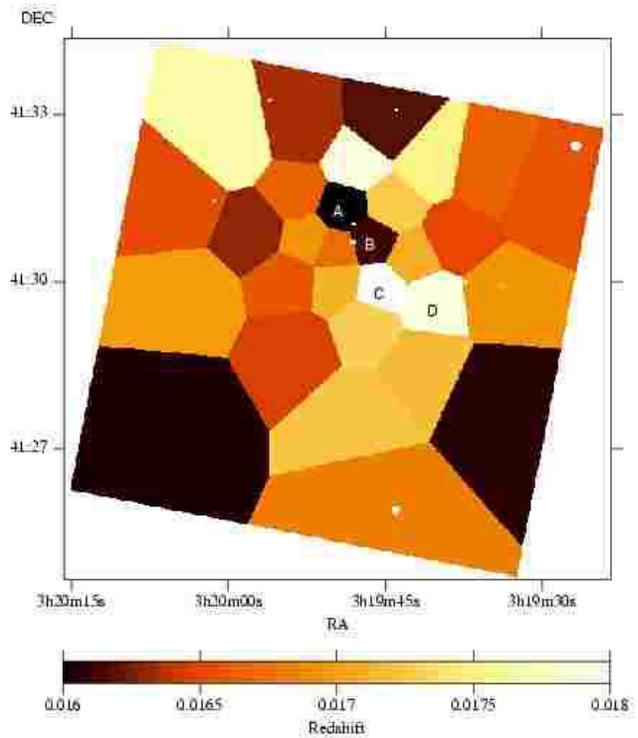}
  \caption{Redshift distribution generated from a $S/N \sim 700$
    map. Typical errors are $\sim 0.0005$. The bins marked $A$, $B$,
    $C$ and $D$ have redshifts of $0.0160 \pm 0.0004$, $0.0161 \pm
    0.0003$, $0.0180 \pm 0.0005$ and $0.0179 \pm 0.0003$,
    respectively.}
  \label{fig:redshiftmap}
\end{figure}

The mean redshift found was $0.0169 \pm 0.0001$
(Fig.~\ref{fig:redshift}~[top]), with a standard deviation,
$\sigma_z$, of $7.7\times10^{-4}$, which is close to the mean error on
each redshift, $6.6 \times 10^{-4}$. $\sigma_z$ corresponds to a
velocity of $230 \kmps$. The distribution of velocities is compatible
with there being a single component. The smallest scale we are
sensitive to is $\sim 30$~arcsec (12~kpc). However, with higher signal
to noise ($S/N \sim 700$, about $4.9 \times 10^5$ counts), we appear
to find significant structure in the redshift map
(Fig.~\ref{fig:redshiftmap}). The bins marked $A$ and $B$ are moving
towards us with respect to regions $C$ and $D$. The relative velocity
of the two pairs is $\sim 570\kmps$. The linear structure is
reminiscent of a galactic rotation curve, but could also be due to the
radio jets from the centre (although there is no simple agreement with
the configuration of the jets).

\begin{figure}
  \includegraphics[angle=-90,width=\columnwidth]{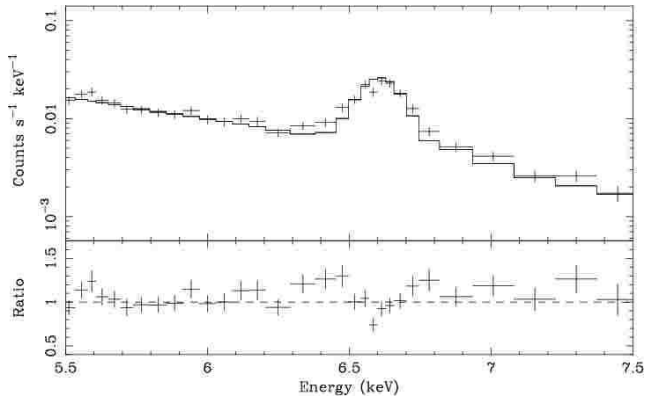}
  \caption{Detail around the Fe-K lines to demonstrate that a
    \textsc{mekal}
    model is too narrow to fit the data without smoothing. The data have
    been rebinned to have a signal-to-noise of 10 in each bin,
    grouping together a maximum of 30 channels.}
  \label{fig:specunsmooth}
\end{figure}

In addition to this simple analysis, we fitted the same \textsc{mekal}
model convolved with a Gaussian (the \textsc{gsmooth} model in
\textsc{xspec} with an energy power index of 1), to test for a
velocity spread on the Fe-K lines.  Neither the \textsc{apec} nor
\textsc{mekal} models include thermal broadening effects on the
emission lines. If there is no turbulence, a 4~keV plasma will show a
broadening of $\sim 2.6$~eV for the Fe-K lines.  Unexpectedly, we
found a significant extra line width in the spectra from most bins
(Fig.~\ref{fig:redshift}~[bottom]), with a mean of $32.2 \pm 0.1$~eV
at 6 keV, corresponding to $1600\kmps$. The $1 \sigma$ error on each
value was $\sim 10$~eV. To demonstrate the strength of the signal we
show a spectrum in Fig.~\ref{fig:specunsmooth} from one of the bins
fitted with a model without smoothing. The residuals in the wings of
the line are easily visible. We also checked that \textsc{apec} models
gave similar deviations. We note that this result may be instrumental,
for example due to charge transfer inefficiency of the detector.

We note that the mean redshift found is not the nominal redshift of
the cluster. The offset corresponds to a velocity of $420 \kmps$. It
is possible that this may be caused by a systematic gain shift,
although we have corrected the gain in the data using
\textsc{corr\_tgain}.

\section{Abundance structure}
\subsection{Fixing abundances at solar ratios}
\label{sect:solarabun}
\begin{figure}
  \includegraphics[width=\columnwidth]{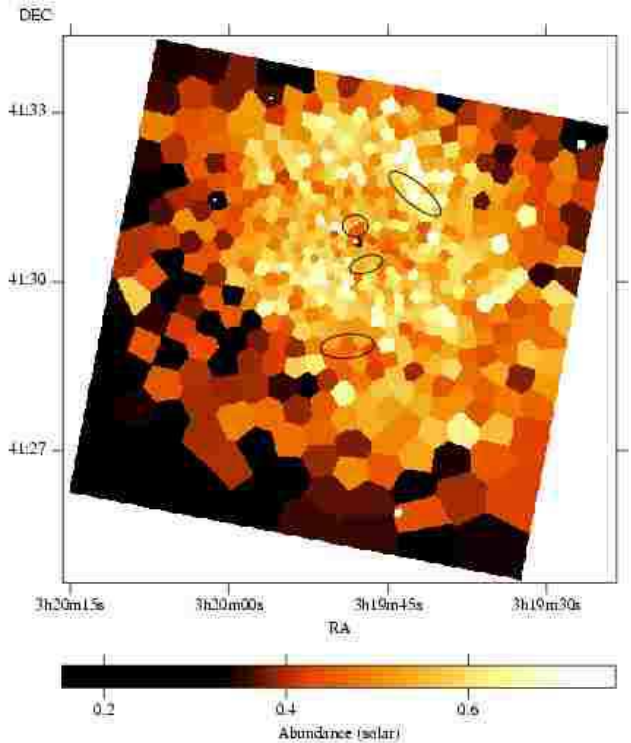}
  \caption{Map of abundance using bins with $S/N \sim 150$. The
    uncertainties on the abundances rise from about $0.04$ in the
    centre to about $0.08 \Zsun$ in the outer regions. The ellipses
    mark the approximate positions of the radio lobes.}
  \label{fig:Z_map}
\end{figure}

Fig.~\ref{fig:Z_map} shows an emission-weighted abundance map produced
by fitting a single temperature model (the same model as in Section
\ref{sect:temperature}) to spectra extracted from a $S/N \sim 150$
binned map.  We did not allow redshifts to vary here, but we do in
Section \ref{sect:indivabun}.  The abundances were measured assuming
solar abundance ratios, although the elemental abundance we are most
sensitive to is Fe. The abundance map shows a substantial amount of
structure. The overall pattern is that the abundance appears to rise
to a peak around 100 arcsec (37 kpc) to the NW and 60 arcsec to the
SE.  There is an abundance enhancement in the outer SW part of the
map. Inside this radius the abundance drops down, especially to the W
side of the cluster.

\begin{figure}
  \centering
  \includegraphics[angle=-90,width=\columnwidth]{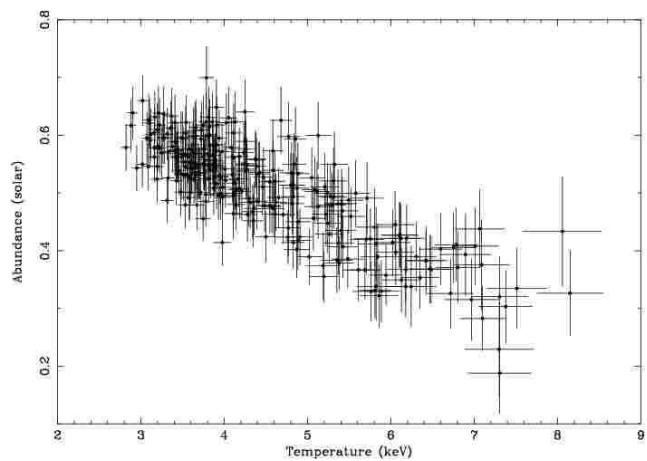}
  \caption{Plot of abundance against temperature for each bin,
    generated from a $S/N \sim 225$ map. The same relation exists for
    the $S/N \sim 150$ map, but with larger uncertainties.}
  \label{fig:T_Z_scatter}
\end{figure}

The morphology of the abundance map follows the temperature map
(Fig.~\ref{fig:T_map}) closely. Those regions with low temperatures
generally have high abundances. We can quantify this somewhat by
plotting the abundance of bins against their temperature, which is
shown in Fig.~\ref{fig:T_Z_scatter}. The uncertainties on the
temperatures and abundances for a particular bin are correlated, but
the correlation is that uncertainties in abundance increase with
increasing uncertainties in temperature, which is the opposite sense
to the trend in the best-fitting values. By faking a dataset with the
same emission-weighted temperature distribution as the real cluster
(measured from a $S/N \sim 225$ map), but with a constant abundance of
$0.5 \Zsun$ everywhere, and repeating the analysis, we verified there
was no significant systematic effect we could find to account for this
trend. There is a minor systematic effect in the procedure, with a
decrease of $0.05\Zsun$ over the temperature range observed.

There are a number of high-abundance `clumps', for example at
(03:19:40.9 +41:31.21) where $Z = 0.68 \pm 0.05 \Zsun$, (03:19:49.3
+41:29:59) where $Z=0.67 \pm 0.04 \Zsun$.  and (03:19:34.6 +41:29:51)
where $Z=0.70 \pm 0.05 \Zsun$ (its surrounds are at $0.47 \pm 0.02
\Zsun$). Many of these regions have scales of around 20 arcsec.  The
existence of small-scale inhomogeneities in the abundance distribution
was hinted at by Schmidt et al (2002). There is also a low abundance
linear ($Z \sim 0.5 \Zsun$) feature from the core to at least 60
arcsec WSW (this is more easily visible using high $S/N$ binning: see
Fig.~\ref{fig:abun300_1}).

One high abundance peak is associated with the position of the outer
NW radio lobe. The material just within the radius of the lobe has a
mean abundance of $\sim 0.53 \pm 0.01 \Zsun$, whilst the material at
the radius of the lobe or just beyond has an abundance of $0.62 \pm
0.015 \Zsun$.  There may also be an enhancement in abundance
corresponding with the position of the inner SW radio lobe. The gas
along the line of sight of the inner NE lobe does not have an obvious
enhancement, but the material further away in radius is enhanced.
There is some enhanced abundance gas lying outside the outer S radio
lobe.

\begin{figure*}
  \begin{tabular}{ll}
    \includegraphics[width=0.85\columnwidth]{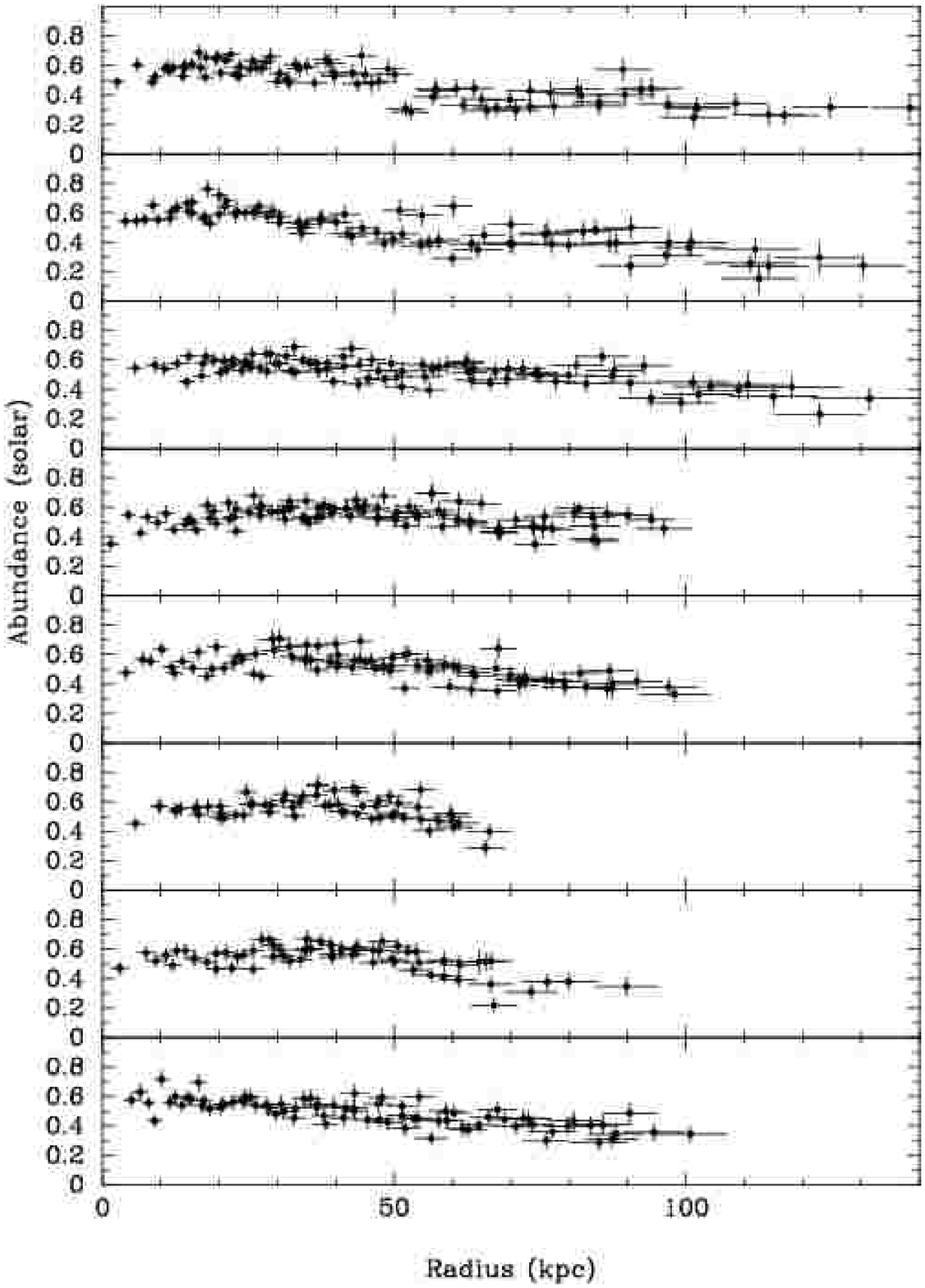}
    &
    \includegraphics[width=0.85\columnwidth]{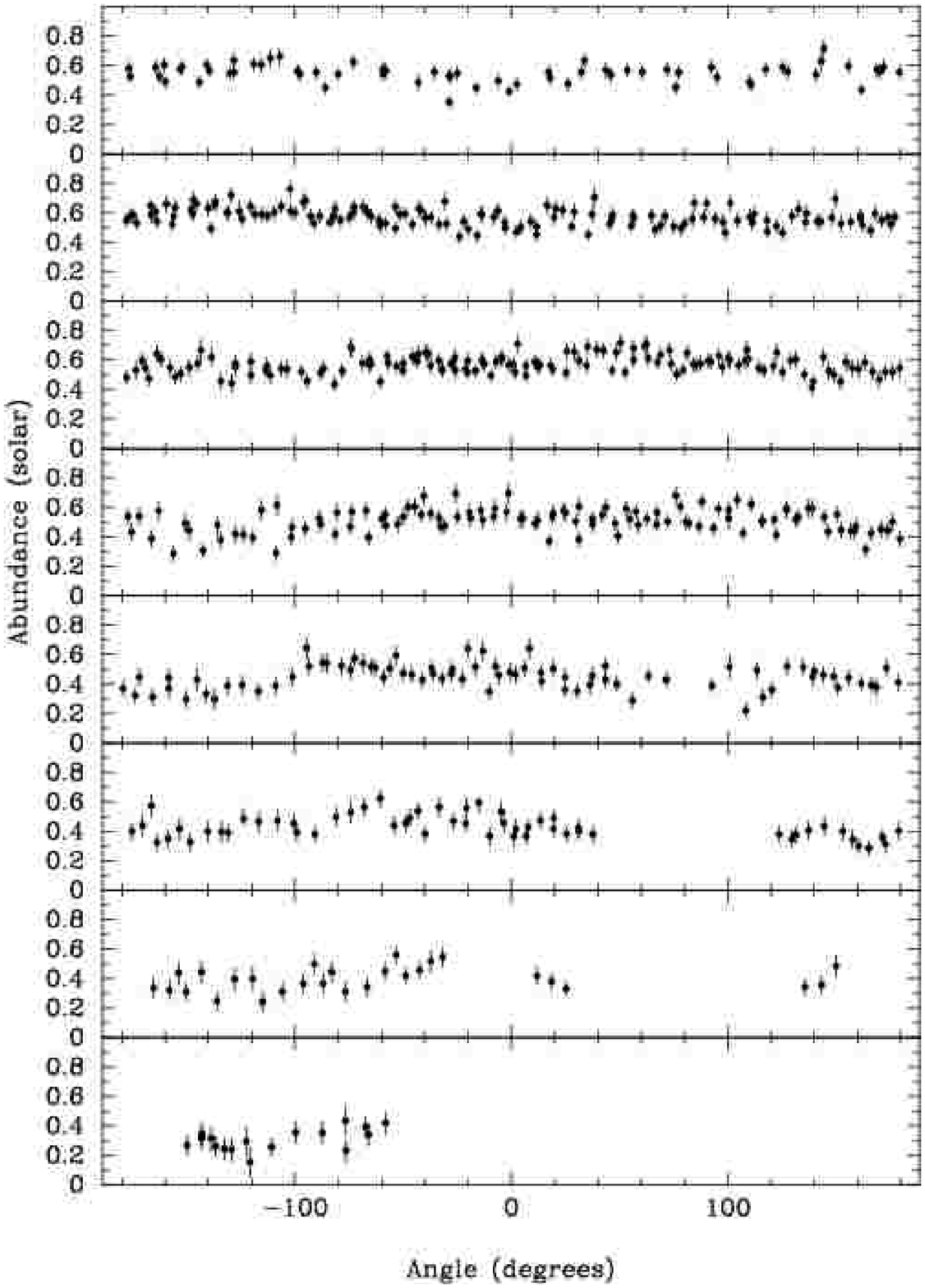}
  \end{tabular}
  \caption{Radial and annular abundance profiles for each bin in an
    abundance map generated from a $S/N \sim 150$ binned image. See
    Fig.~\ref{fig:T_100_plots} for an explanation of the coordinate
    system.}
  \label{fig:Z_150_plots}
\end{figure*}

In Fig.~\ref{fig:Z_150_plots} we plot the abundance as a function of
radius and angle.  The plots, in particular the abundance as a
function of angle, highlights that the metals are not uniformly
distributed.  The abundance dips and peaks by magnitudes of up to
$0.1-0.2\Zsun$.

The abundances we measure may be subject to further systematic errors.
The gas in the cluster may be optically thick to resonantly scattered
lines (e.g. Gilfanov, Sunyaev \& Churazov 1987), however Dupke \&
Arnaud (2001), Churazov et al (2003b), Gastaldello \& Molendi (2003)
observe no evidence for resonant scattering in this cluster.  Another
effect is known as `Fe-bias' (Buote 2000), where one underestimates
the abundance of the emitting gas if it consists of more than one
phase. However we test for the presence of multiple components in
Section \ref{sect:multicomp}, and allow for projection effects in
Section \ref{sect:projection}.

\subsection{Allowing individual abundances to vary}
\label{sect:indivabun}
\begin{figure}
  \includegraphics[angle=270,width=\columnwidth]{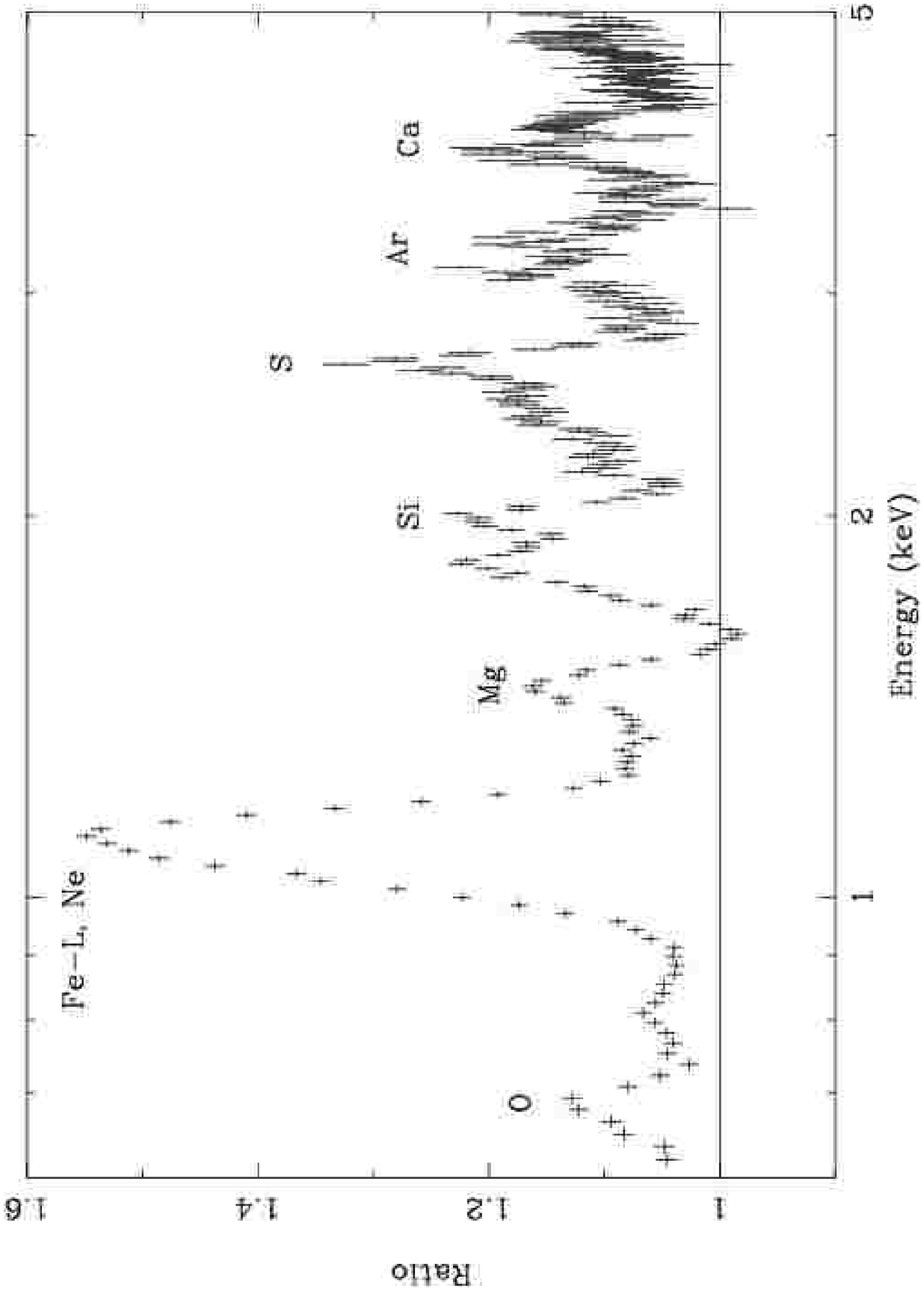}\\[2mm]
  \includegraphics[angle=270,width=\columnwidth]{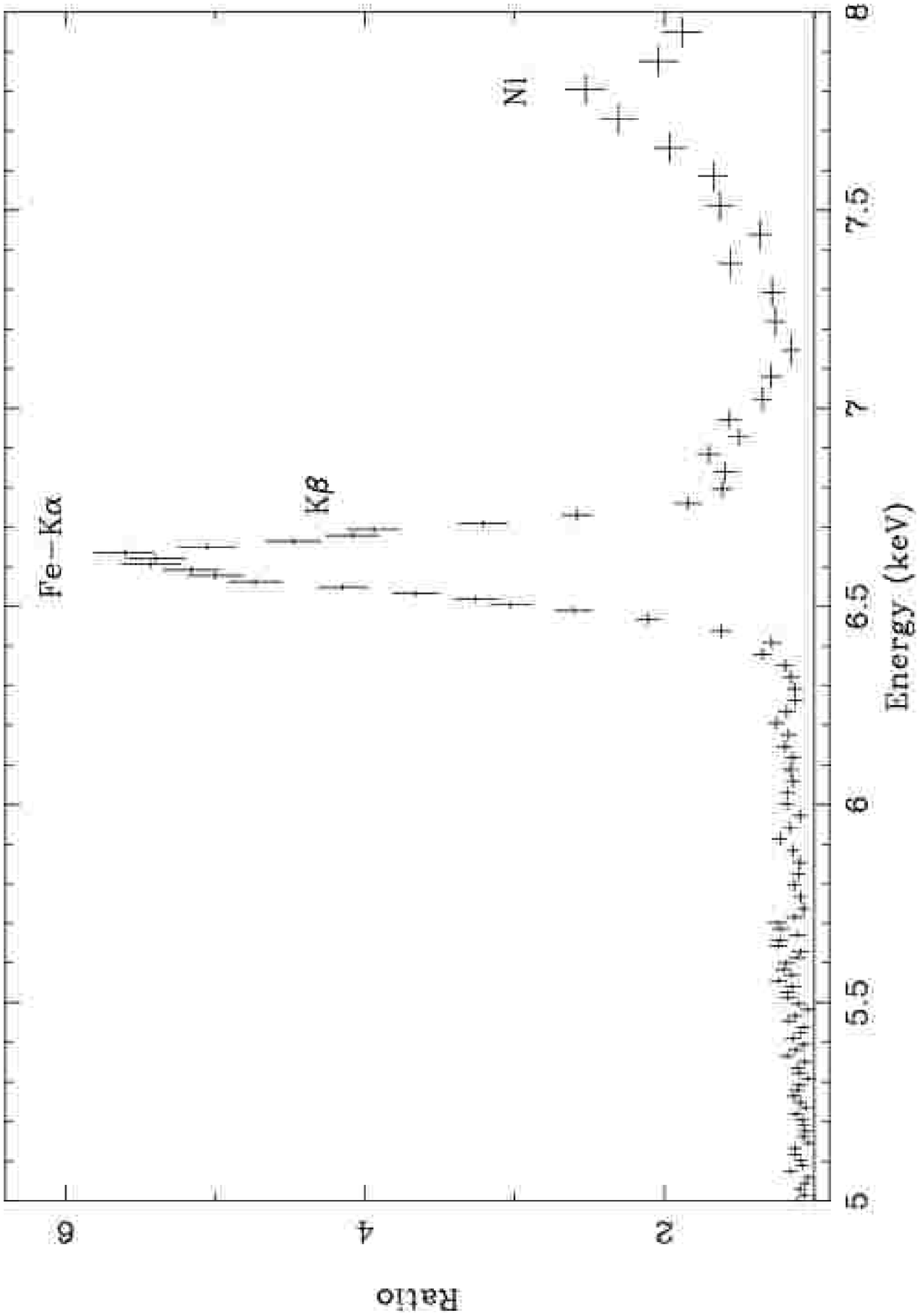}
  \caption{Ratio of a best fitting \textsc{vmekal} model
    (after setting the O, Ne, Mg,
    Si, S, Ar, Ca, Fe and Ni abundances to zero) to a spectrum
    extracted from an annulus from radius 130 to 200 arcsec (26~kpc
    wide). Visible are O, Fe-L, Mg, Si, S, Ar, Ca, Fe-K and Ni lines.
    Both the hydrogenic and helium-like lines of S, Ar and Ca are
    seen.}
  \label{fig:ratioZ}
\end{figure}

By fitting a \textsc{vmekal} or \textsc{vapec} model to the spectra
extracted from each bin we can make maps of individual elemental
abundances. To demonstrate the excellent signal-to-noise of the data,
we show in Fig.~\ref{fig:ratioZ} the contributions to a spectrum by
the various metals (We note that residuals are also seen in
Fig.~\ref{fig:ratioZ} where emission lines due to Ti and Cr are
expected, but only Cr has a plausible strength).  By excluding the
energy range 1.3--2.3 and below 0.6 keV (where we are most uncertain
about the calibration) the fits are not sensitive to Mg or Si
abundances.  The elements we fit for are Fe, Ar, Ca, Ne, O, Ni and S.
Other abundances remain frozen at solar values.  Excluding effects
such as resonant scattering and Fe-bias, the Fe abundance is the most
secure.  O abundances could be problematic as O emission lines lie
close to 0.6 keV, and could be affected by systematics errors near
that limit. Ni and S abundances have larger uncertainties than could
be obtained from the full energy range, as some of their lines are in
the 1.3-2.3 keV region.

\begin{figure*}
  \includegraphics[width=.33\textwidth]{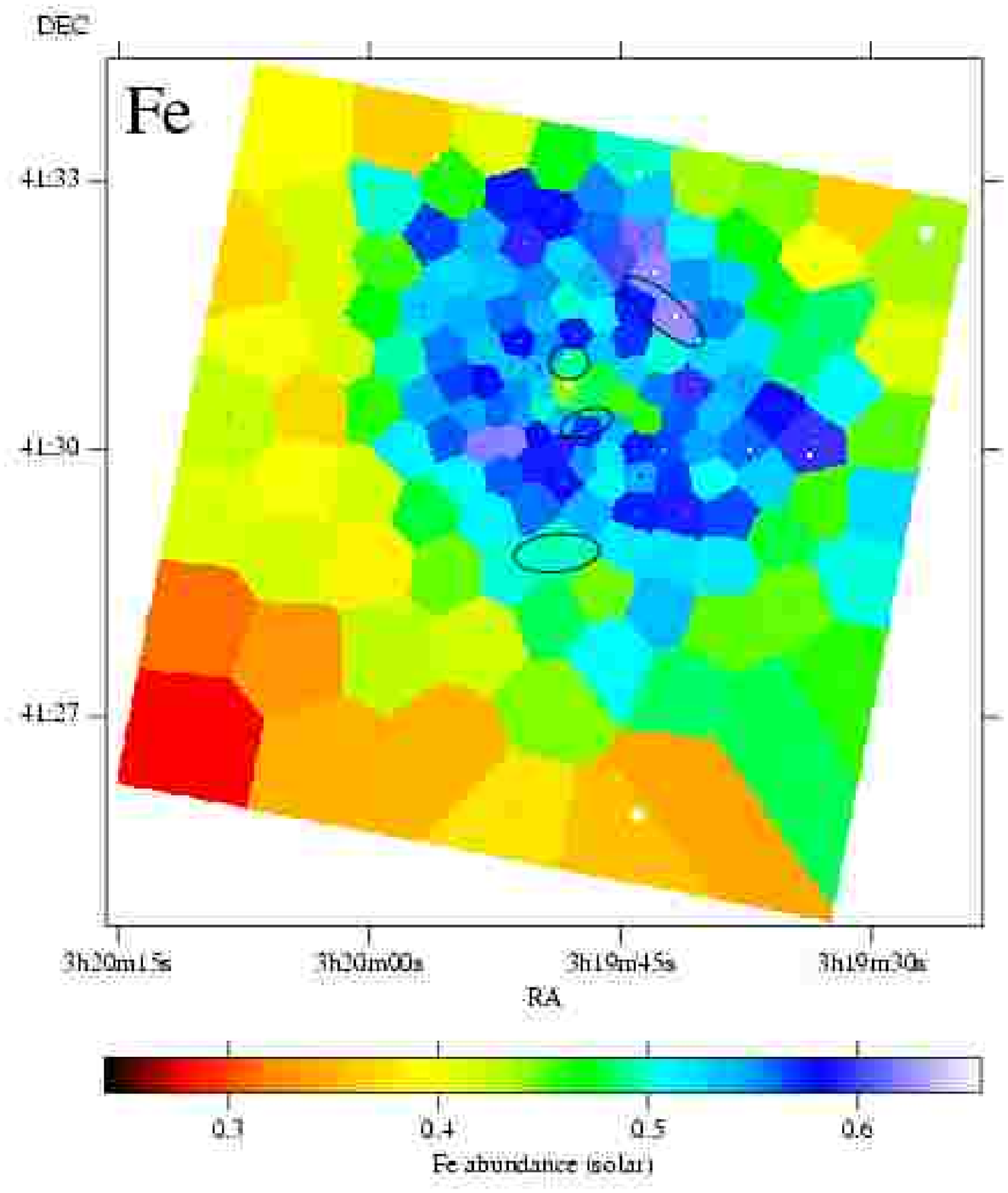}
  \includegraphics[width=.33\textwidth]{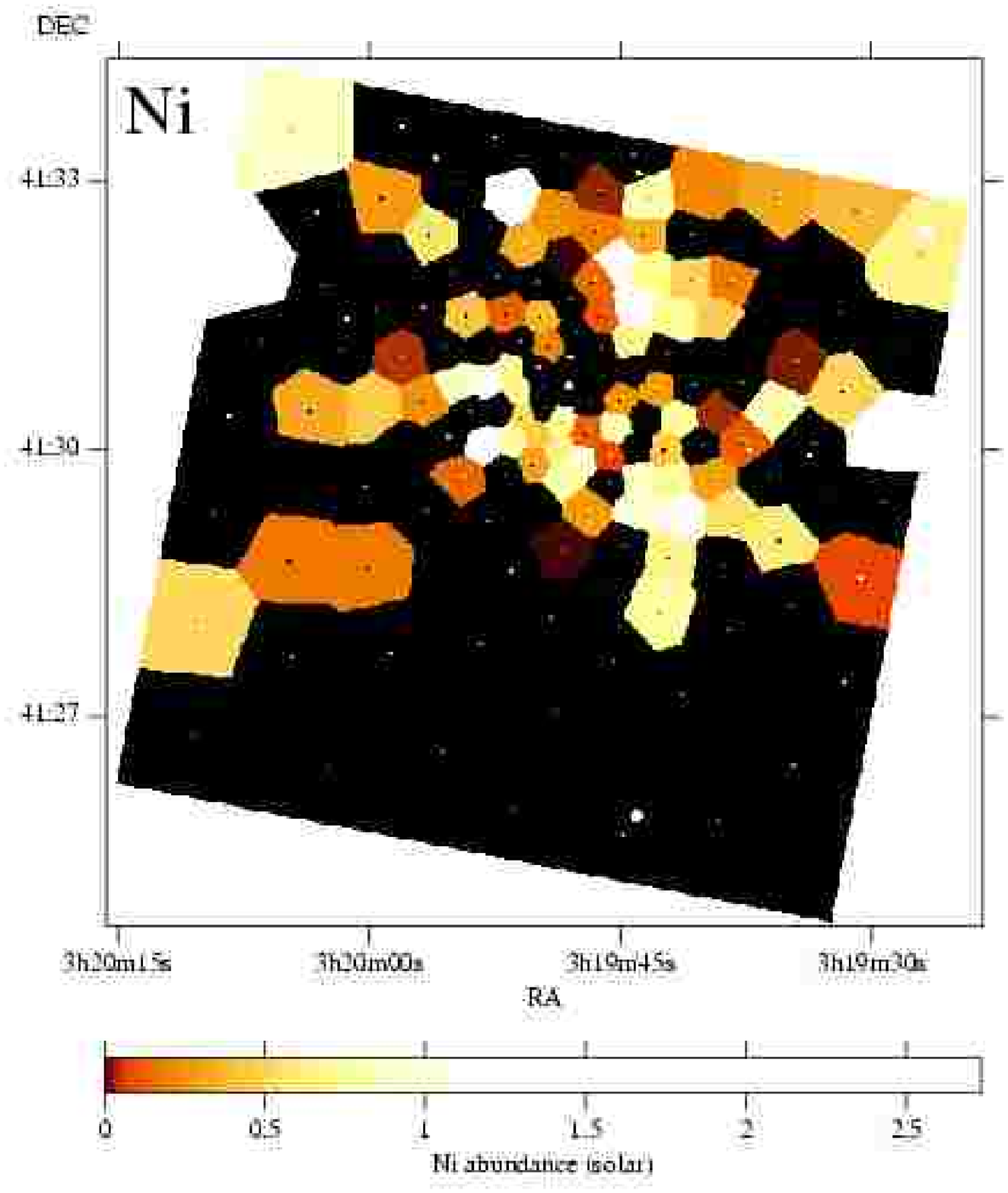}
  \includegraphics[width=.33\textwidth]{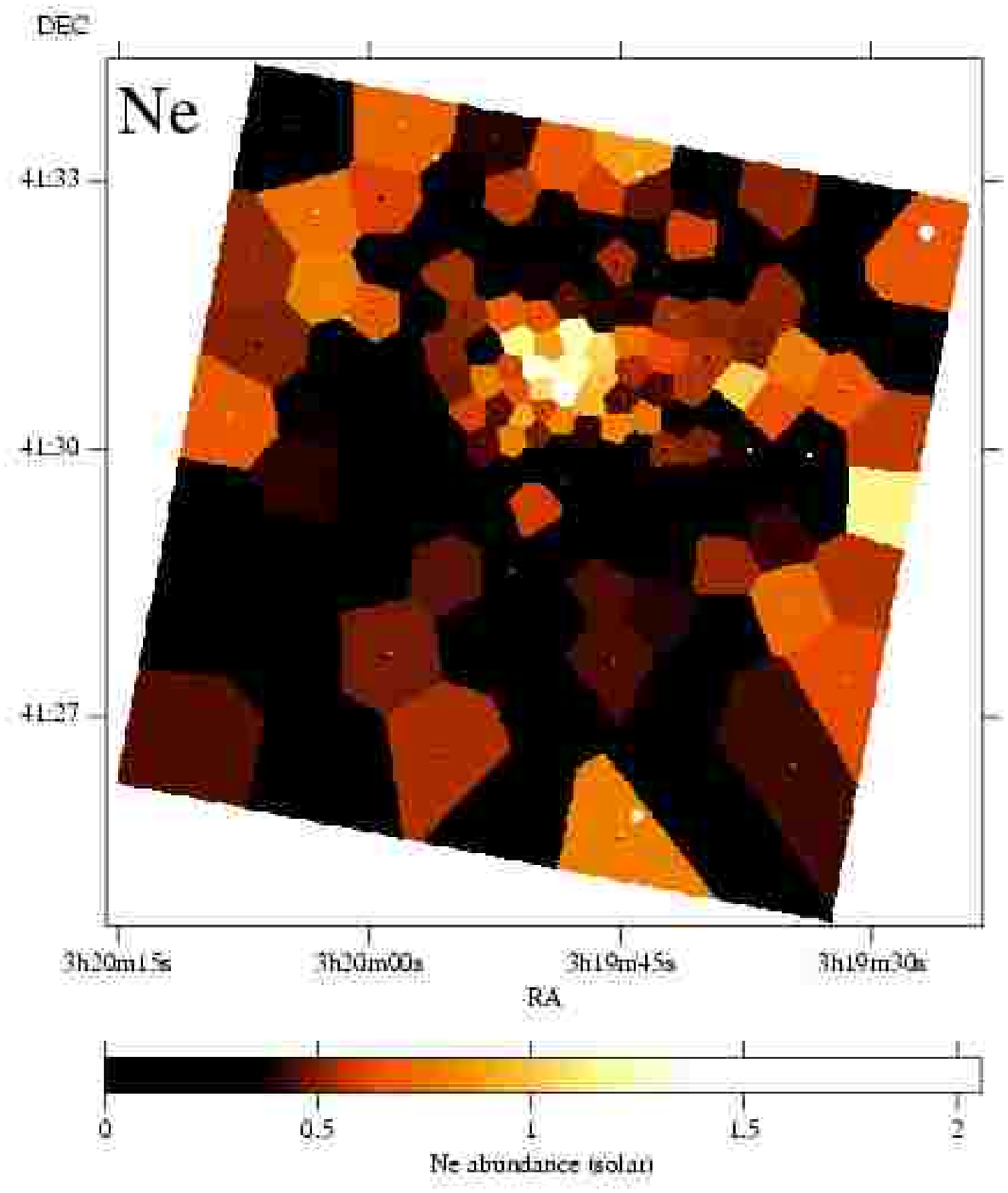}
  \includegraphics[width=.33\textwidth]{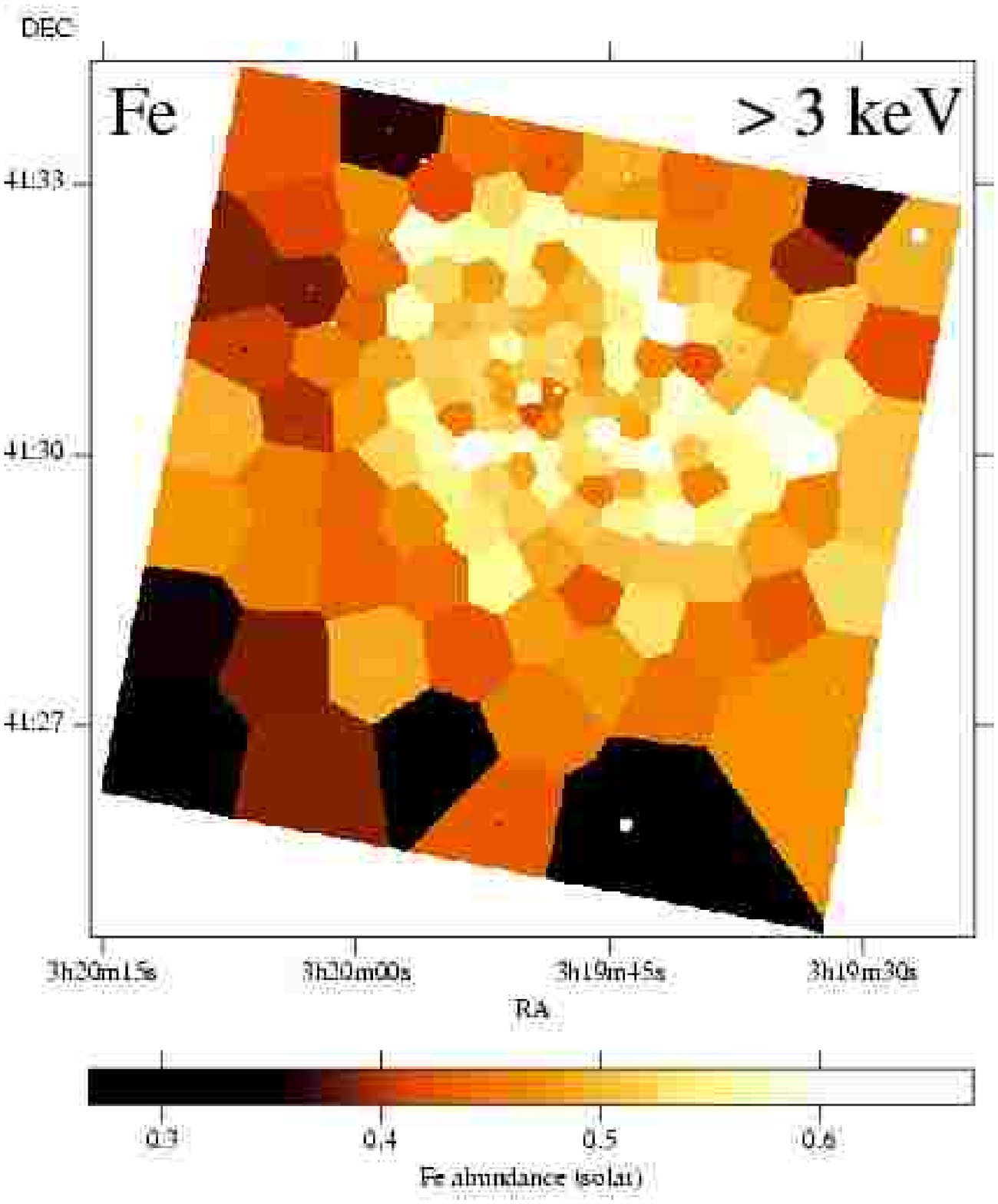}
  \includegraphics[width=.33\textwidth]{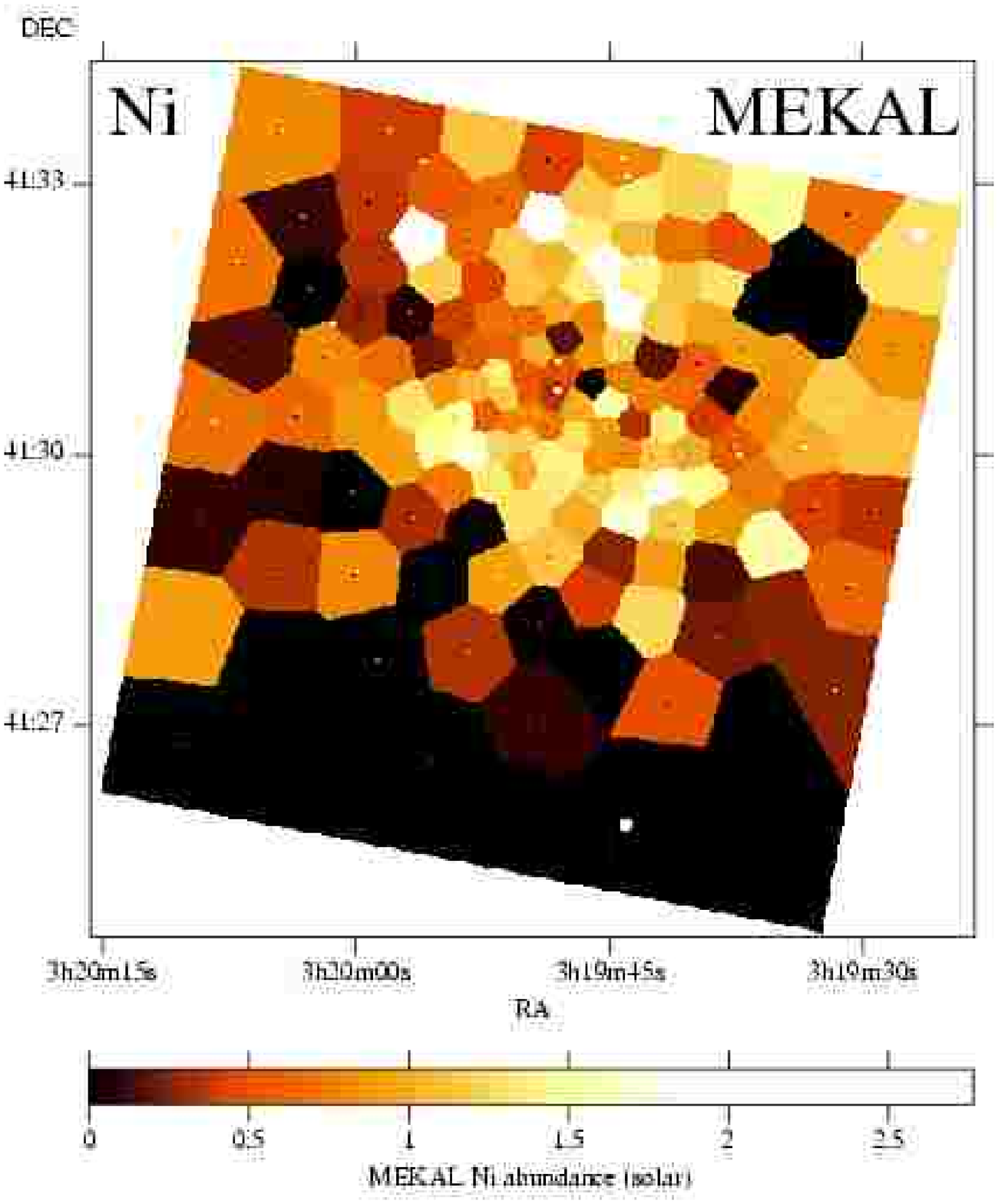}
  \includegraphics[width=.33\textwidth]{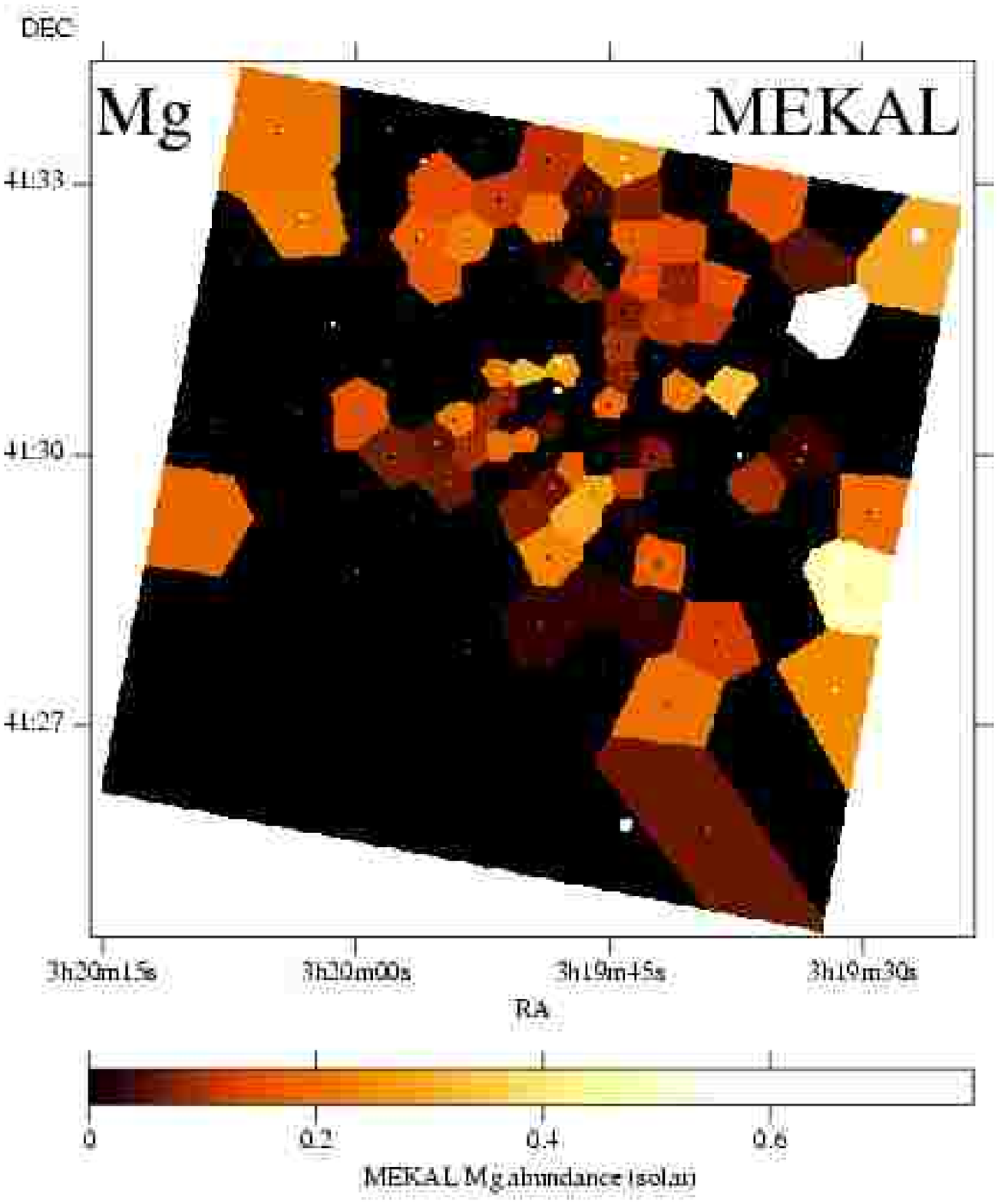}
  \includegraphics[width=.33\textwidth]{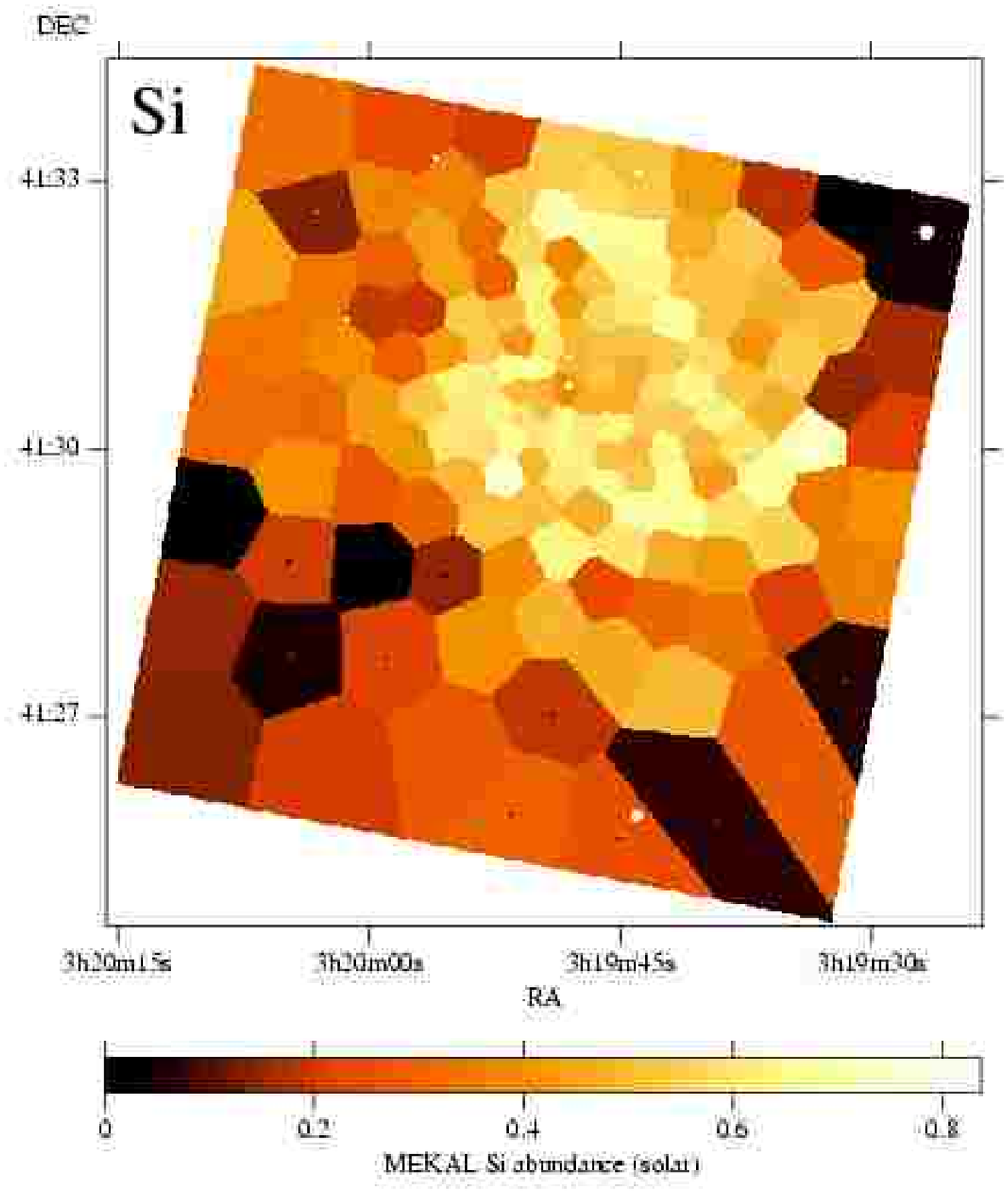}
  \includegraphics[width=.33\textwidth]{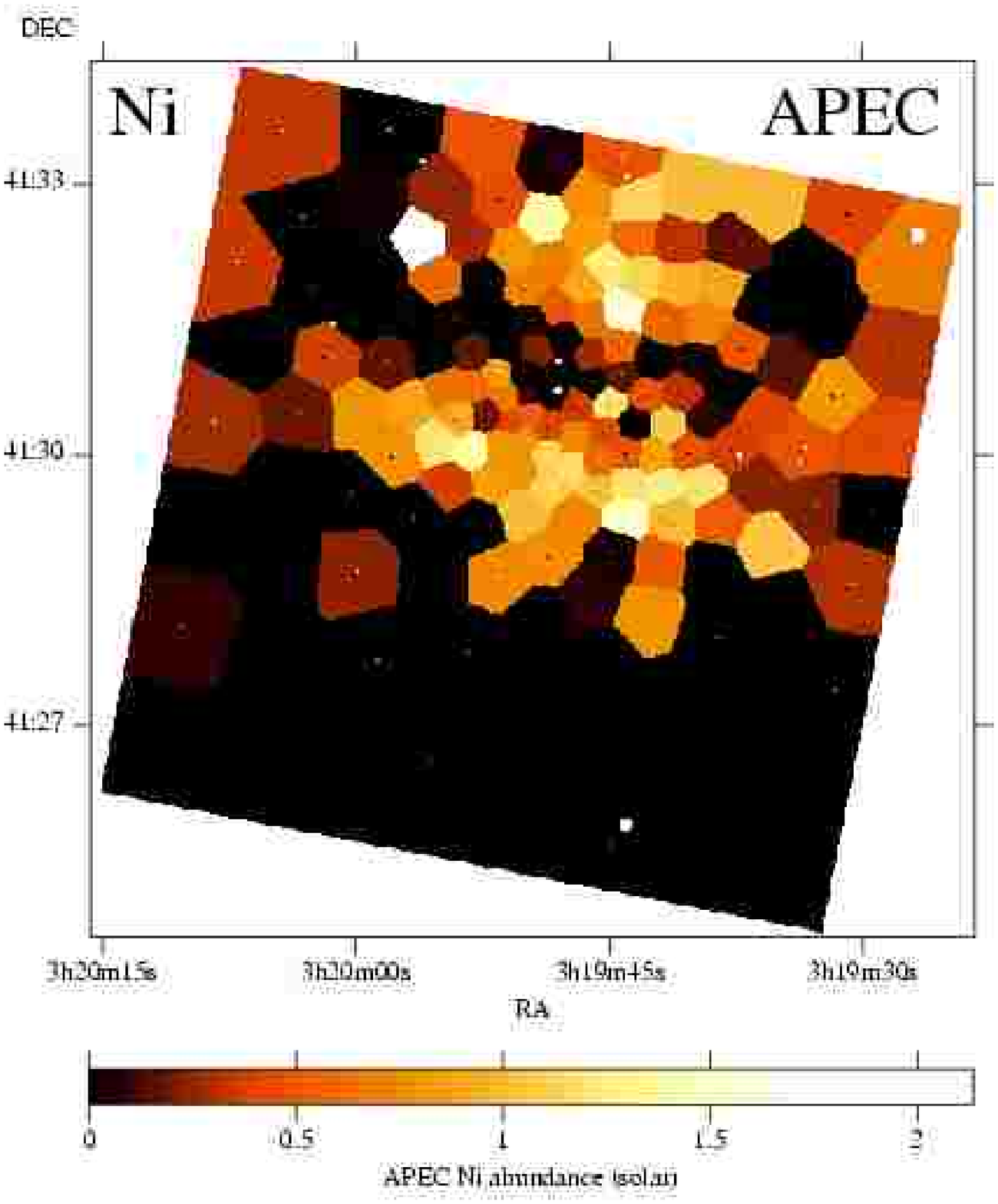}
  \includegraphics[width=.33\textwidth]{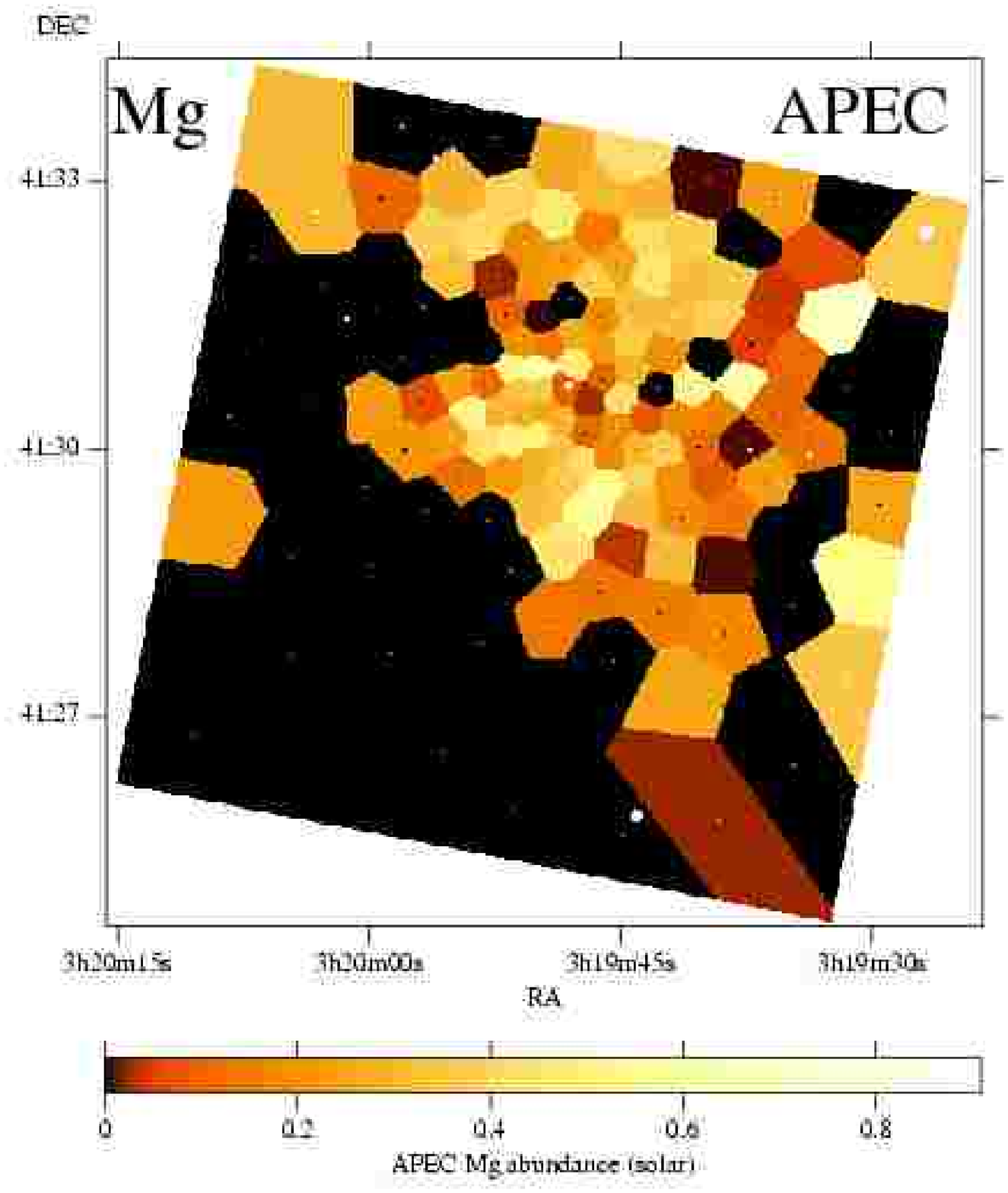}
  \caption{Maps generated by fitting spectra with $S/N \sim 300$.
    (Top Row) Fe, Ni and Ne maps, generated by fitting a
    \textsc{vmekal} model excluding the energy range 1.3-2.3 keV.
    (Centre Left) Fe-K map, generated by fitting spectra above 3 keV.
    (Centre Centre, Centre Right, Bottom Left) Ni, Mg and Si maps
    created by fitting \textsc{vmekal} models to 0.6-8 keV. (Bottom
    Centre, Bottom Right) Ni and Mg maps from fitting \textsc{vapec}
    models to 0.6-8 keV. The boxes within each bin mark the $1\sigma$
    uncertainties.}
  \label{fig:abun300_1}
\end{figure*}

We plot the Fe, Ni, and Ne abundances obtained from a \textsc{vmekal}
fit to data extracted from a $S/N \sim 300$ map in
Fig.~\ref{fig:abun300_1}~(top row). We also created maps of O, Ca, Ar
and S abundances, which we do not show here.  When fitting the data
the temperature, absorption, abundances and redshift of the gas were
allowed to be free. The redshift was allowed to vary as the best
fitting redshift is not the nominal redshift of the cluster.  We also
tested the approach fitting the data with fixed redshift, and using
$C$ statistics instead, yielding very similar results to those
presented here.

The Fe abundance map shows a very similar pattern to the overall
abundance map (Fig.~\ref{fig:Z_map}), highlighting the apparently low
abundance extension to the W, the abundance peak near the outer NW
radio lobe and to the SE, and the drop in abundance to the centre.

The Ni map indicates there is very little evidence of Ni enrichment
in the S of the cluster; the black areas on the map are all
upper-limits. Around the core of the cluster there is a ring-like
feature of high abundance. Near the core the Ni abundance is
apparently zero again.

The Ne abundance shows a simpler pattern. The abundance rises towards
the centre. Most of the bins at the centre have supersolar Ne
abundances. There are a number of features, though, that appear
significant. There is a low abundance feature to the SE of the core
(centred near 03:19:58.4, +41:30:09), and another to the bottom SE of
the map.

The other abundance maps have less-clear patterns to them. The Ca
abundance appears to be quite patchy. Binning up the data further
identifies an arc of high abundance and low statistical significance
to the W and S of the cluster, about 90 arcsec from the nucleus. The
Ar abundance map is similarly patchy, with a high abundance to the NW.
The O abundance map has a rather strange morphology. There are a
number of low and high abundance regions, some apparently linear in
shape. The O abundance appears low near the core. The odd O abundance
pattern suggests it is the result of a systematic error, particularly
as its emission lines lie at low energy.

We overplotted the abundance maps with positions of cluster members.
Curiously the locations where Ni is enhanced appears to coincide with
the positions of galaxies in the cluster. It is difficult, however, to
justify this physically as the galaxies (except for NGC 1275) are
probably moving rapidly and so there should be little significant
enrichment to their current surroundings. A galaxy is likely to only
remain in one of the bins for $\sim 10^7$ yr.

We fitted the abundances at the position of the outer NW radio lobe
and within that region, yielding the values in Table
\ref{tab:lobeabundance}. It appears the projected abundances of O, Ar,
Ca, Fe and Ni are smaller within the lobe, whilst Ne is larger (as
expected from the maps above). There is no trend in S abundance.

\begin{table*}
  \centering
  \begin{tabular}{llllllll}
    Element & O & Ne & S & Ar & Ca & Fe & Ni \\
    \hline
    Within & $0.37\pm0.06$ & $0.70 \pm 0.07$ & $0.61 \pm 0.05$ &
    $0.35 \pm 0.12$ & $0.25 \pm 0.12$ & $0.54 \pm 0.01$ &
    $<0.14$ (1$\sigma$)\\
    Without & $0.53\pm0.08$ & $0.31 \pm 0.08$ & $0.60 \pm 0.06$ &
    $0.95 \pm 0.15$ & $0.81 \pm 0.16$ & $0.62 \pm 0.015$ &
    $0.54 \pm 0.28$ \\
  \end{tabular}
  \caption{Abundances just within the position of the outer NW radio
    lobe, and at the position of the lobe and just beyond.}
  \label{tab:lobeabundance}
\end{table*}

In addition to the fits to the energy range 0.6-1.3, 2.3-8 keV, we fit
the data in other energy bands. In Fig.~\ref{fig:abun300_1}~(centre
left) we show an Fe-K map, created by fitting the data above 3 keV
with Fe, Ar, Ca and Ni free. The map is similar to the one based on
Fe-L and Fe-K lines but there are some differences. The abundances are
higher by about $0.05\Zsun$ in the outer regions to the S, lower
towards the centre (by about $0.04 \Zsun$), except to the W of the
nucleus where the abundances are higher (by $\sim 0.03 \Zsun$). To the
W of the nucleus the low-abundance feature is not as prominent.

We also computed maps for Mg, Si and Ni, using the full band between
0.6 and 8 keV and ignoring the systematic effects. Since we do not
understand the nature of these residuals the maps must be treated with
caution.  We show the \textsc{vmekal} Si map in
Fig.~\ref{fig:abun300_1}~(bottom left). The \textsc{vapec} map is
similar. The Si abundance follows the Fe abundance closely, rising to
a peak away from the nucleus and dropping down at the centre. We show
separate \textsc{vmekal} and \textsc{vapec} maps for Mg and Ni as
there are substantial differences between the models. \textsc{vapec}
has stronger Mg than \textsc{vmekal}, but weaker Ni.

\section{Multiple temperature component analysis}
\label{sect:multicomp}
As we mentioned in Section \ref{sect:solarabun}, if we use the
incorrect number of temperature components in our fits it may lead to
an underestimate of the abundance. It may be the case that in the
centre of the cluster, where there is most likely to be multiphase
gas, the abundance is being underestimated. In particular the solar,
Fe and Ni abundance drops seen in Fig.~\ref{fig:Z_map} and
Fig.~\ref{fig:abun300_1} may be systematic effects. We therefore have
attempted to fit multiple temperature components to the extracted
spectra.

If there are multiple phases, and assuming they are in pressure
equilibrium, an expression giving the volume-filling-fraction (VFF;
see equation 1 of Sanders \& Fabian 2002 for the two-phase version) of
the $i$th phase ($f_i$) is given by
\begin{equation}
  f_i = \frac{ K_i \: T_i^2 } { \sum_j K_j \: T_j^2 },
\end{equation}
where $K_i$ is the normalisation of the $i$th phase (proportional to
its emission measure), and $T_i$ is its temperature.

We fitted a model made up of a number of fixed temperature components
with variable normalisations, tying their abundances together. The
temperature components started with one at 0.5 keV, increasing in
temperature by factors of two until 16 keV. This approach has the
advantage that we can test for the presence of gas close to a
particular temperature. In addition the temperatures of a component do
not jump to very high or low values in the fit depending on initial
fit parameters; in our model the normalisation of a component simply
drops to zero if there is no gas near that particular temperature.  If
the gas is single-phase but not at the temperature of one of our
components, then we will detect it at the temperature of the
neighbouring components.

We also fit a model consisting of a temperature component, its
temperature free in the fit, and components at fixed fractions of that
temperature ($1/2$, $1/4$ and $1/8$). The results from this
alternative model appear to be consistent with those from this model.

\begin{figure*}
  \includegraphics[width=0.33\textwidth]{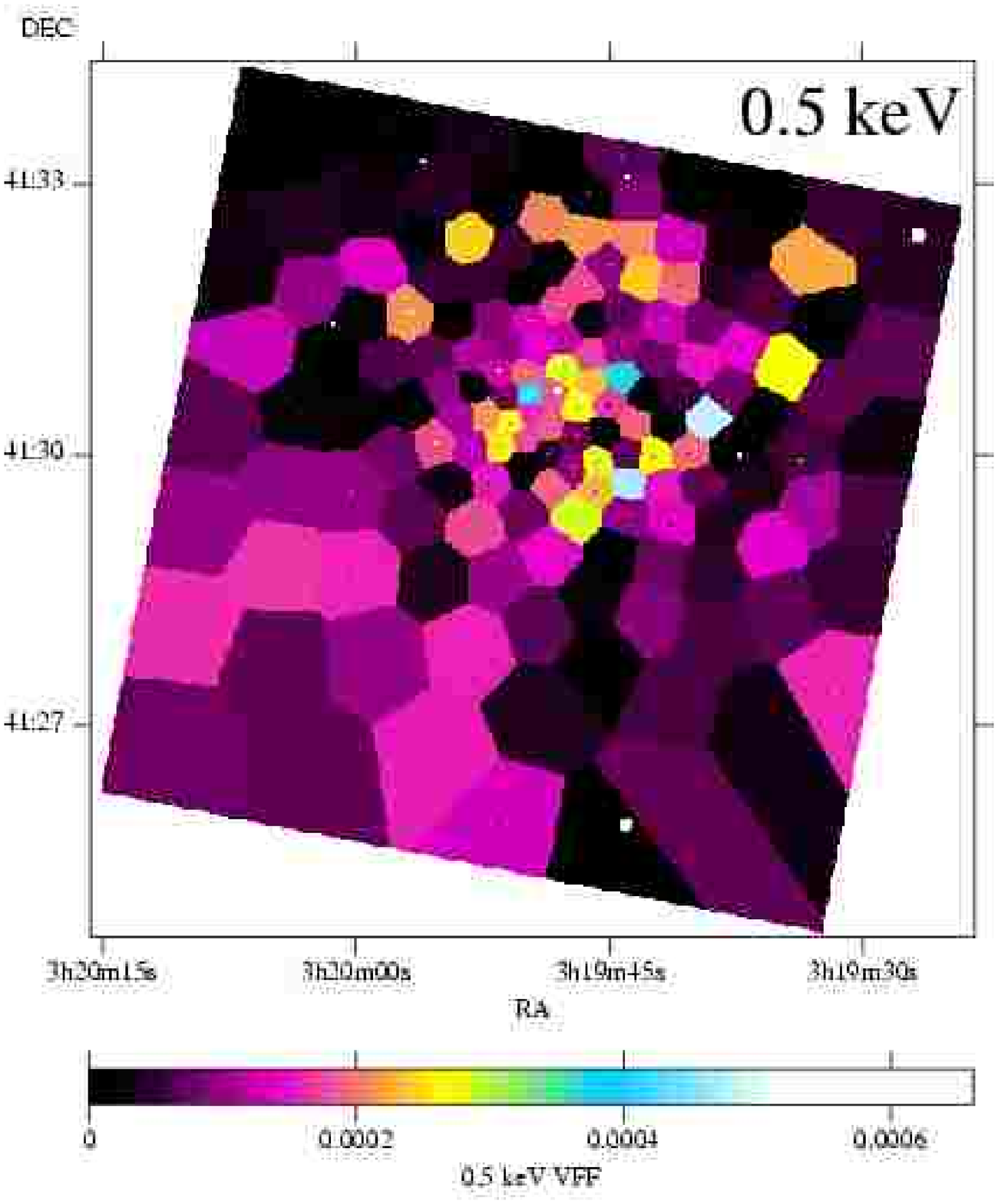}
  \includegraphics[width=0.33\textwidth]{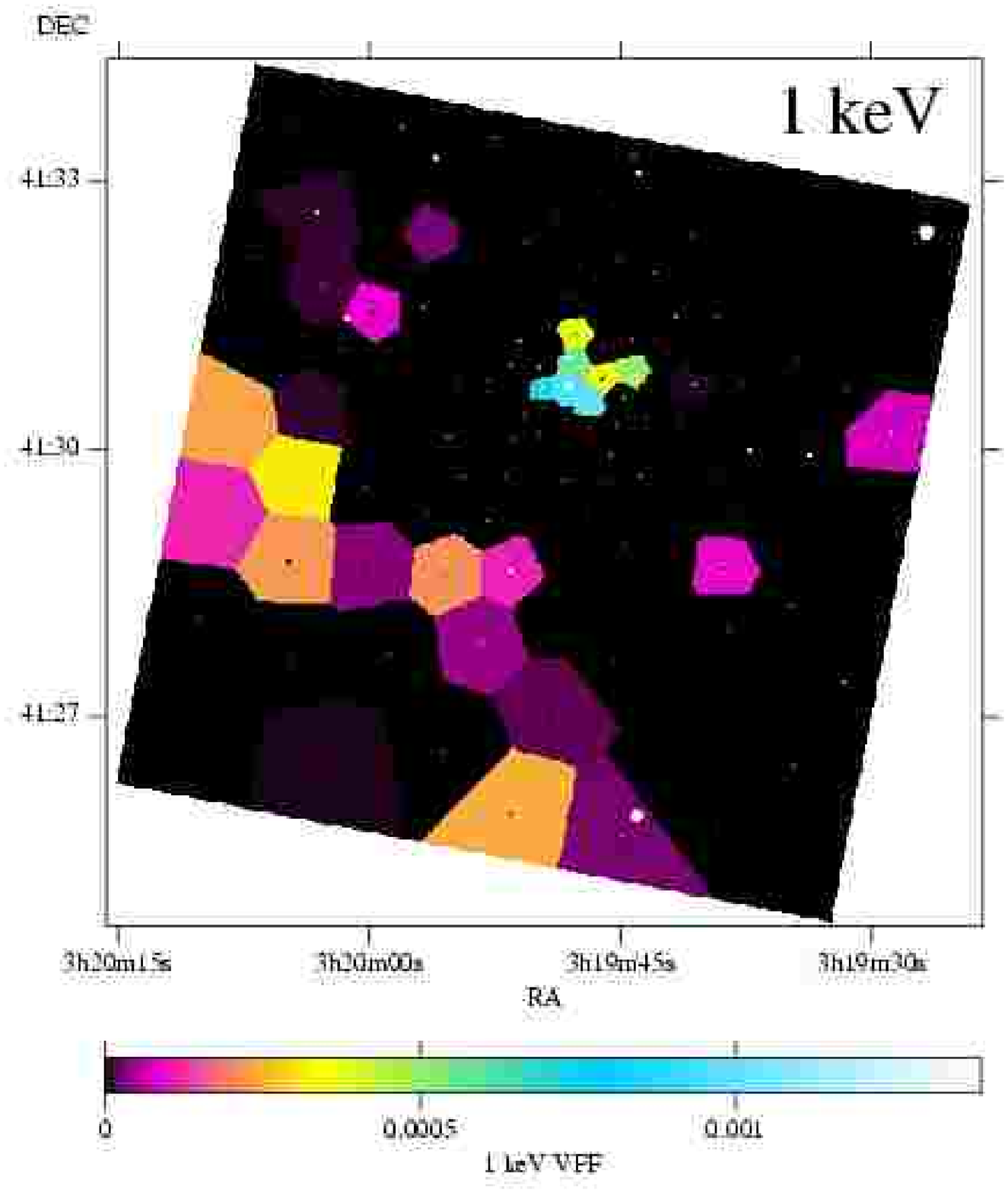}
  \includegraphics[width=0.33\textwidth]{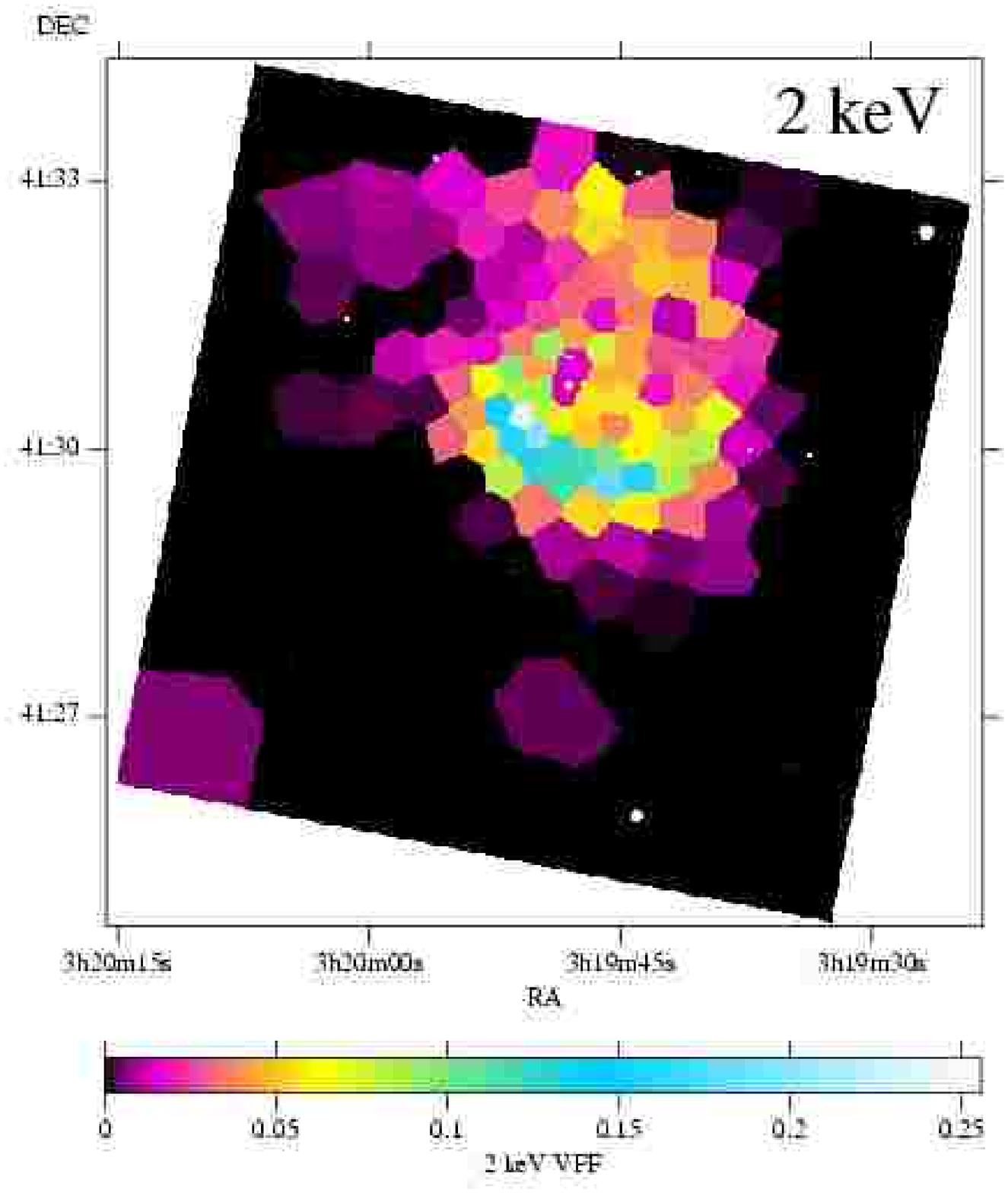}
  \includegraphics[width=0.33\textwidth]{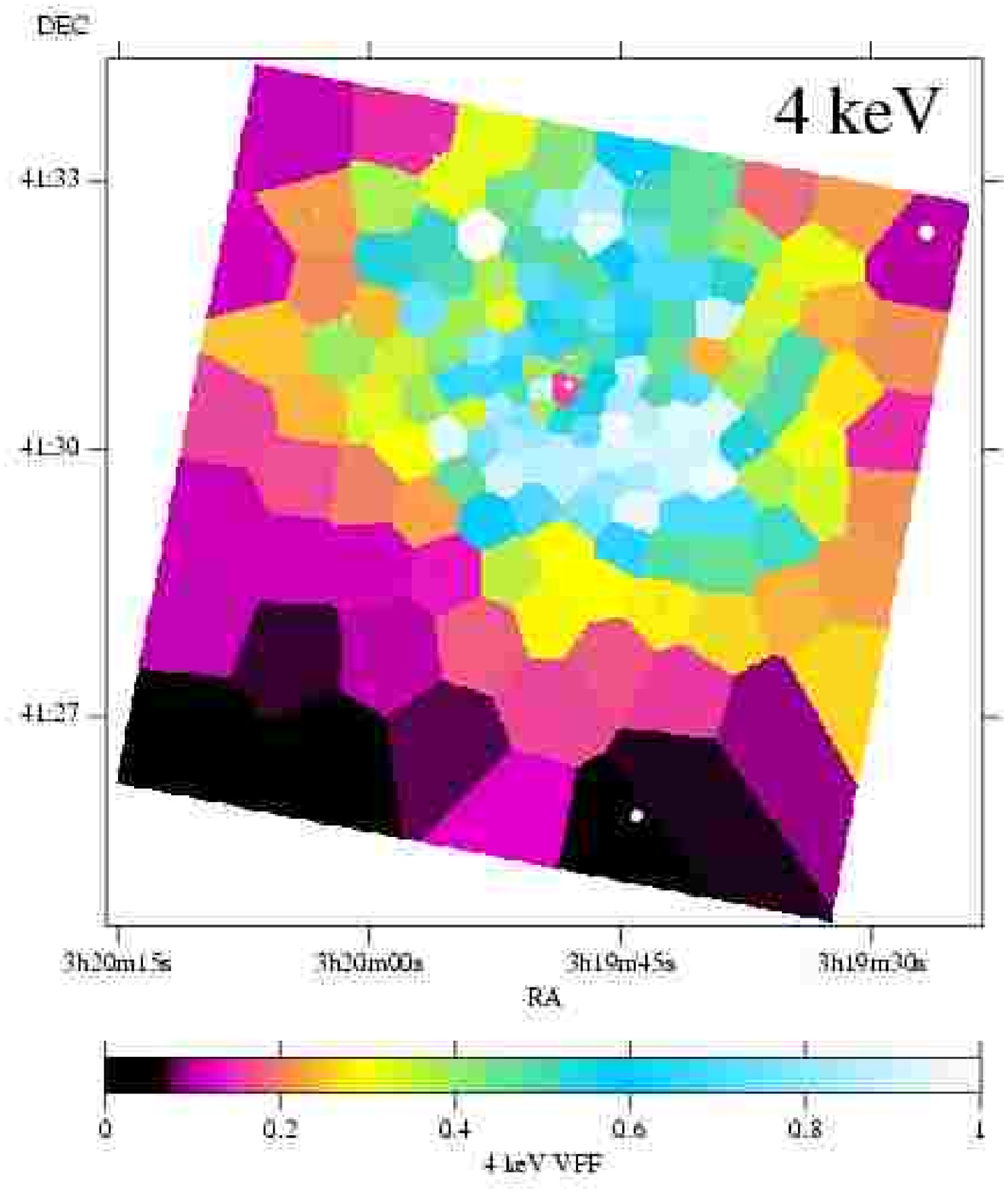}
  \includegraphics[width=0.33\textwidth]{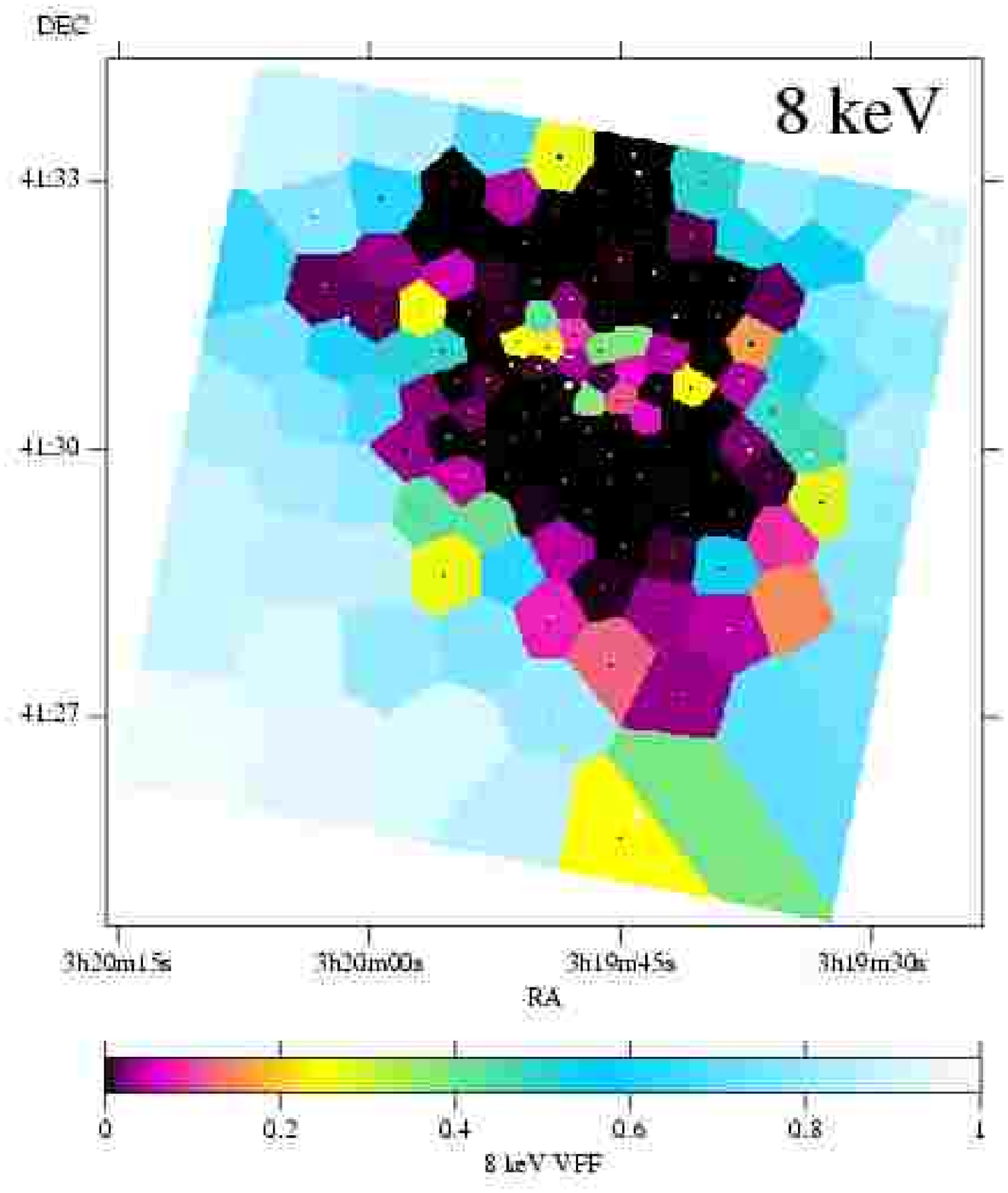}
  \includegraphics[width=0.33\textwidth]{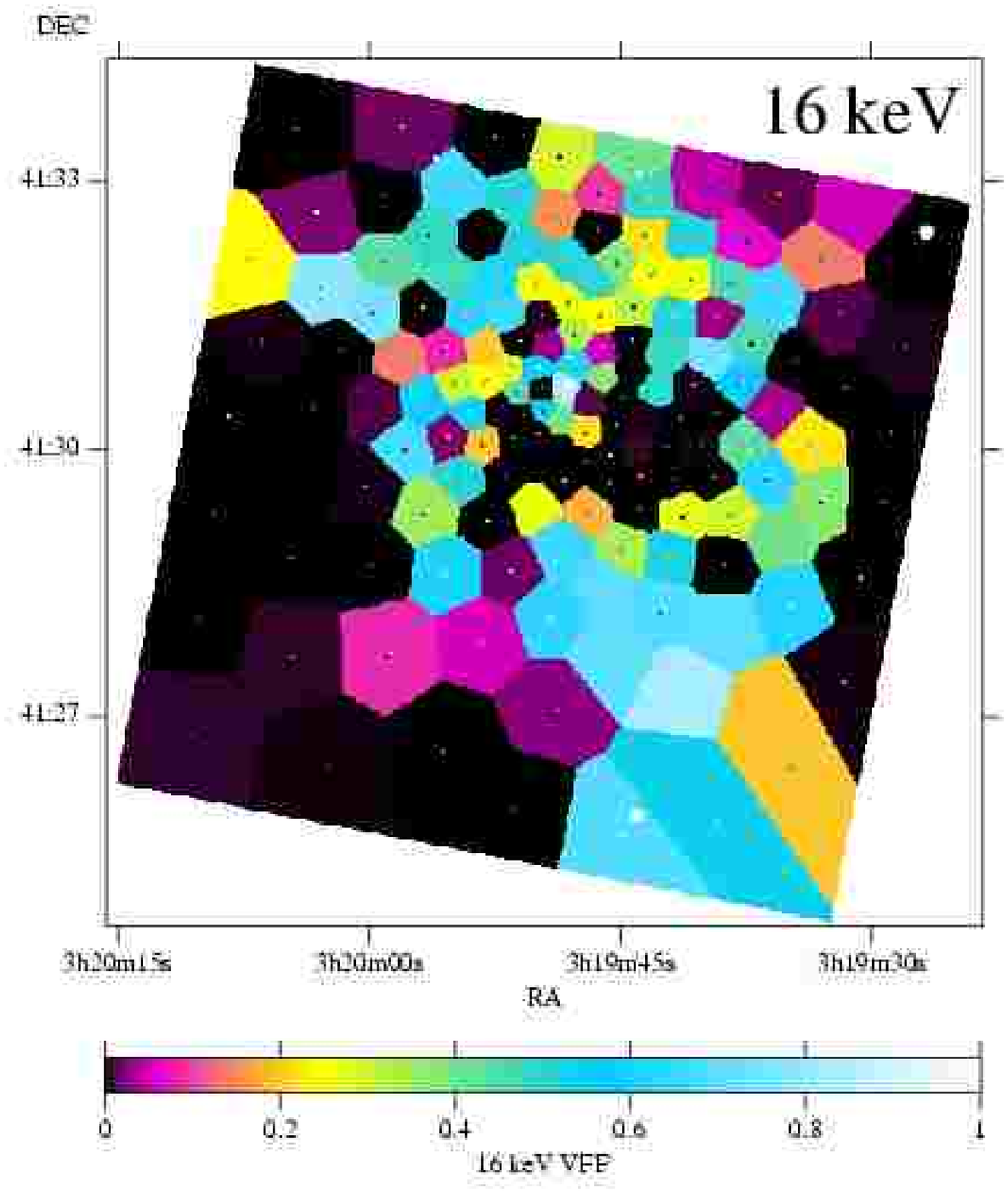}
  \caption{Volume-filling-fractions of components starting at 0.5 keV,
    and increasing in temperature by factors of two. The maps were
    generated by fitting spectra extracted from a $S/N \sim 300$ map
    with a \textsc{vmekal} model with multiple temperature components.
    The boxes within each bin mark the $1\sigma$ uncertainties on the
    VFF, assuming there is only uncertainty on the normalisation of
    that component.}
  \label{fig:multicomp}
\end{figure*}

In Fig.~\ref{fig:multicomp} is shown the VFF of each temperature
component in each bin. The 0.5 keV map shows a diffuse very low VFF
component covering the image. There is some enhancement towards the
centre of the cluster. It is possible this is a real effect, but this
map is sensitive to the calibration at low energies and could be
plagued by systematic errors.  At 1 keV, there are only upper limits
in most bins, except in the very core and in a band running in a NE-SW
direction in the SE of the image. At 2 keV there is a great deal of
gas. Most of this gas follows the morphology as the cool gas in the
single-phase temperature map (Fig.~\ref{fig:T_map}). Looking at the 4
keV map, most of that gas with a high VFF occupies the region of the
low-temperature swirl, but there is a large clump to the SW of the
nucleus. In addition there is a diffuse halo of hotter gas outside of
the inner core, corresponding to what we would expect from the mean
temperature of the gas there.  When we reach 8 keV, we only see gas in
the outer parts of the cluster, except for a W-E band to the N of the
nucleus, corresponding in position with the high-velocity system. The
16 keV map is rather unusual in that it shows a region of hot emission
around the core, but not so much in the outer regions where the mean
temperature is hottest, or to the S of the nucleus where it is
coolest.

We also repeated the analysis on a faked dataset. The faked dataset
was generated from real $S/N\sim 225$ projected temperature,
abundance, normalisation and absorption maps, faking a spectrum using
\textsc{xspec} for each detector pixel, and populating an events file
with the events necessary to create each detector pixel spectrum.  The
faked data has a single temperature component at all location, and
includes no projection effects. The 0.5 keV faked map shows some
points at the values observed in the real data, but the majority of
bins have smaller VFFs than the real data.  The 1 keV faked map shows
no concentration of VFFs at the centre. The 2 and 4 keV maps are
similar to their real counterparts. The 8 keV map resembles the real
map, but does not have the band of high VFFs running across the N of
the nucleus. The 16 keV map shows a few random points with significant
VFFs, and does not show similar structures to those seen in the real
data.

\begin{figure}
  \includegraphics[width=\columnwidth]{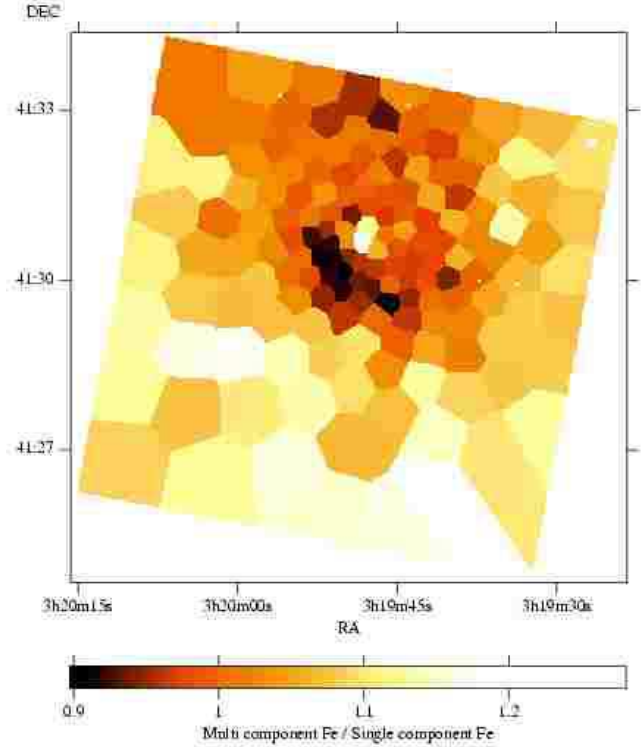}
  \caption{Ratio of the multiple component Fe abundance to the single
    component Fe abundance}
  \label{fig:multicompFe}
\end{figure}

If we examine the Fe abundance obtained by a multiple component fit,
the Fe abundance still declines in the centre, and along the low
abundance SW extension from the core. We can divide the Fe abundance
map with the single-component map (Fig.~\ref{fig:abun300_1}), the
result of which is shown in Fig.~\ref{fig:multicompFe}. It appears that
introducing multiple temperature components does not increase the
abundance over much of the central region, except in the innermost few
bins. If we fit the data with the temperature components separated by
a ratio of 1.5 instead of 2 then we do not see a significant change
the abundance map. We also do not see a difference using the
alternative model that consists of a free upper fitted temperature
with fractional temperatures below that.

\subsection{Cooling flow model}
\label{sect:cflow}
\begin{figure*}
  \includegraphics[width=0.33\textwidth]{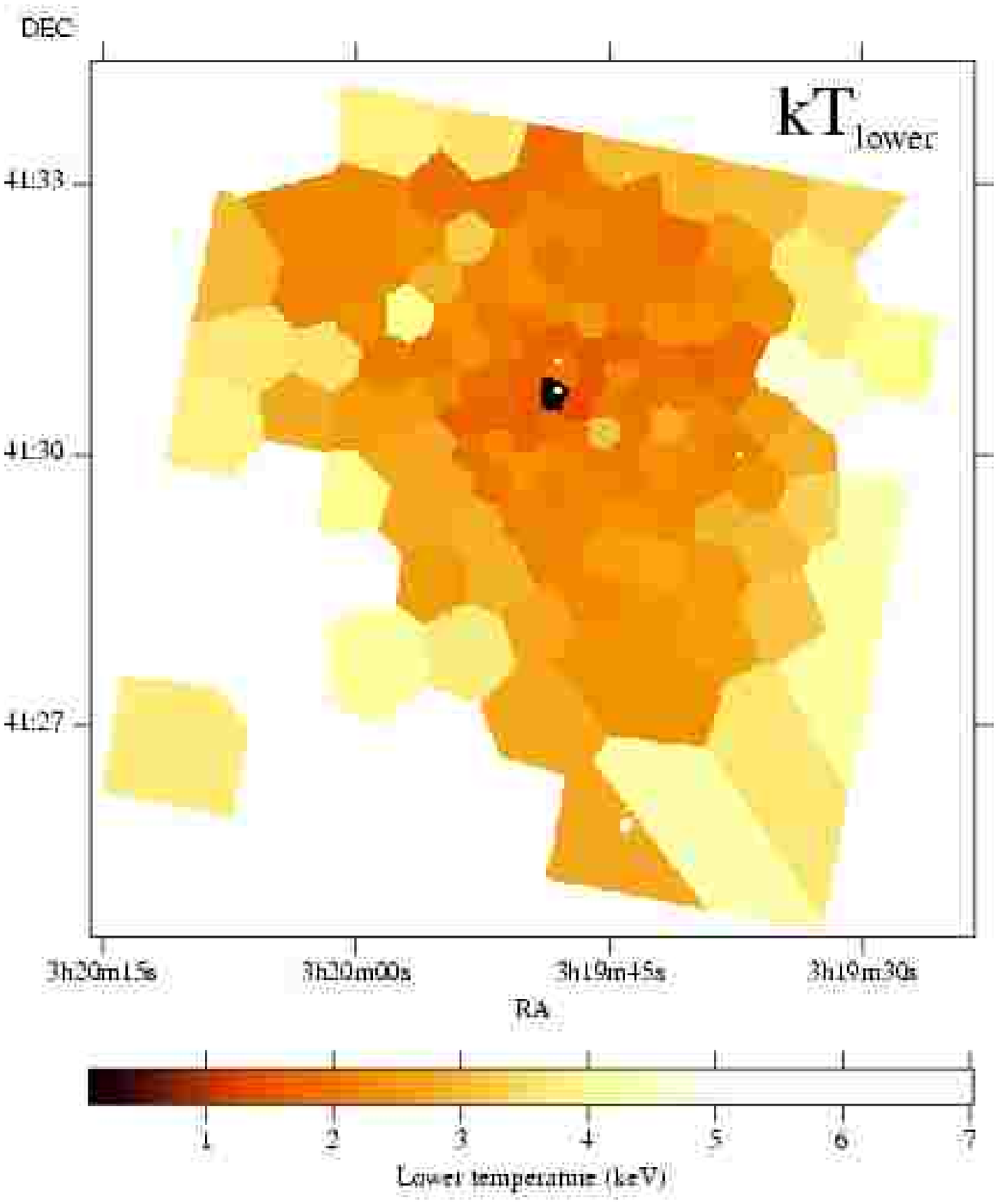}
  \includegraphics[width=0.33\textwidth]{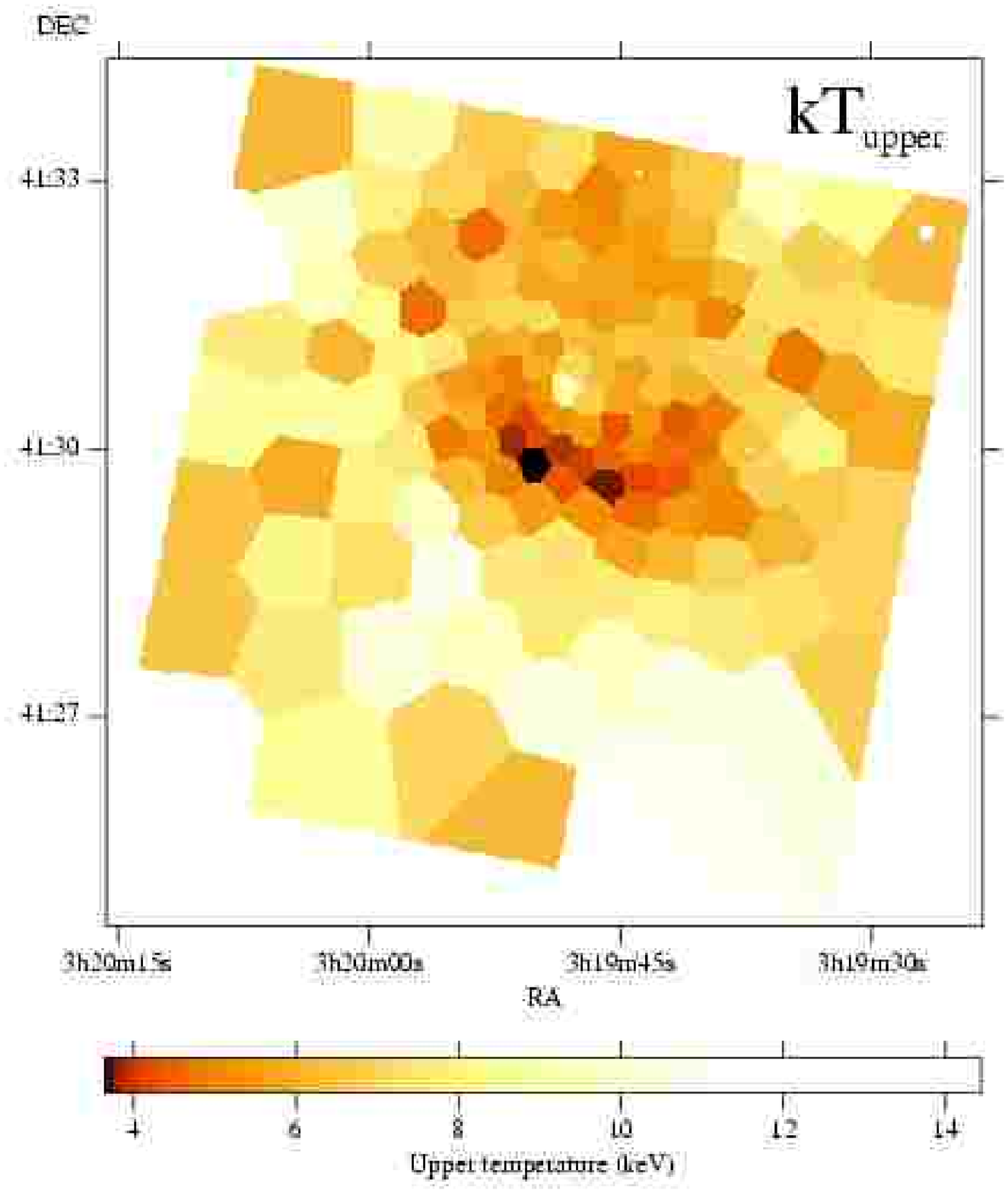}
  \includegraphics[width=0.33\textwidth]{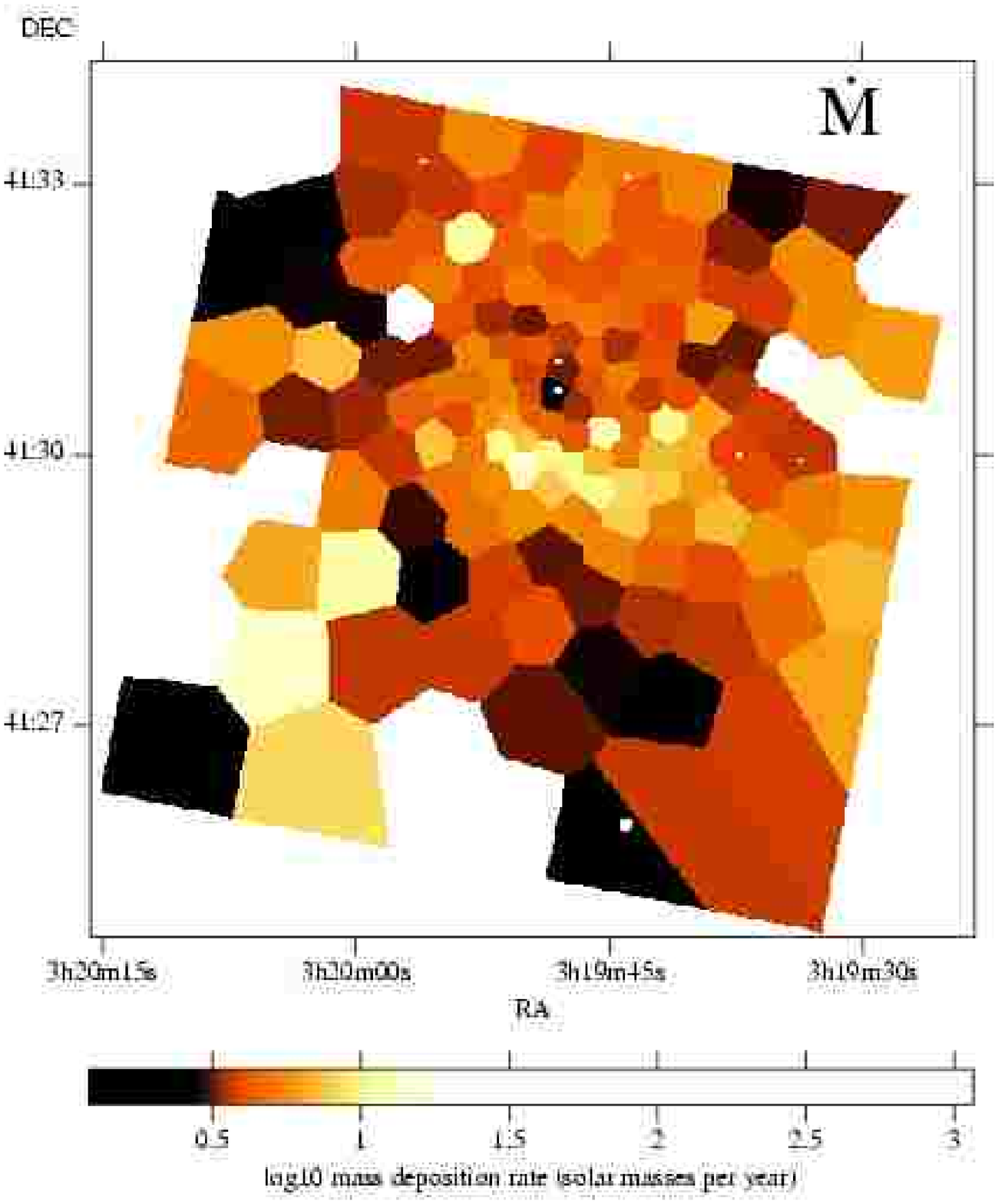}
  \caption{Results of fitting cooling flow model to the spectra
    extracted from the $S/N \sim 300$ bins. Left panel is the lower
    temperature of the cooling flow, the middle panel is the upper
    temperature, and the right panel is the mass deposition rate
    \emph{in each bin}.}
  \label{fig:cflow}
\end{figure*}

We can fit an isobaric cooling flow model (Fabian 1994; Johnstone et
al 1992) in which the gas is cooling from an upper temperature
($kT_\mathrm{upper}$) to a lower temperature ($kT_\mathrm{lower}$), at
a certain rate ($\dot{M}$). The cooling flow model we use is based on
the \textsc{vmekal} spectral model, with variable abundances. We
excluded the range 1.3-2.3 keV in the spectral fitting.
Fig.~\ref{fig:cflow} shows $kT_\mathrm{lower}$, $kT_\mathrm{upper}$
and $\dot{M}$.

Those regions where $kT_\mathrm{upper}$ is close to
$kT_\mathrm{lower}$ at the edges of the image should be ignored, as
the mass deposition rate is meaningless. The $kT_\mathrm{lower}$
distribution is fairly flat over the core of the cluster, but lowest
near the central source. The $kT_\mathrm{upper}$ distribution is
lowest where the projected temperature is lowest. The most significant
mass deposition rates are where the projected temperature is lowest.
If we sum the total mass deposition rate where the projected
temperature is lowest, we find $\dot{M} \sim 255 \Msunpyr$. However
$kT_\mathrm{lower}$ is 2 to 2.5 keV over much of this region, in
agreement with the results of on the lack of low energy gas seen in
other clusters (e.g. Peterson et al 2001, 2003).

\section{Accounting for projection}
\label{sect:projection}
The above analyses do not take projection effects into account. In
order to account for projection, we have to assume some symmetry.  The
maps in the previous sections show that the cluster is not symmetric
in its core, and so it is not possible to perfectly account for
projection effects. To account for the variation as a function of
angle, we divided the cluster into a number of sectors.

We divided each sector into a regular set of partial-annuli, and
performed spectral fitting accounting for projection effects using the
\textsc{xspec} \textsc{projct} model. The spectra were fit from the
outermost annulus inwards, fitting the parameters for an annulus, and
freezing its parameters when fitting interior annuli. Annuli were fit
without including in the fit the data from any interior annuli. We fit
the absorption for each shell as Fig.~\ref{fig:NH_map} reveals that
there is a large deal of systematic variation and possibly intrinsic
variation. The value of $N_H$ from this projection model may not mean
very much if the absorption is intrinsic to the object.

This procedure has the advantage that parts of the cluster which do
not fit correctly assuming spherical symmetry or for which our
spectral model does not fit the data well, do not affect the
parameters of exterior annuli. It has the disadvantage that the
statistical weight of interior annuli does not feed into the fitting
of a spectrum.  Additionally, the uncertainties we calculate using the
usual $\Delta \chi^2$ approach do not include the uncertainties on the
parameters for exterior annuli. A result of our approach may be
oscillatory profiles, where a parameter is underfit on one annulus,
and subsequently overfit on the next innermost annulus. Indeed the
results from outermost annuli must be taken with caution as the
projection model does not know what volume the gas emitting in the
shell occupies, as we do not account for projection to the edge of the
cluster. Our analysis will fail for regions where spherical symmetry
within a sector cannot be assumed.

\subsection{Simple single-temperature model}
\begin{figure*}
  \includegraphics[width=\columnwidth]{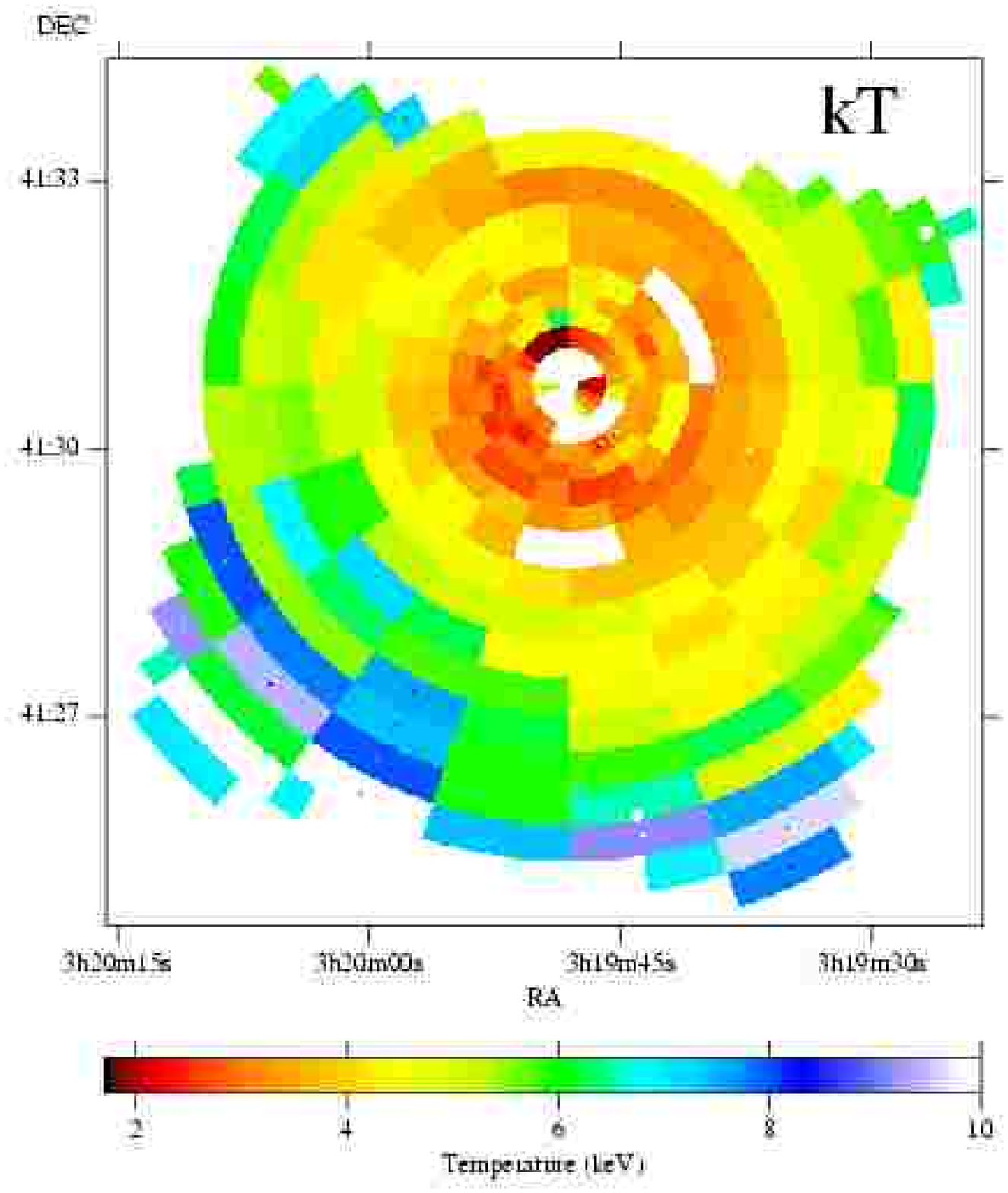}
  \includegraphics[width=\columnwidth]{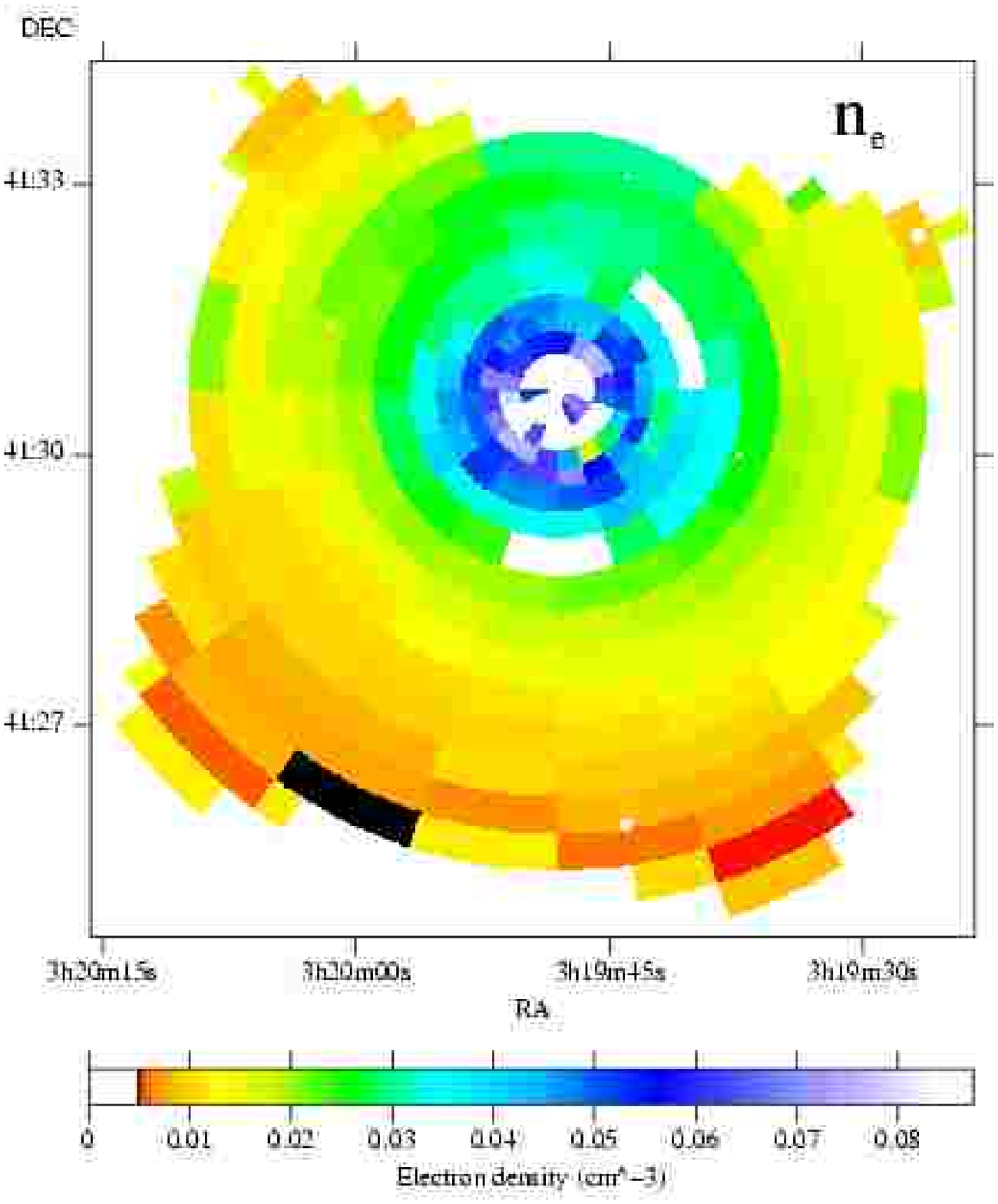}
  \includegraphics[width=\columnwidth]{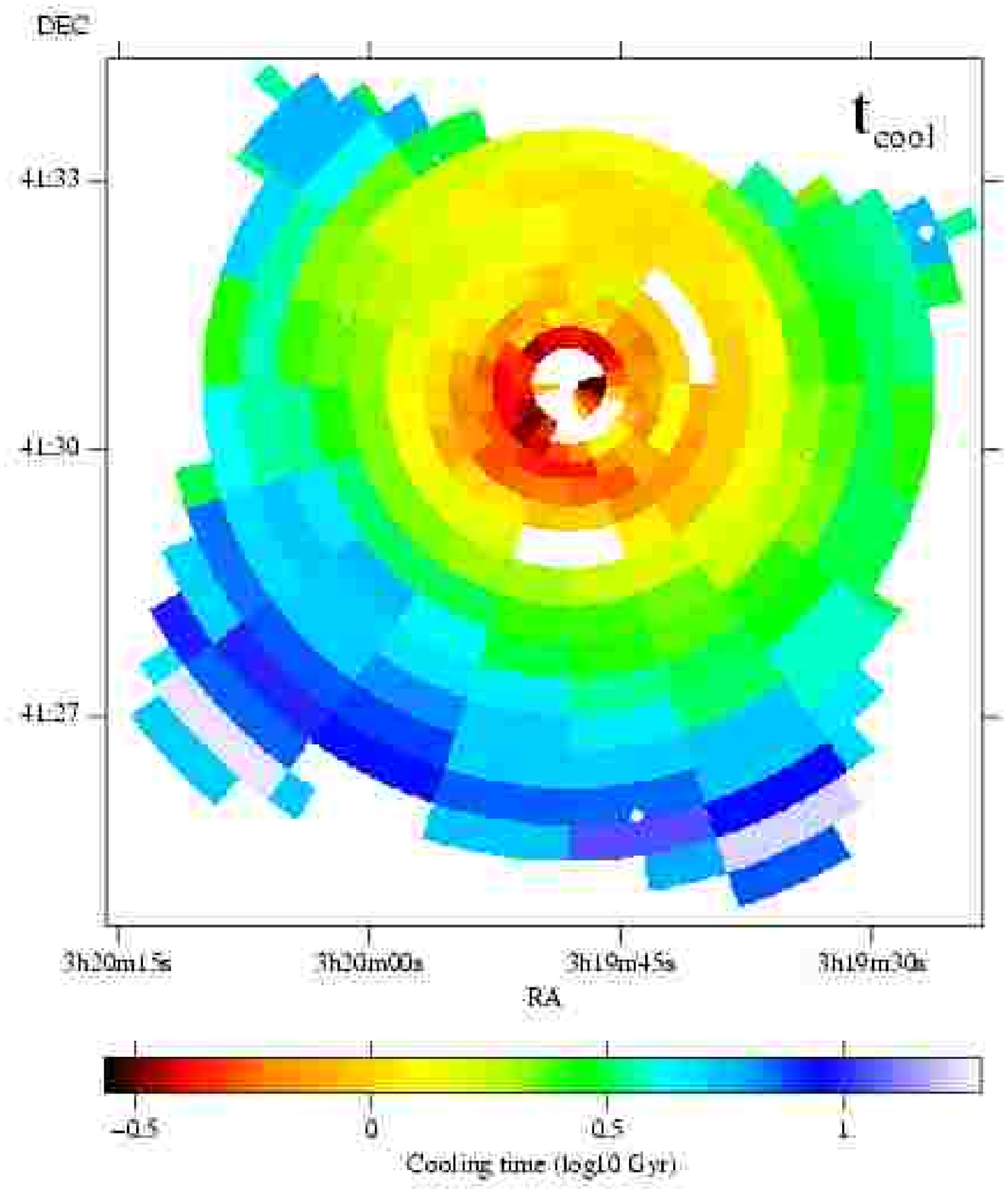}
  \includegraphics[width=\columnwidth]{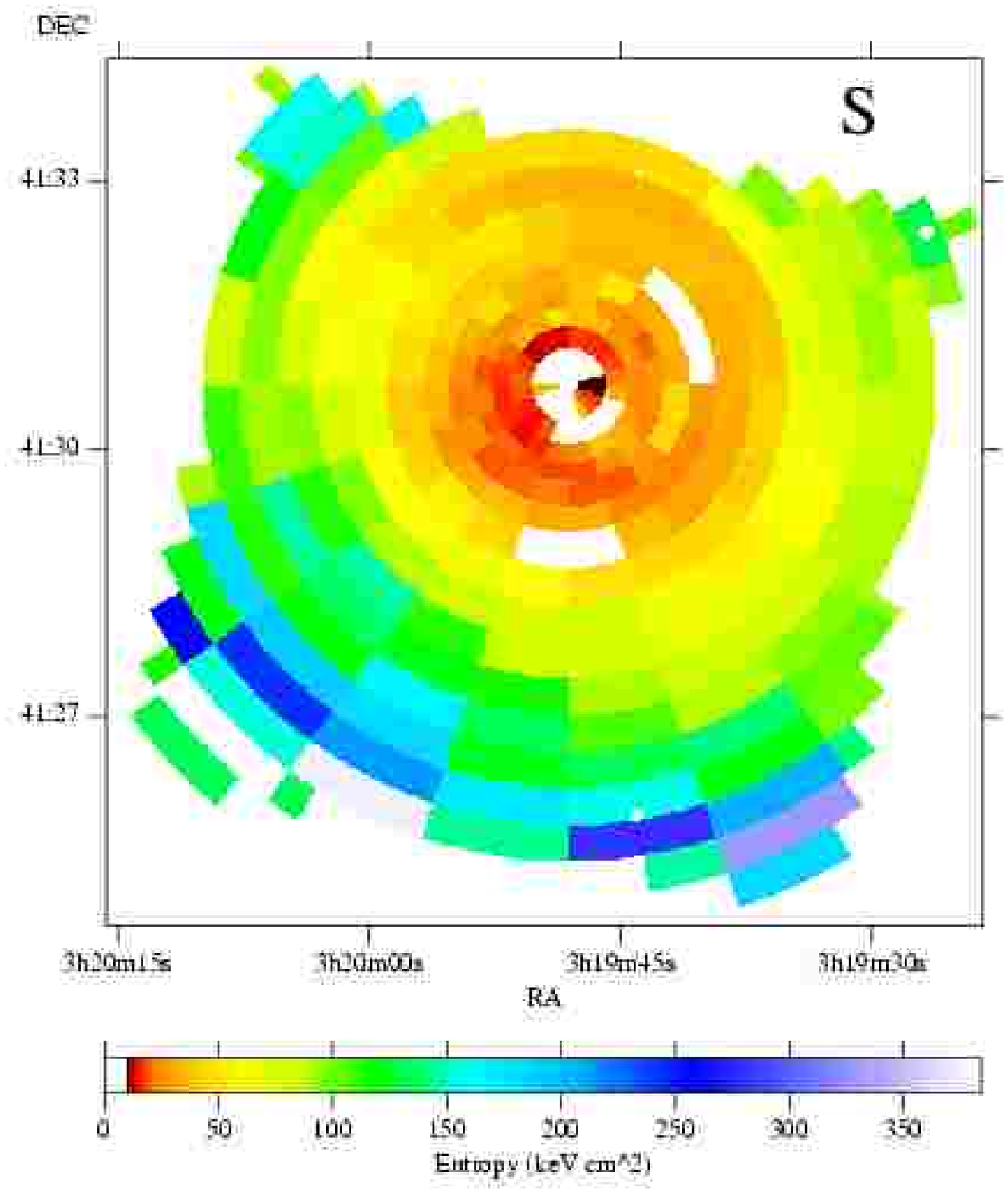}
  \caption{Temperature, density, cooling time and entropy plots,
    generated by fitting spectra in twenty sectors, accounting for
    projection effects. The boxes within each bin on the temperature
    plot mark the $1\sigma$ uncertainties. Sectors in which the radio
    lobes are located or projection effects failed are not shown.
    These values are plotted in Fig.~\ref{fig:deproj20_plots}.}
  \label{fig:deproj20}
\end{figure*}

\begin{figure*}
  \includegraphics[width=\columnwidth]{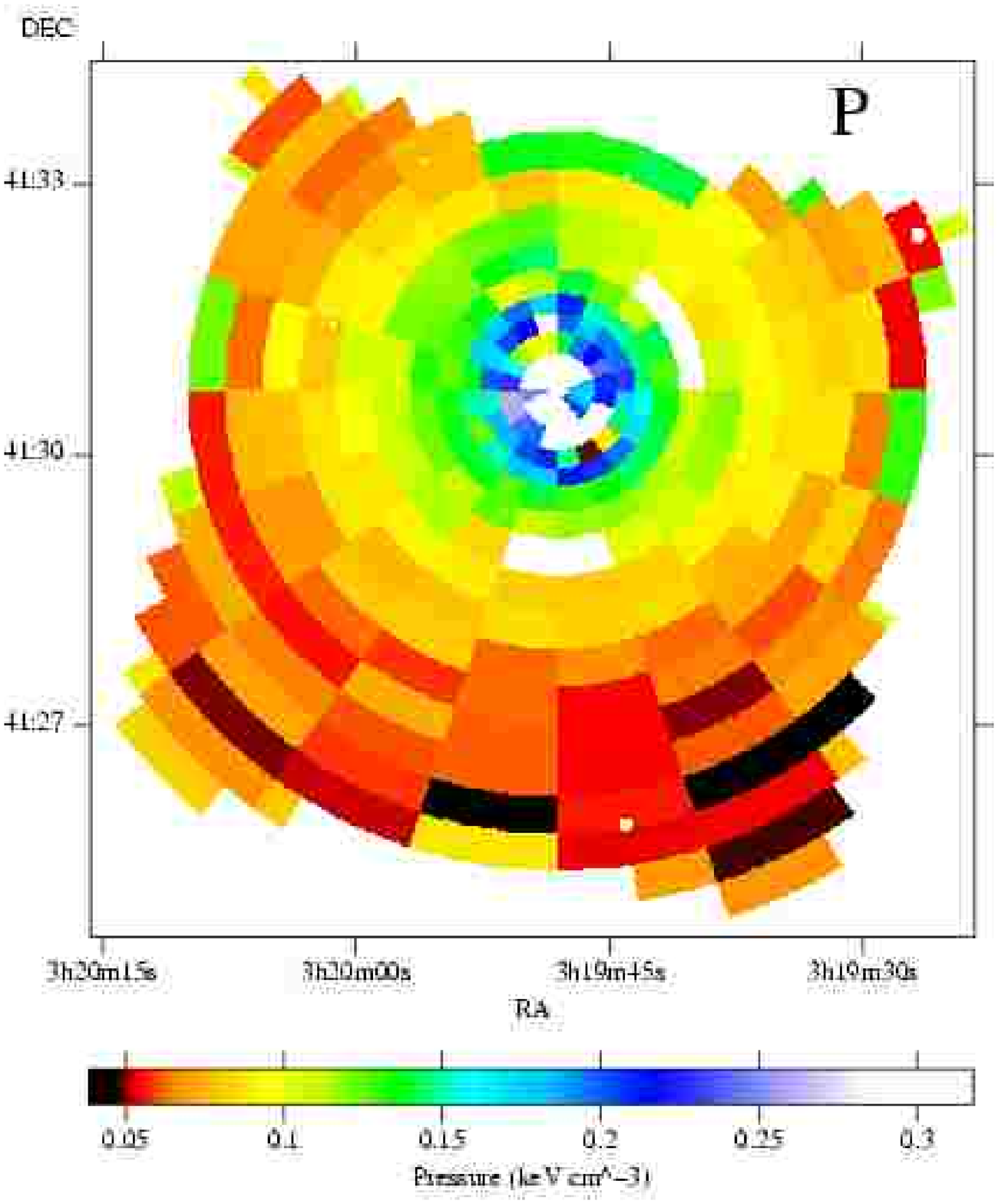}
  \includegraphics[width=\columnwidth]{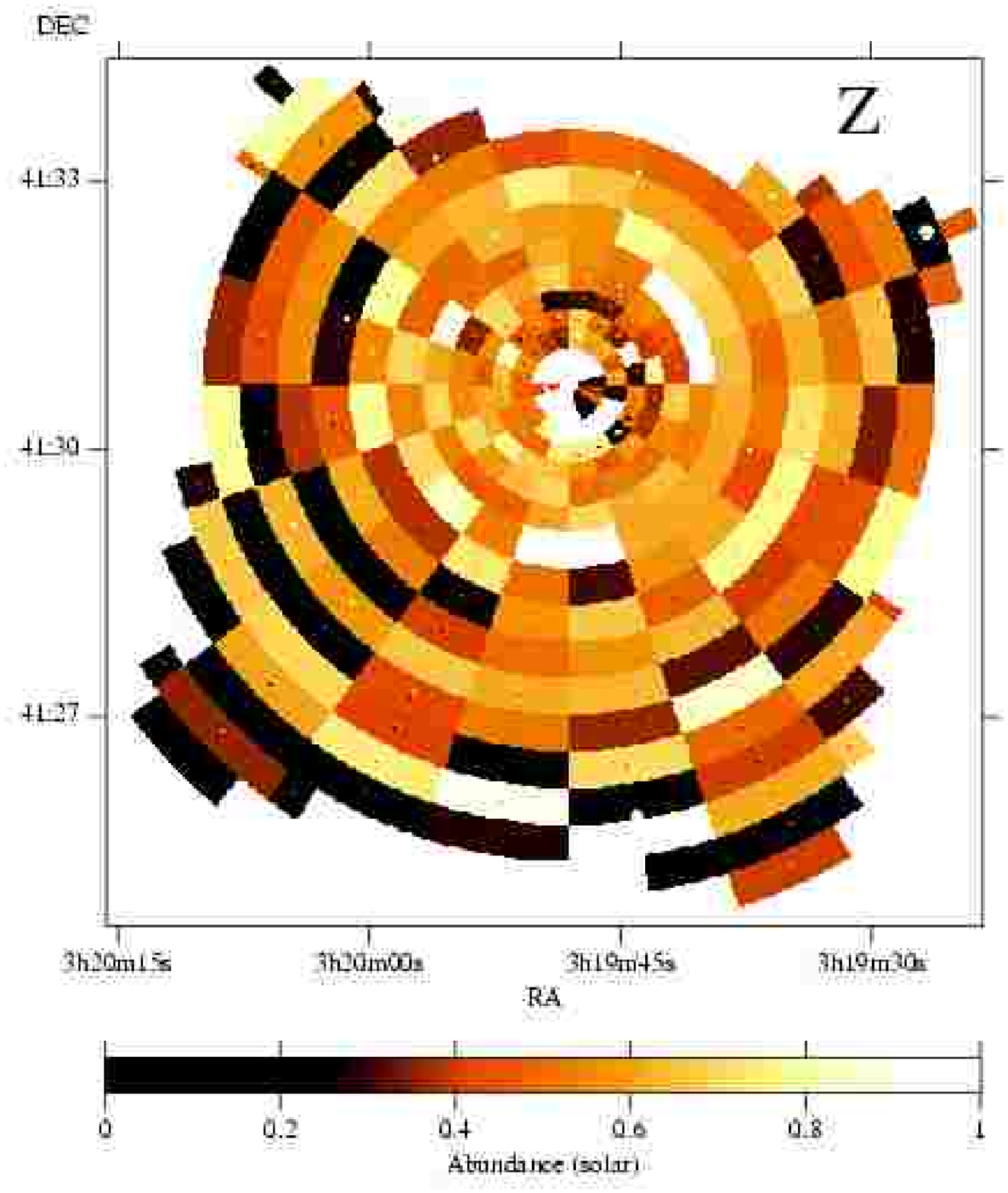}
  \caption{(Left) Pressure map of the cluster, accounting for
    projection. (Right) Abundances of sectors. The boxes within each
    sector mark the $1\sigma$ uncertainties.}
  \label{fig:deproj20_press_abun}
\end{figure*}

\begin{figure}
  \includegraphics[width=\columnwidth]{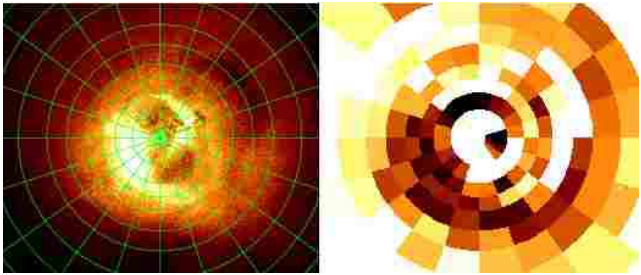}
  \caption{(Left) Regions used in the projection analysis on an intensity
    image of the centre of the core. (Right) A temperature map of the
    same sectors, from $\sim 2$ keV in black to $\sim 4$ keV or more
    in white (or where the spherical symmetry assumption fails).}
  \label{fig:deproj20_zoomT}
\end{figure}

\begin{figure*}
  \includegraphics[angle=-90,width=\columnwidth]{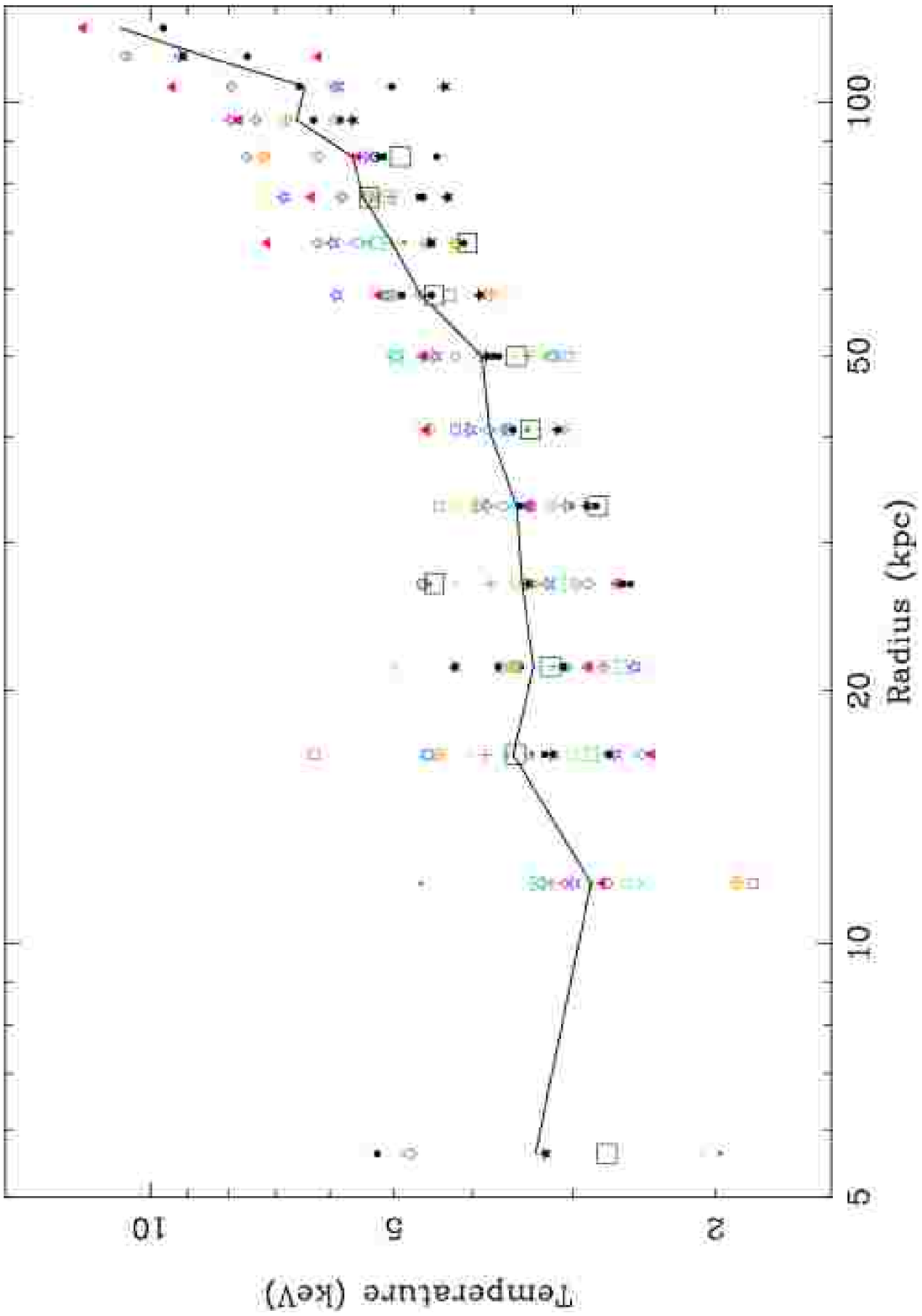}
  \hspace{2mm}
  \includegraphics[angle=-90,width=\columnwidth]{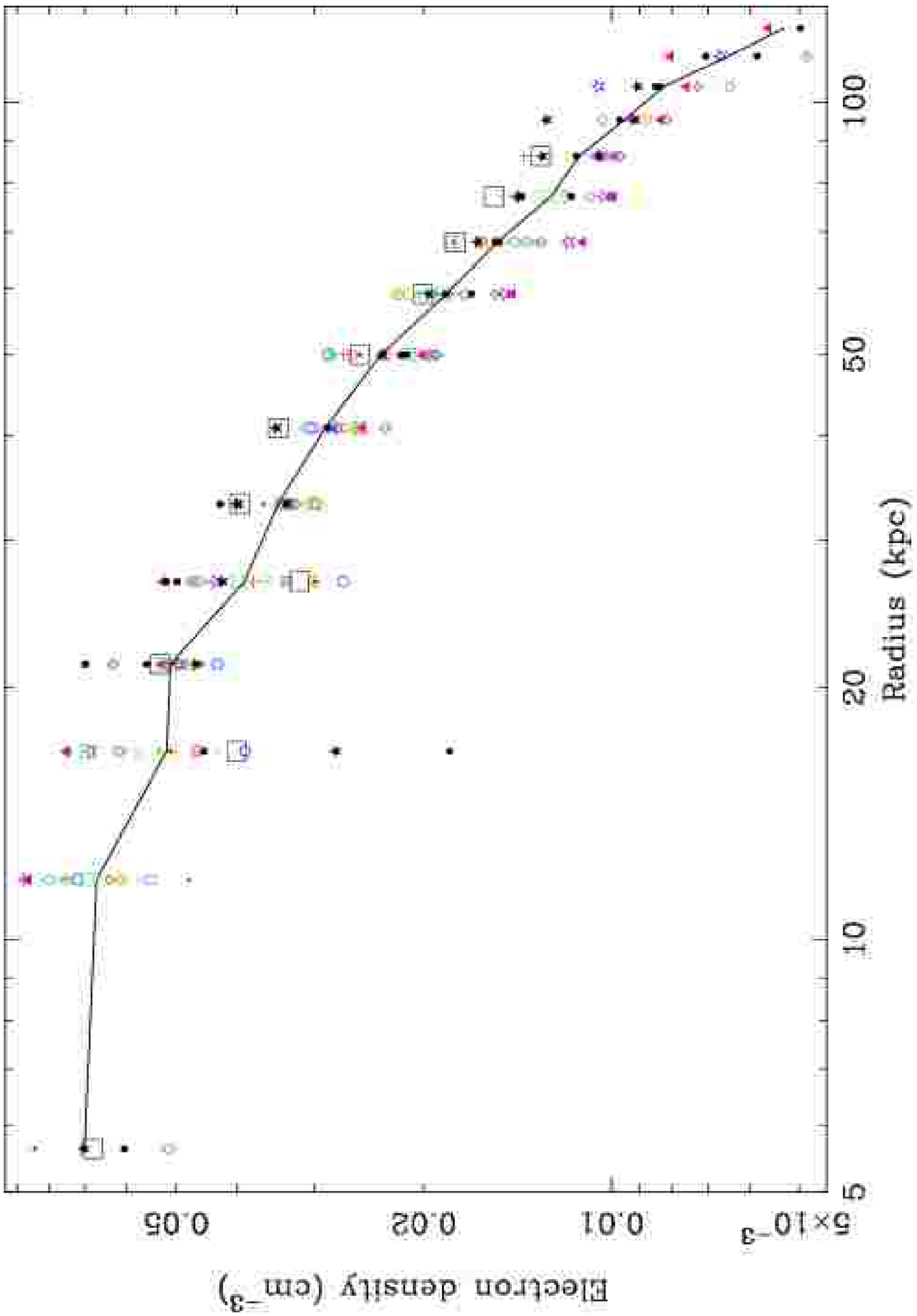} \\[2mm]
  \includegraphics[angle=-90,width=\columnwidth]{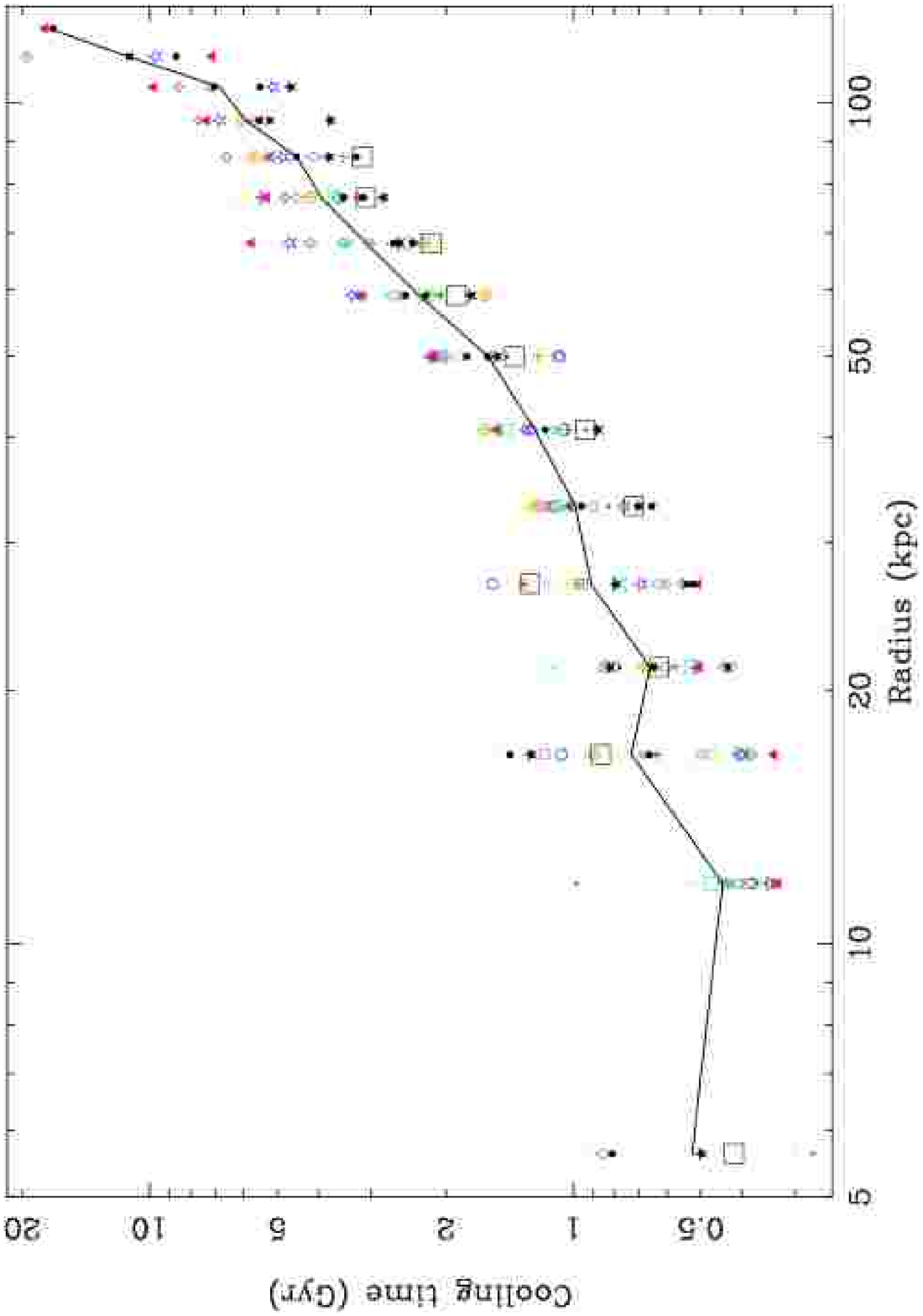}
  \hspace{2mm}
  \includegraphics[angle=-90,width=\columnwidth]{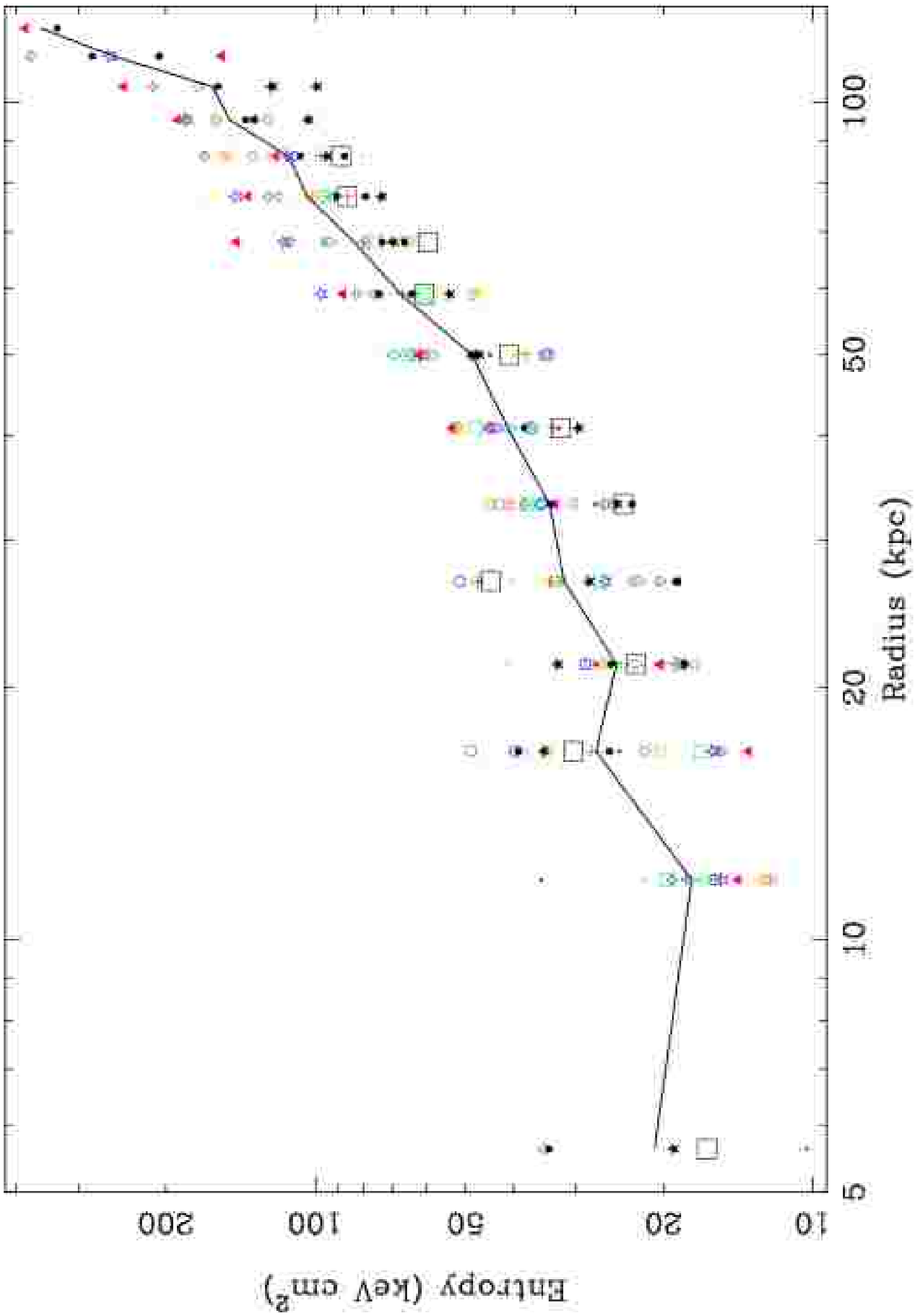}\\[2mm]

  \includegraphics[angle=-90,width=\columnwidth]{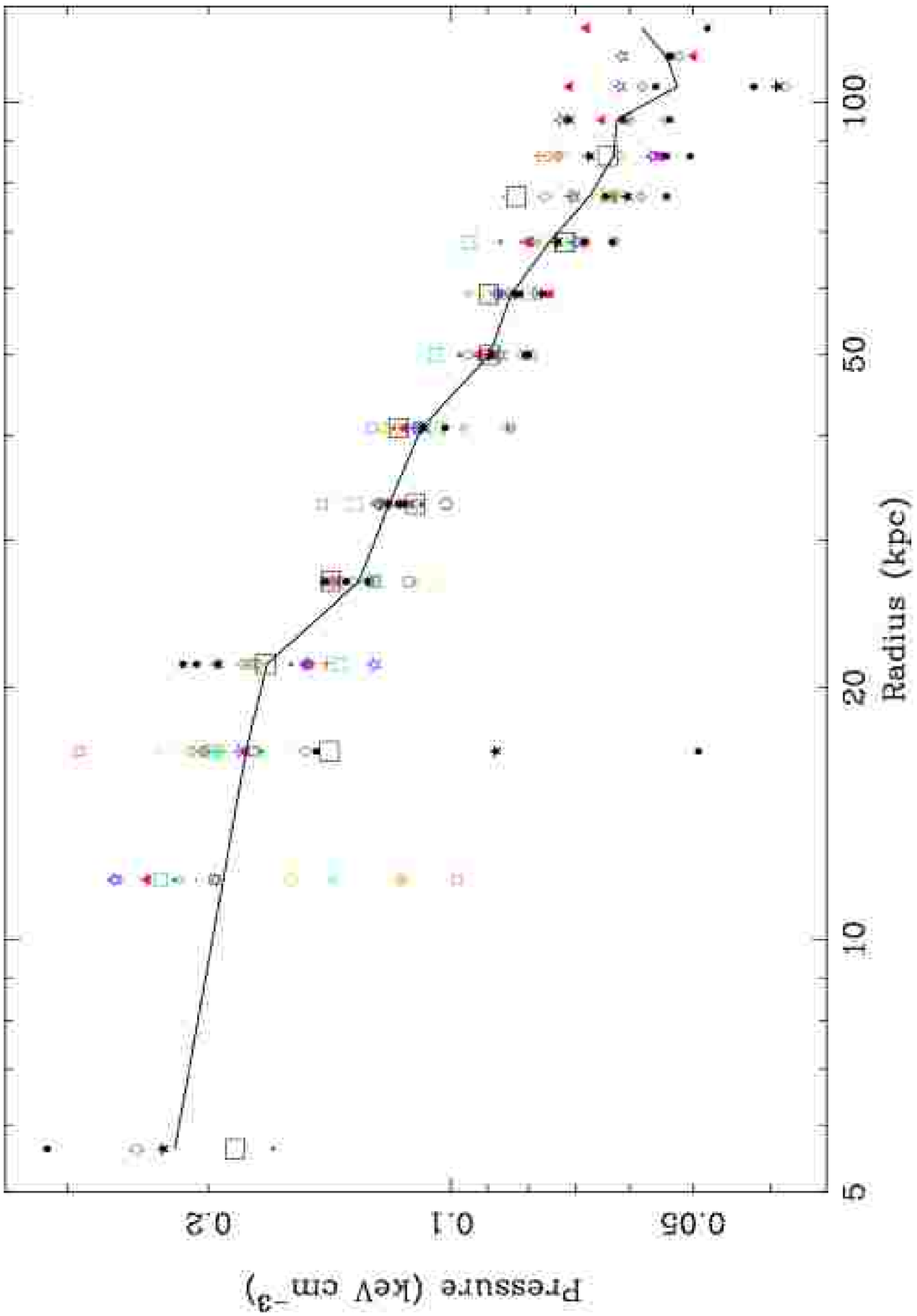}
  \hspace{2mm}
  \includegraphics[angle=-90,width=\columnwidth]{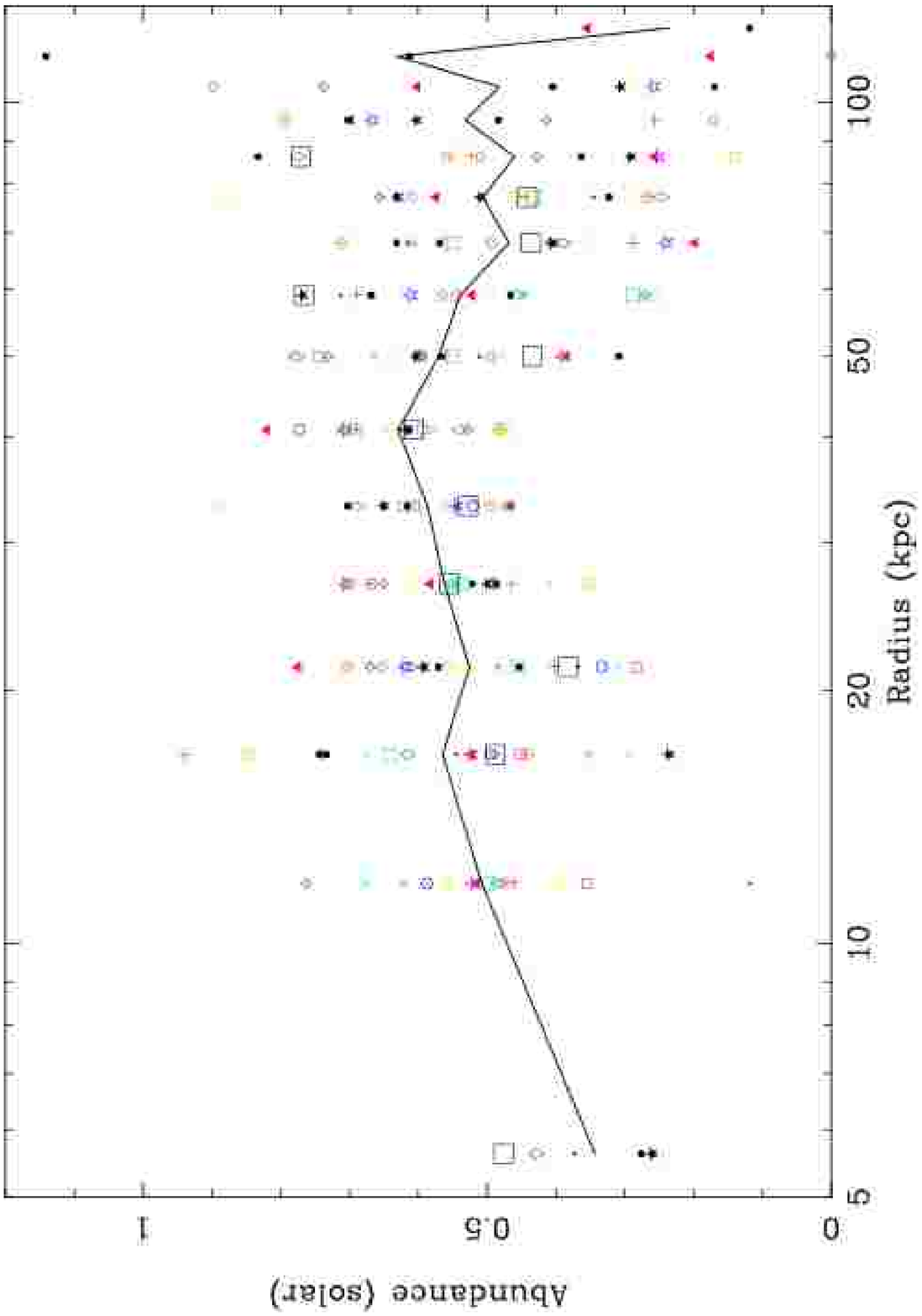}
  \caption{Combined radial profiles of the values in
    Fig.~\ref{fig:deproj20} and Fig.~\ref{fig:deproj20_press_abun}.
    Each sector is marked by its own unique plot symbol. The mean
    profiles are shown by the lines. The outermost points in each
    sector are not plotted. The radii shown are the mean radii for
    individual sectors.}
\label{fig:deproj20_plots}
\end{figure*}

Firstly we conducted a simple \textsc{mekal} fit accounting for
projection, to fit for the temperature, absorption and abundance. In
Fig.~\ref{fig:deproj20} is shown the temperature of each sector, its
electron density, the mean radiative cooling time, and its entropy.
Sectors occupied by the radio lobes or where the projection model
failed are not shown.  The electron density was calculated from the
emission measure of the model in each partial annulus. The mean
radiative cooling time was found by taking the ratio of the thermal
energy of the gas to its luminosity, and the entropy was calculated
using the relation $S = k T \: n_e^{-2/3}$. By multiplying the
temperature of each sector by its electron density we created a
pressure map, shown in Fig.~\ref{fig:deproj20_press_abun}~(Left). To
convert the electron pressure to total pressure, multiply by a factor
of 1.9. The azimuthal symmetry seen in this map, with for example no
`swirl' evident, argues against significant non-hydrostatic pressure
sources such as bulk motions. If we compute the ratio of the
standard-deviation to the mean of the pressure in the annulus where
the low temperature swirl is most prominent (sector 8 from the
centre), we find a spread of 6.7~per~cent. This is much less than the
spread of the density (14~per~cent) or of the temperature
(15~per~cent). As inspection of Fig.~\ref{fig:deproj20_plots} reveals,
local offsets in the temperature are anti-correlated with local
offsets in density.  The low temperature swirl does not therefore
indicate recent bulk motion in the gas (although the denser gas must
presumably sink).  A detailed temperature map of the core is shown in
Fig.~\ref{fig:deproj20_zoomT}. The radii of the sectors used were
matched to the radii of the rims and other features.

Fig.~\ref{fig:T_100_plots} shows that we do not see temperatures below
3 keV in the projected temperature analysis. Accounting for
projection, we see some temperatures below 2 keV on the N edge of the
radio lobe. In the innermost 40 arcsec of the cluster the analysis
accounting for projection effects fails and the temperatures become
unconstrained. This is because there is too little emission than
expected from projection alone. Therefore there is more emission from
gas on the plane of the sky than along the line-of-sight, and the
structures we observe are not spherically symmetric.

Accounting for projection by assuming spherical symmetry also appears
to fail in the outer NW radio lobe, which we discuss separately in
Section \ref{sect:radiolobes}. However, for much of the cluster the
morphology of temperature follows the morphology of the projected
temperature map closely (Fig.~\ref{fig:T_map}).

In Fig.~\ref{fig:deproj20_plots} are shown radial plots containing the
values for all of the sectors. The values in each sector generally
vary smoothly from annulus to annulus, except in the centre and at the
position of the outer NW radio lobe. In addition, although there is
variation from sector to sector, the sectors appear to follow the same
trends. There are some regions for which the entropy appears to
increase inwards away from the lobes and the cold rims (e.g. 230
arcsec from the core to the SE), suggesting there are convectively
unstable volumes in the core. In these plots we do not plot the
outermost sectors as the density is overestimated as we cannot
subtract the contribution from gas that lies outside that sector, nor
do we plot points were the projection model fails.

\begin{figure}
  \includegraphics[angle=-90,width=\columnwidth]{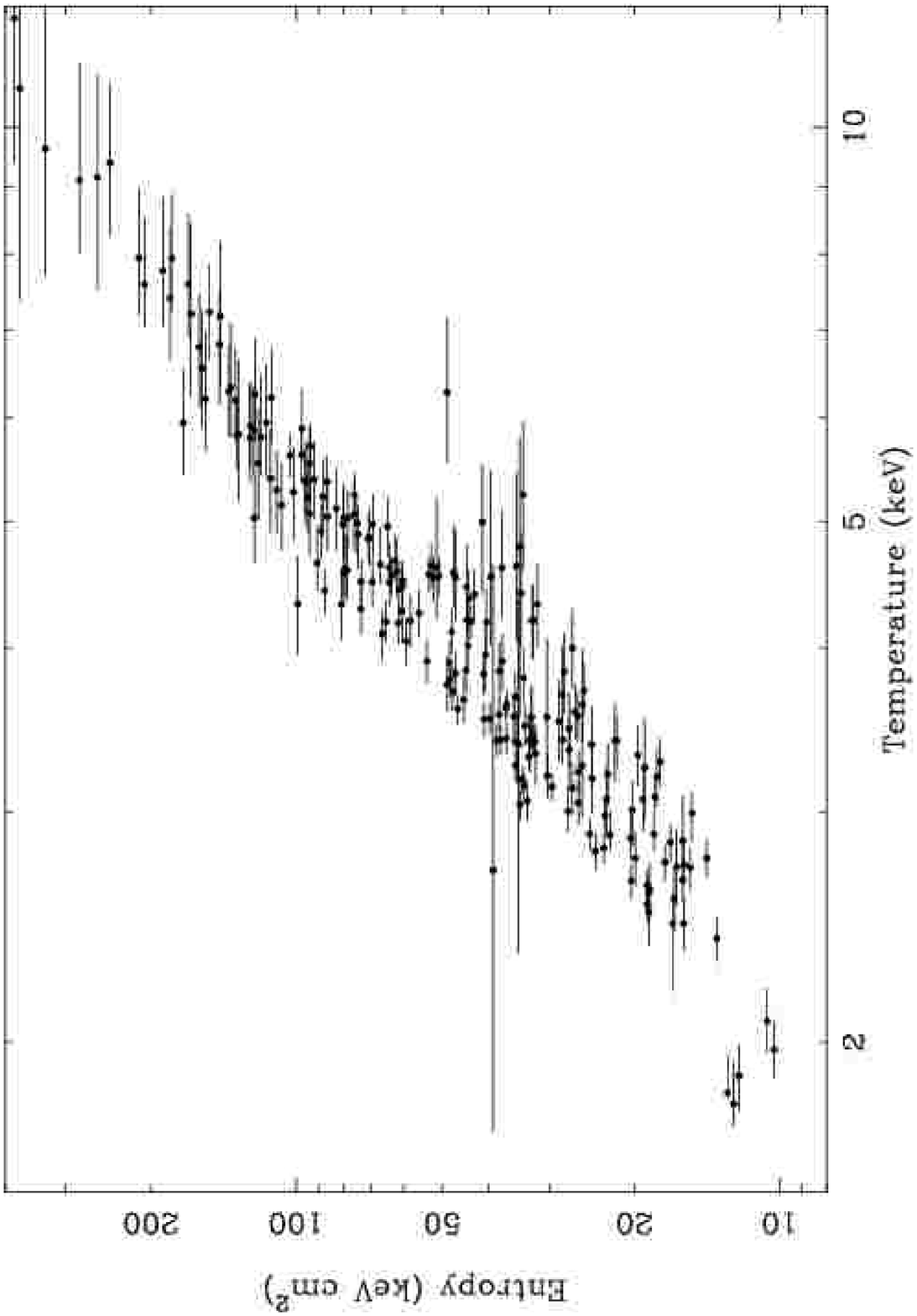}
  \\[2mm]
  \includegraphics[angle=-90,width=\columnwidth]{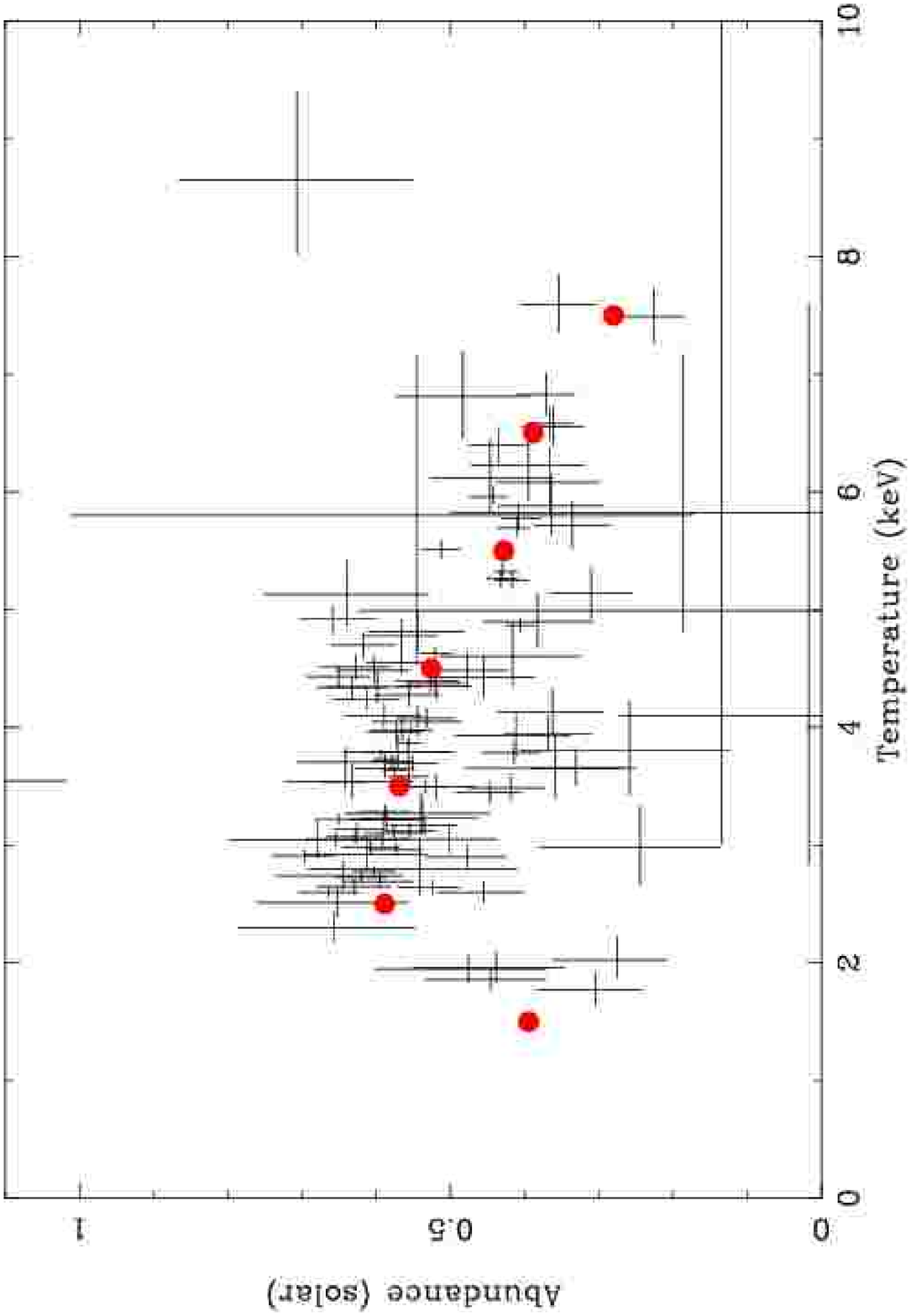}
  \caption{(Top) Entropy of the partial-annuli in Fig.~\ref{fig:deproj20}
    plotted against the temperature. Outer annuli, and sectors
    occupied by the radio lobes or where projection failed were
    excluded.  (Bottom) Abundance of each partial-annulus plotted
    against its temperature. The solid points mark the weighted mean
    abundance in 1 keV bins between 1 and 8 keV. Fewer radial sectors
    were used to create this plot than in the rest of this section,
    reducing the uncertainties on the abundance measurements.}
  \label{fig:T_varplots}
\end{figure}

We can plot the temperatures of individual bins against their
entropies (Fig.~\ref{fig:T_varplots}~[top]). The distribution is well
fit with the relation $S = 1.5 (kT)^{2.5}$, with $S$ in $\keVcmsq$ and
$kT$ in keV. There appear to be separate clumps of points in the plot
which are separate in temperature.

In Fig.~\ref{fig:deproj20_press_abun}~(right) we plot the abundances
of each of the sectors. There is a reasonable correspondence with the
projected abundance map in Fig.~\ref{fig:Z_map}. The abundance profile
peaks away from the centre like the projected abundance
(Fig.~\ref{fig:deproj20_plots}), although the points are more noisy.
We can also plot the abundance of sectors against their temperature
(Fig.~\ref{fig:T_varplots}~[bottom]) to test the projected relation in
Fig.~\ref{fig:T_Z_scatter}. Although the abundances have larger
uncertainties, when they are averaged in temperature bins, we see a
similar relation to Fig.~\ref{fig:T_Z_scatter}, but with a dip in
abundance at low temperatures.

\begin{figure}
  \includegraphics[angle=-90,width=\columnwidth]{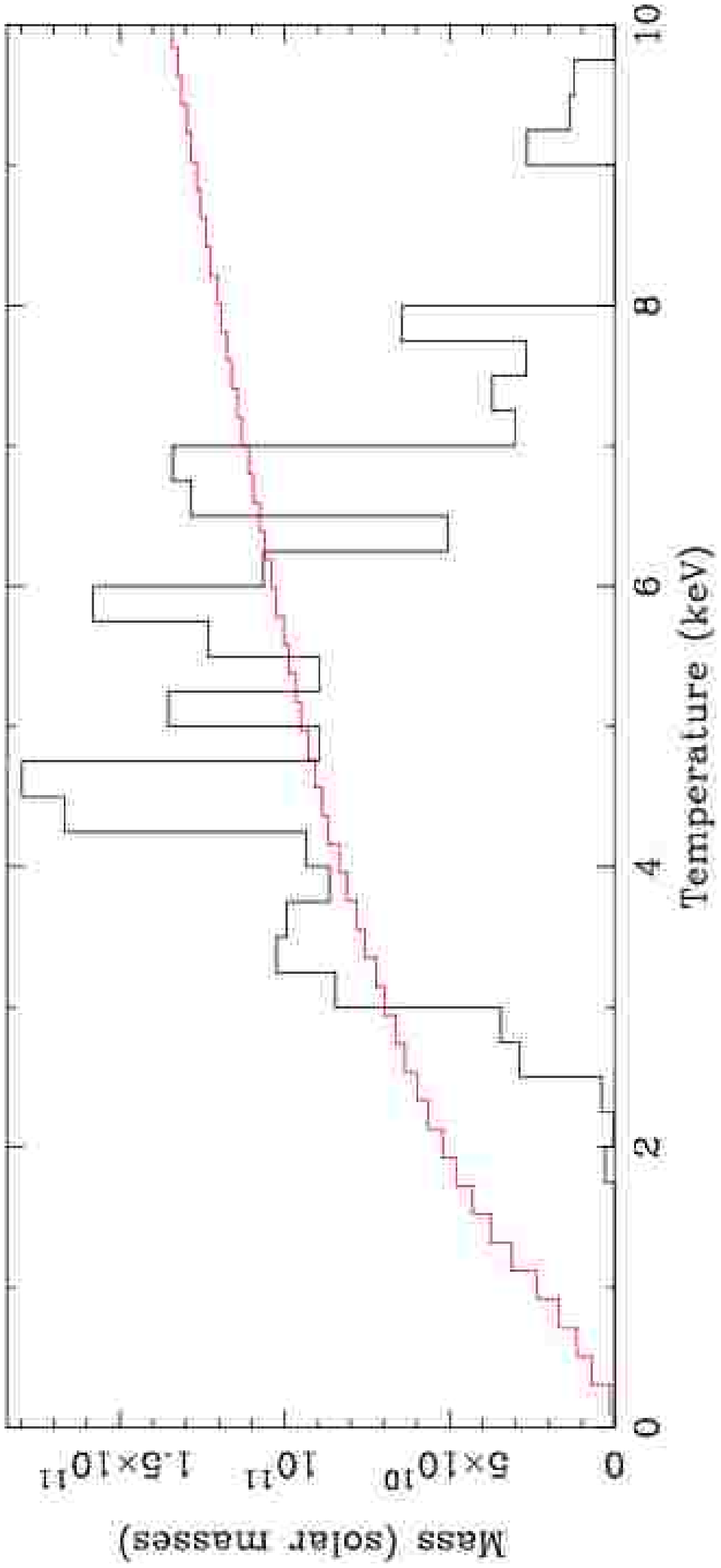}
  \\[3mm]
  \includegraphics[angle=-90,width=\columnwidth]{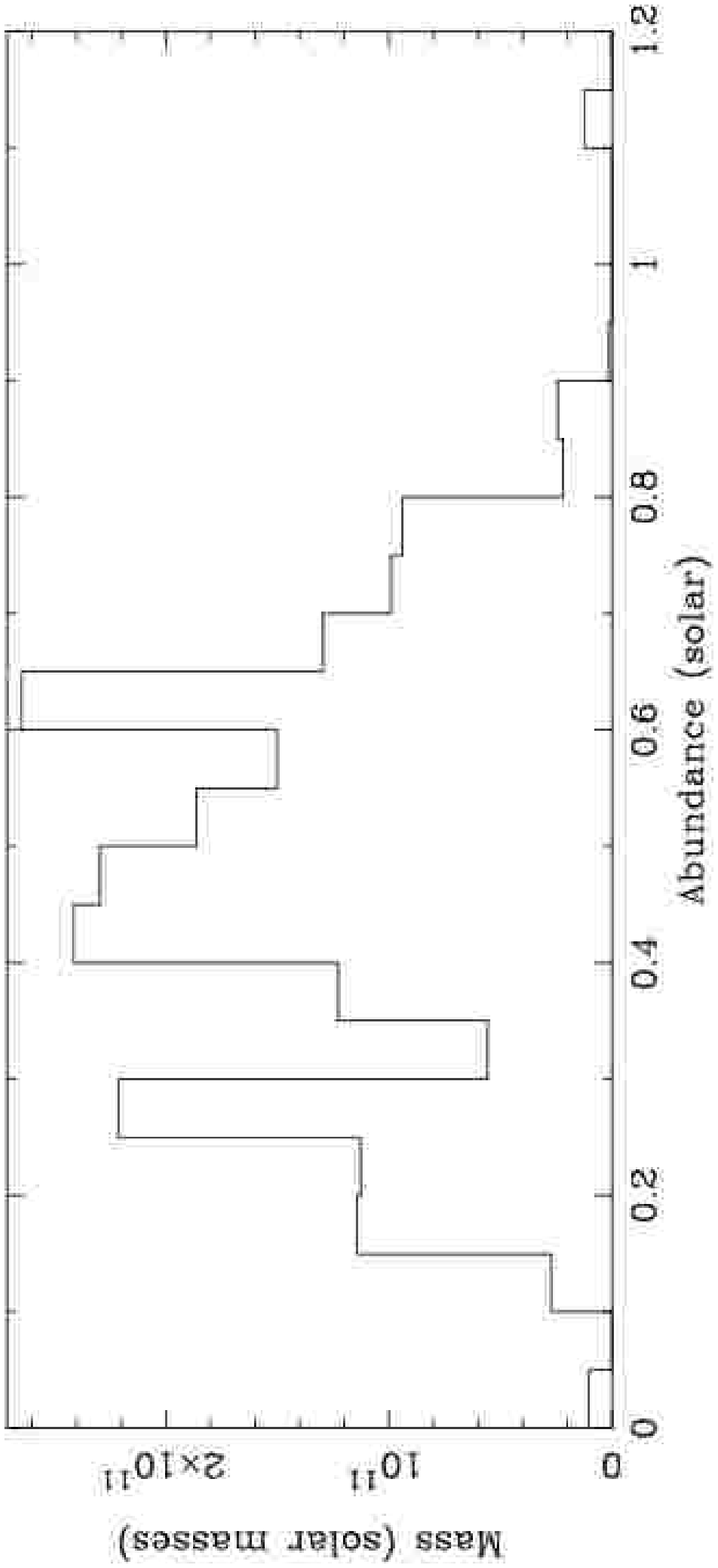}
  \caption{The gas mass as a function of temperature (summed in
    0.25 keV bins), and as a function of abundance (in $0.05 \Zsun$
    bins), computed from Fig.~\ref{fig:deproj20} and
    Fig.~\ref{fig:deproj20_press_abun}~(right). The dotted line on the
    temperature plot shows the distribution as expected in a cooling
    flow with $\dot{M}=300 \Msunpyr$, $P =5 \times 10^5 \Kpcmc$ and
    $Z=0.4\Zsun$.}
  \label{fig:deproj20_mass_temp}
\end{figure}

It is physically interesting to plot the quantity of gas present as a
function of temperature, which we can compare to what would be
expected from a standard cooling flow, and as a function of abundance.
Fig.~\ref{fig:deproj20_mass_temp} shows these distributions,
calculated using the temperatures, densities and abundances from
Fig.~\ref{fig:deproj20_plots}. Sectors in which the emission measures
were consistent with being zero were ignored in this analysis. On the
temperature-mass plot we plot a line representing the amount of mass
we would expect as a function of temperature in an isobaric cooling
flow, with the mass deposition rate $\dot{M}=300 \Msunpyr$, the
pressure $P = n_e T =5 \times 10^5 \Kpcmc$, and abundance $0.4 \Zsun$.
The amount of mass in a temperature interval, $\Delta M \propto
\dot{M}/(P \Lambda)$.  Increasing the abundance or pressure (which can
be functions of temperature) will decrease the mass in a temperature
interval.

Gas is fairly uniformly distributed by mass over a factor of 3 range
in temperature, from 2.5 to 8~keV, with little gas detected below
2~keV. The shape of this observed distribution is the essence of the
`cooling flow problem' (Peterson et al. 2003; Fabian 2003).

\subsection{Fitting individual abundances}
\label{sect:proj_abun}
\begin{figure*}
  \includegraphics[angle=-90,width=0.33\textwidth]{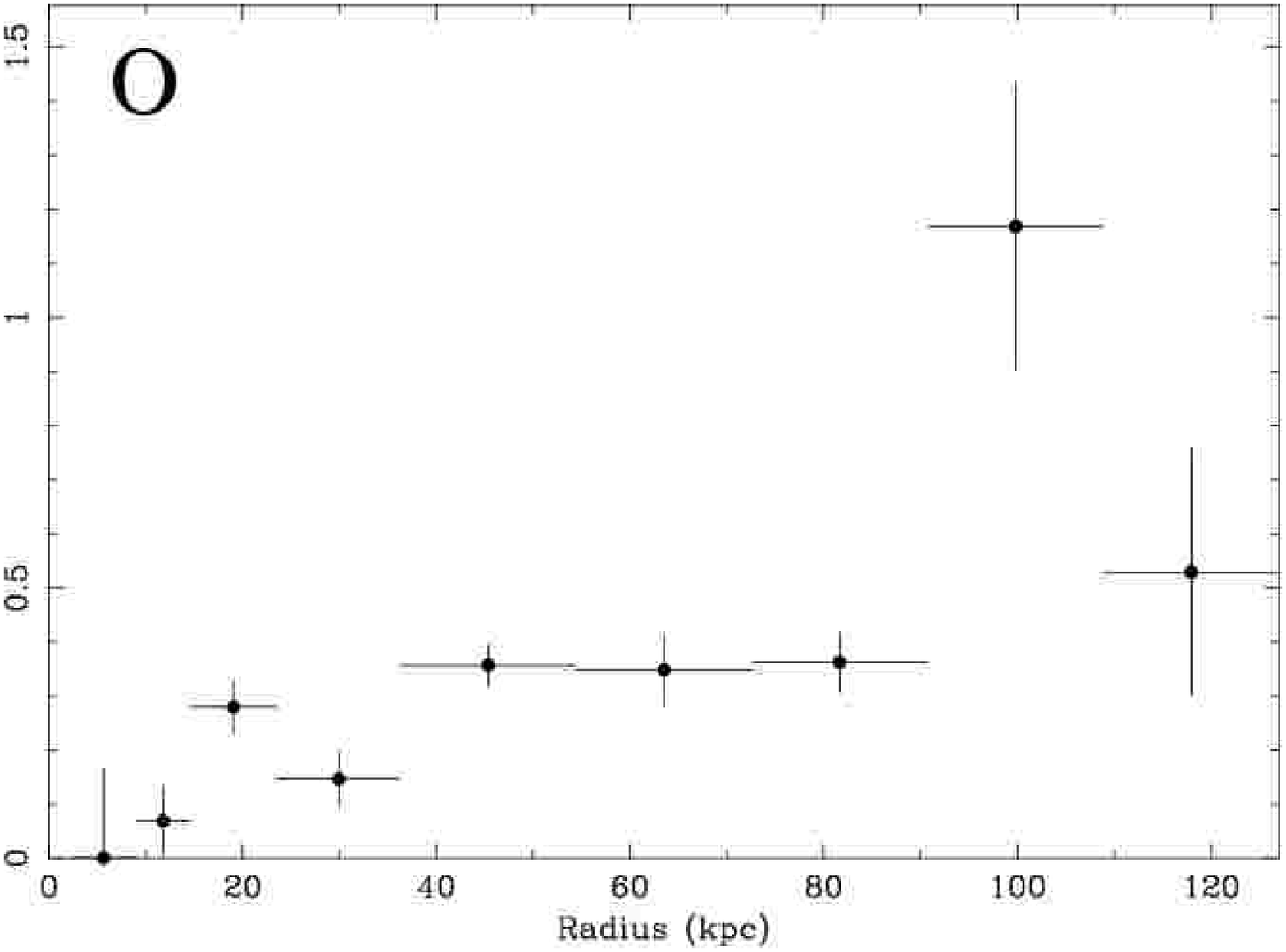}
  \includegraphics[angle=-90,width=0.33\textwidth]{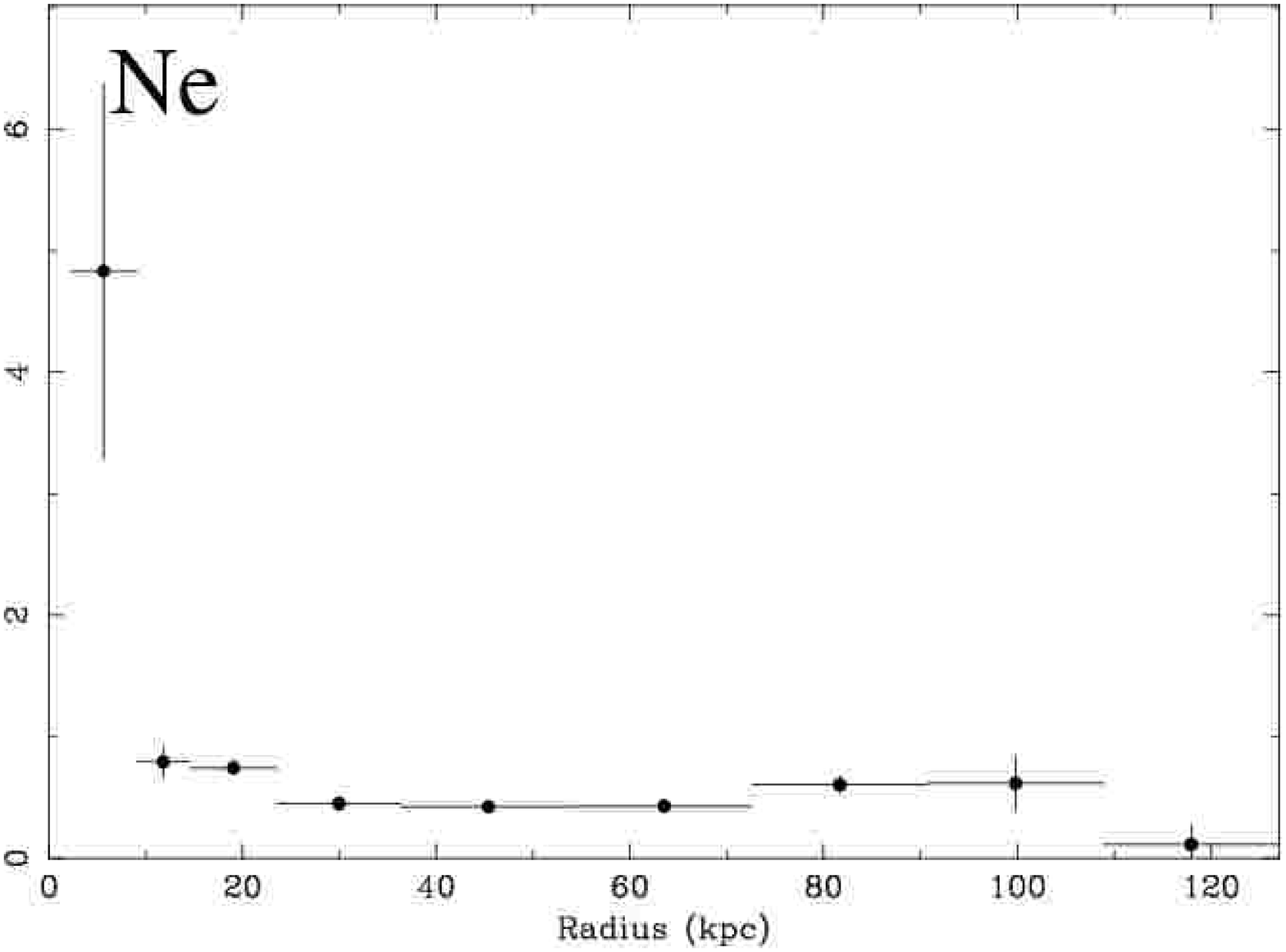}
  \includegraphics[angle=-90,width=0.33\textwidth]{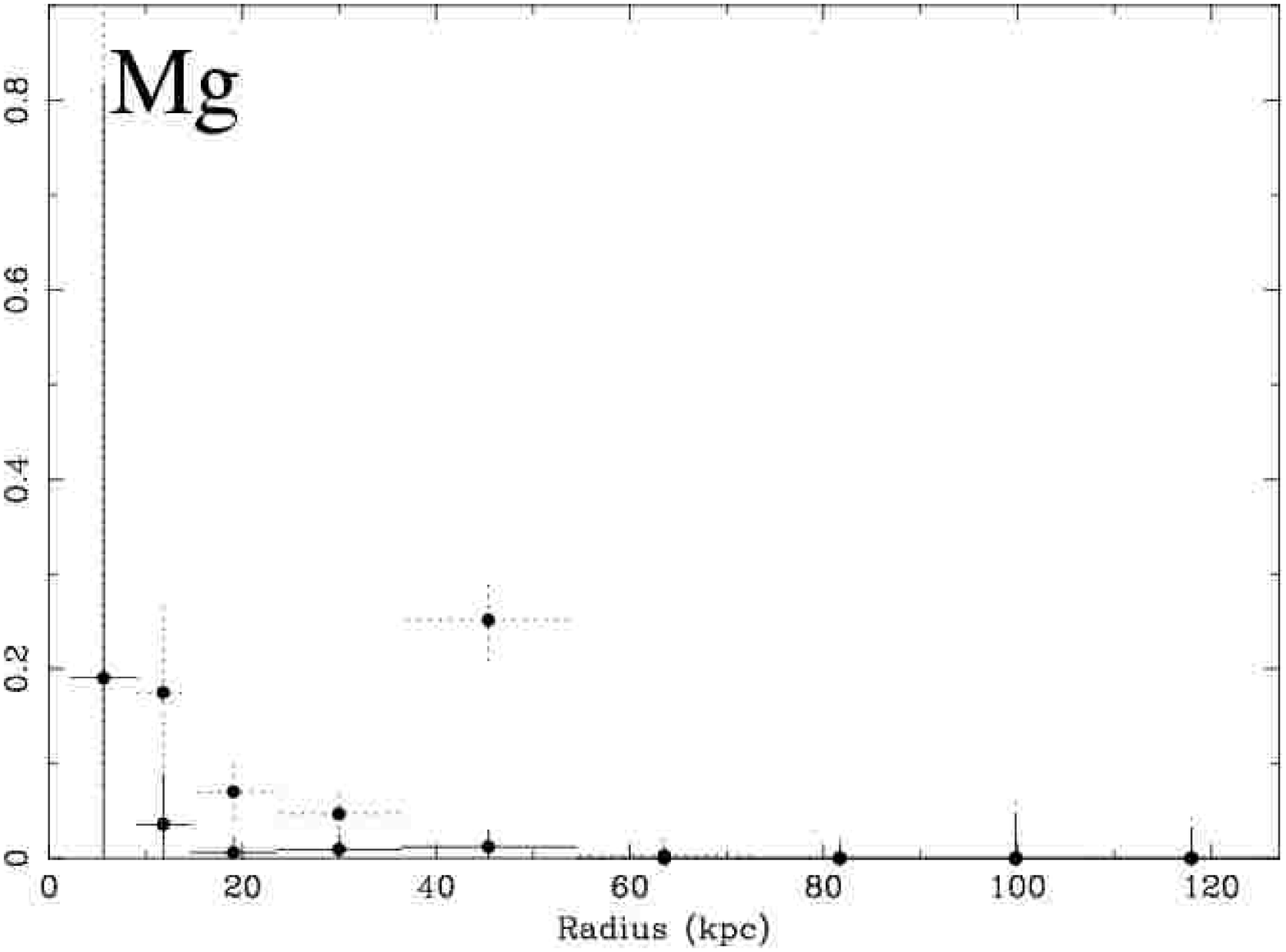}
  \includegraphics[angle=-90,width=0.33\textwidth]{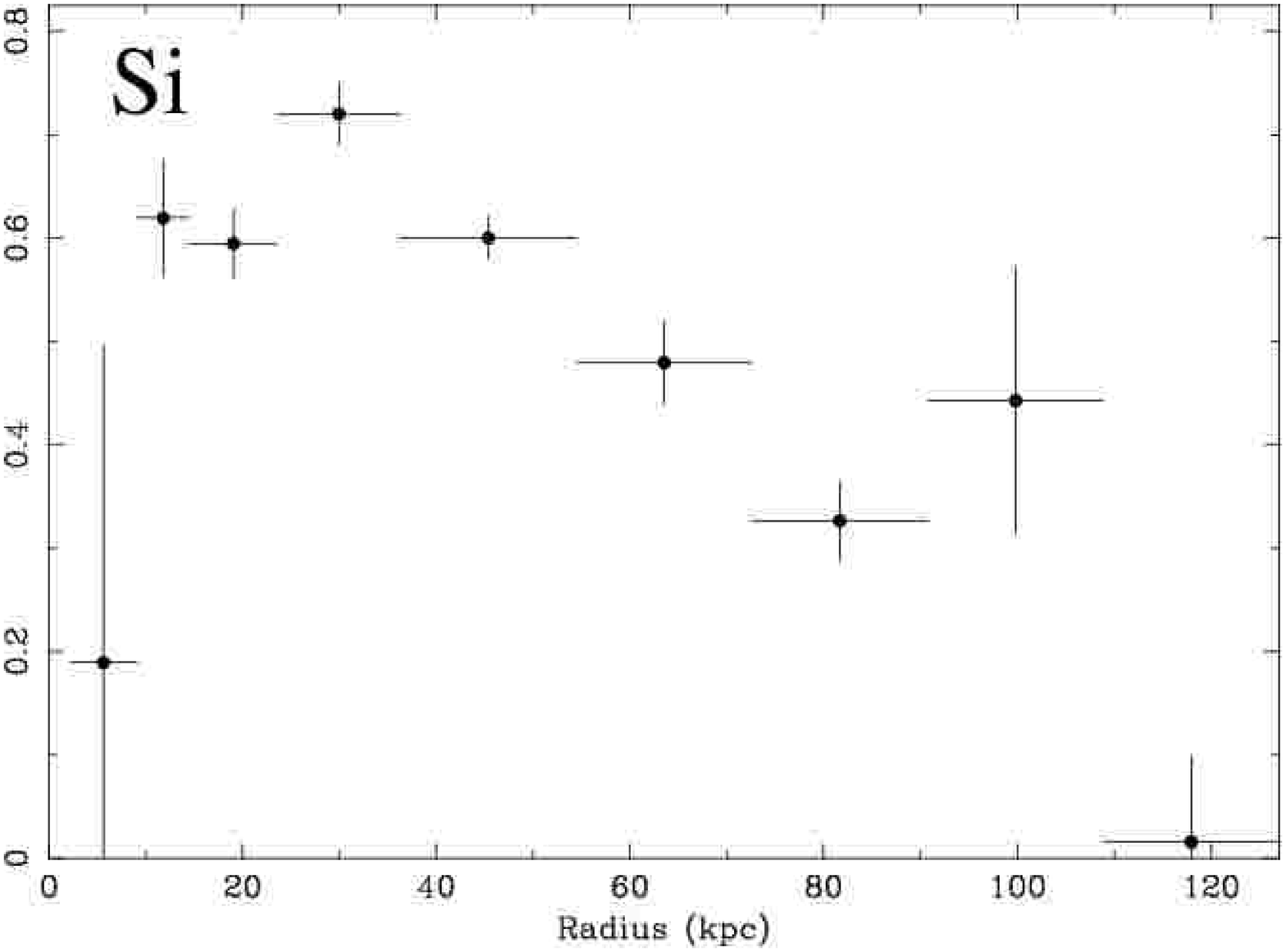}
  \includegraphics[angle=-90,width=0.33\textwidth]{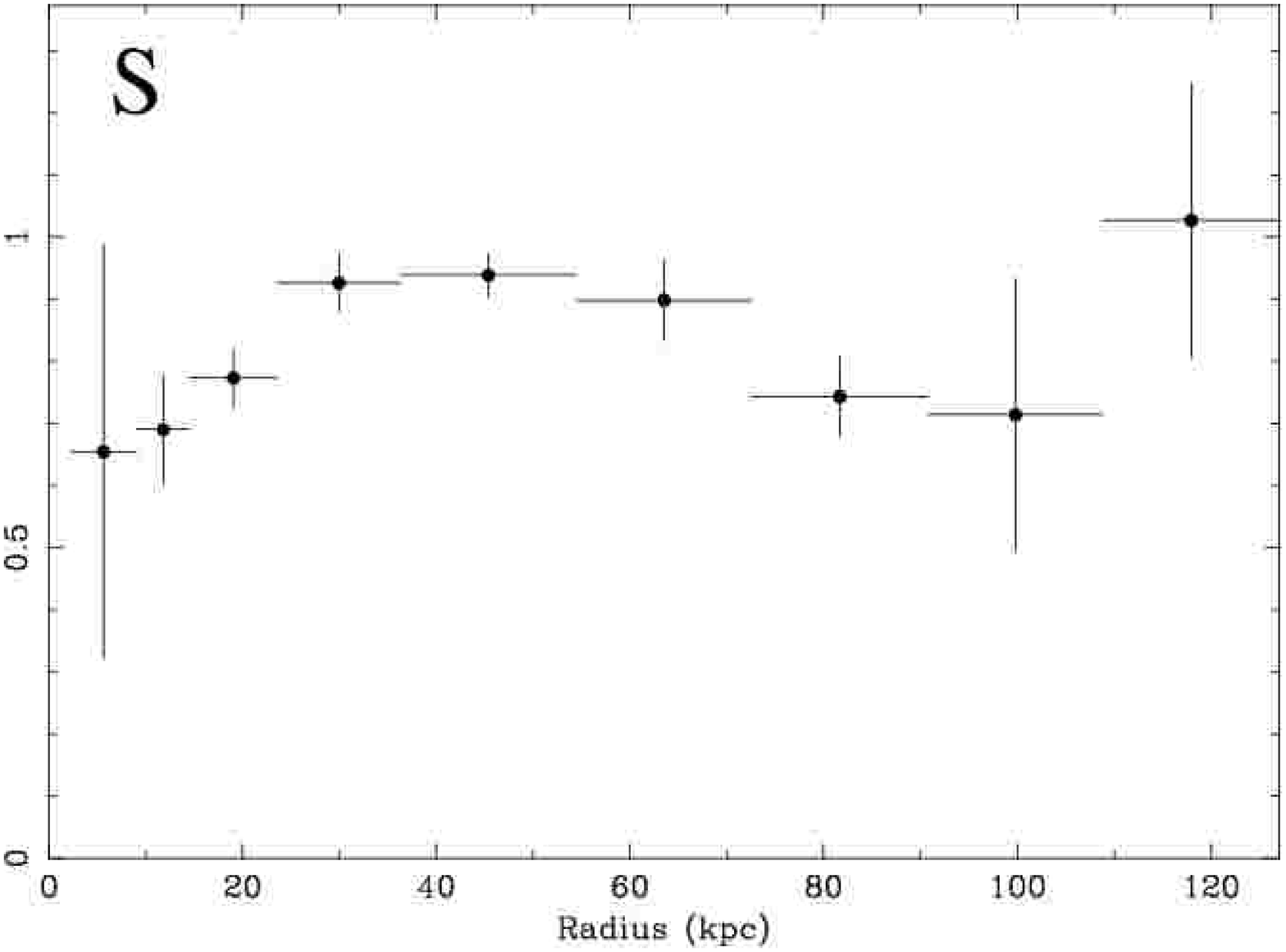}
  \includegraphics[angle=-90,width=0.33\textwidth]{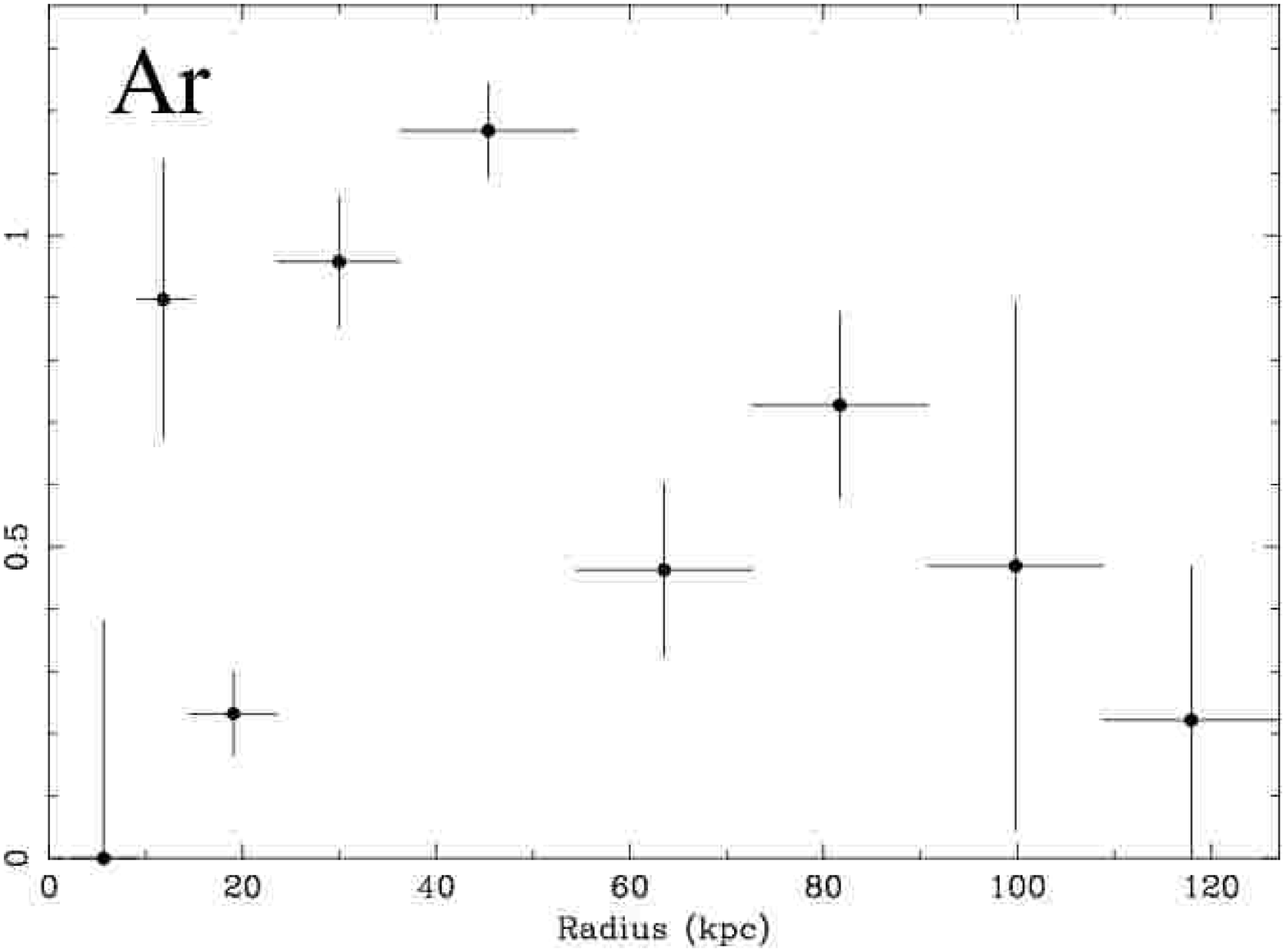}
  \includegraphics[angle=-90,width=0.33\textwidth]{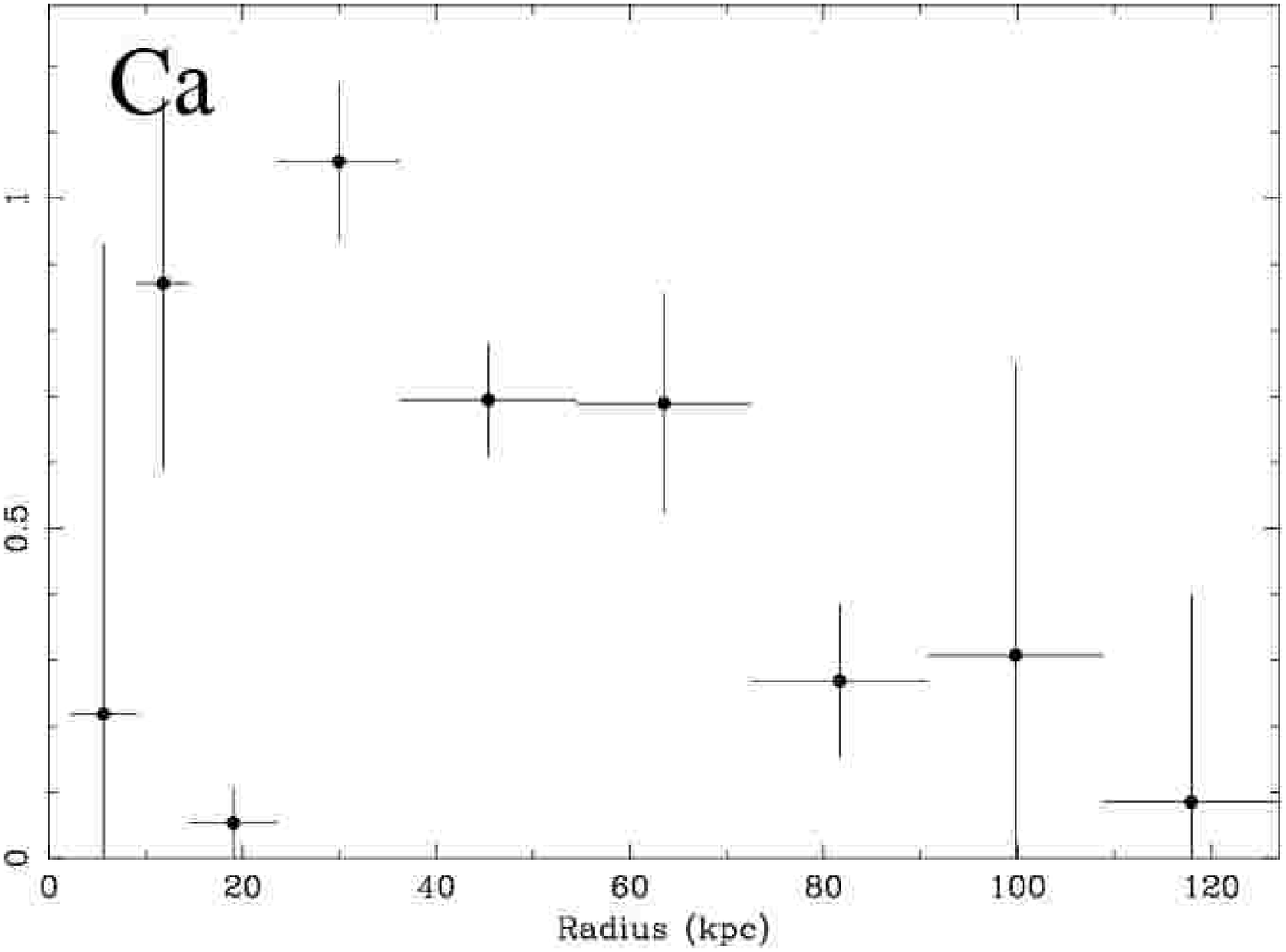}
  \includegraphics[angle=-90,width=0.33\textwidth]{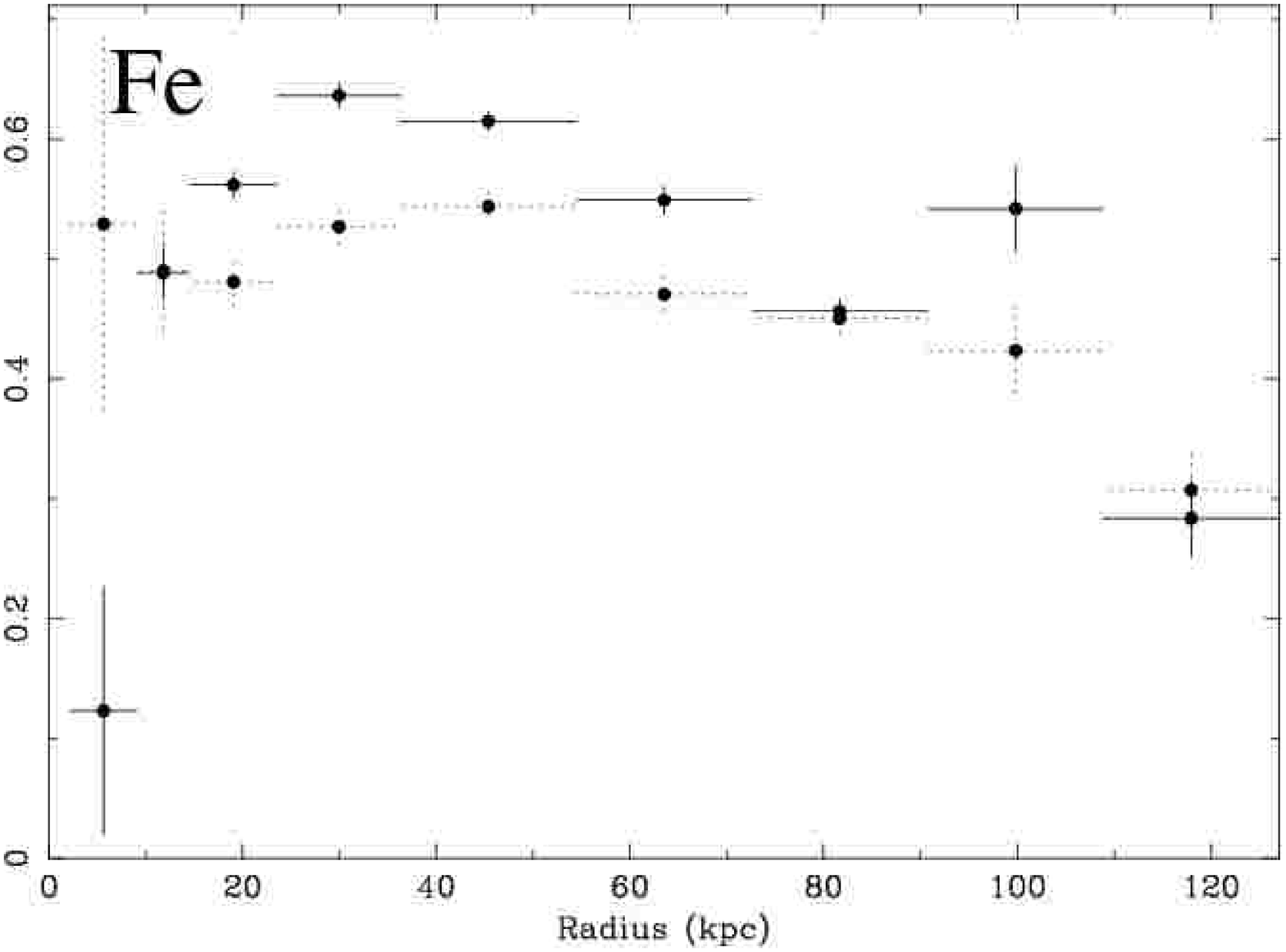}
  \includegraphics[angle=-90,width=0.33\textwidth]{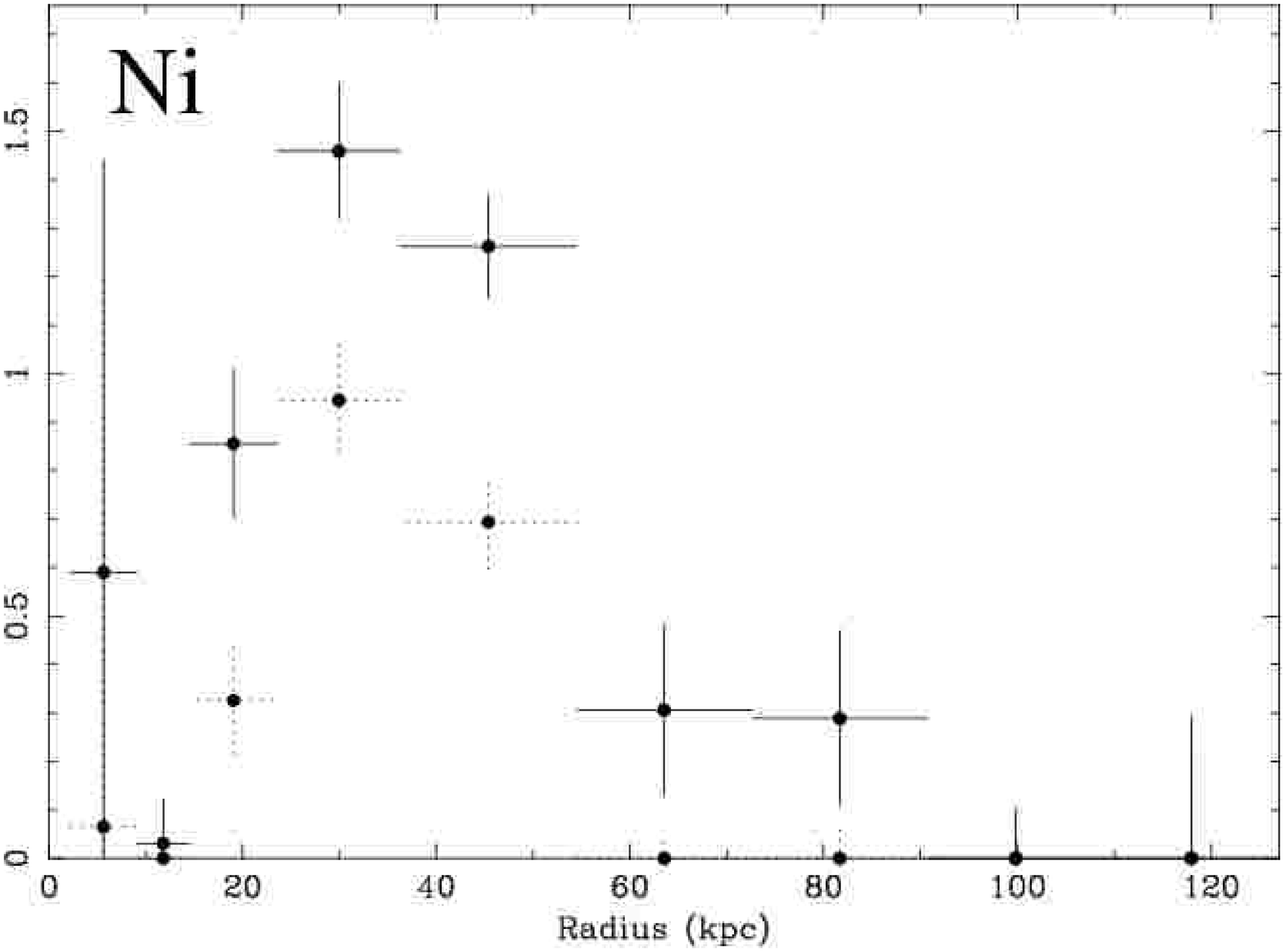}
  \caption{Average abundance profile in the four sectors, produced by
    fitting the full 0.6-8 keV band with a \textsc{vmekal} model
    accounting for projection. The values are the weighted mean of the
    four sectors (symmetrising uncertainties on the sectors and
    excluding any sectors where the temperature is undefined) The
    uncertainties on the points are the uncertainty on the mean. The
    dotted points on the Ni and Mg plots show the results using a
    \textsc{vapec} model, and the dotted points on the Fe plot show
    the Fe-K abundances.}
  \label{fig:abun_profiles}
\end{figure*}

We fit \textsc{vmekal} models with variable abundances to the spectra
extracted from partial-annuli, accounting for projection, and using
the whole band 0.6 to 8 keV. We divided the cluster up into only four
equal sectors, starting the first sector $25^\circ$ to the NW from the
N. The model was fit to the whole band, and the O, Ne, Mg, Si, S, Ar,
Ca, Fe and Ni abundances were thawed. The Ni, Mg, Si and S abundances
may be contaminated with the residuals near 2 keV.
Fig.~\ref{fig:abun_profiles} shows the elemental abundance profiles,
produced by taking the best fitting values in the four sectors, and
computing a weighted mean of the values. This procedure was done to
allow for the variation in temperature in each sector. The
uncertainties of each point are the uncertainties on the weighted
mean. The uncertainties on the best fitting abundances were
symmetrised using the RMS uncertainty. Any sectors where the
temperature was compatible with less than 1 or greater than 14 keV
were excluded. The Ni and Mg results using a \textsc{vapec} model are
also shown. Ni abundances are substantially smaller using this model.
The Fe plot also shows a profile only using the Fe-K lines. The
central values should be taken with caution as the deprojection
analysis appears to fail in three of the four sectors.

The fact that the Si, S, Ar, Ca, Fe and Ni profiles all drop in the
centre of the cluster strongly suggest that `Fe-bias' is not the cause
of the decrease.

\subsection{Multiple temperature components}
\begin{figure}
  \includegraphics[width=\columnwidth]{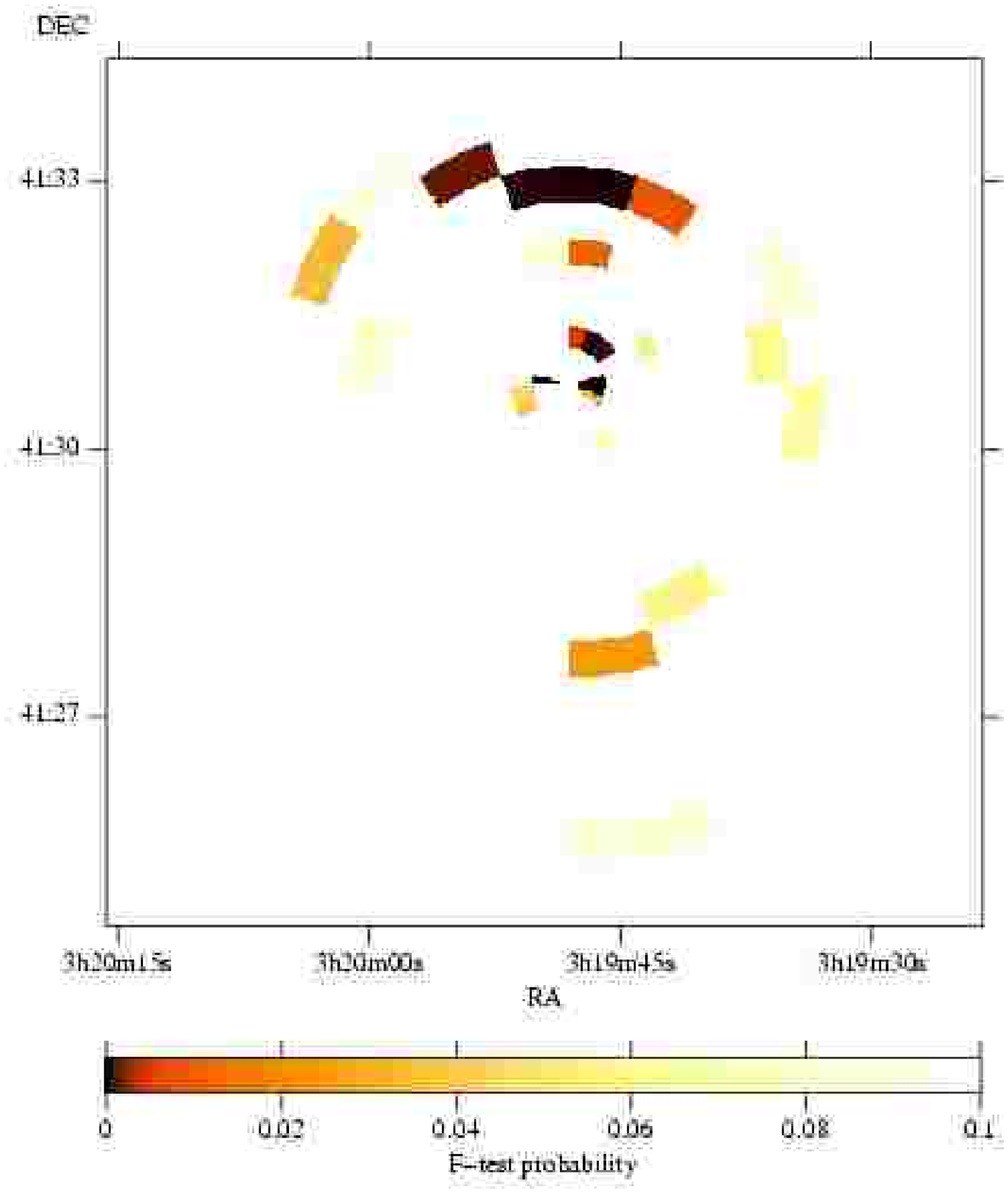}
  \caption{F-test probability of improvement of $\chi^2$ given by
    second component being by chance.}
  \label{fig:deproj_2tftest}
\end{figure}

A simple test for the presence of multiple temperature components is
to add a component at one half of a fitted temperature. It is
difficult to distinguish a temperature component less than a factor of
two away. In Fig.~\ref{fig:deproj_2tftest} we show the probability
that the improvement in $\chi^2$ obtained in adding a second component
is purely due to chance. The plot was generated by fitting a single
temperature model to each annulus, accounting for projection,
measuring $\chi^2$ of the fit, and then adding a second component, and
again measuring $\chi^2$. Then the F-statistic was generated and used
to find the probability. Annuli outside the annulus being fitted
always have two components, so the probability is purely measured in
the annulus.

We find that there is not a substantial improvement of $\chi^2$ given
by multiple temperature fits, except on the N rim of the radio lobe
(which is spatially narrower than the bin used), and to the W of the
nucleus.

We also repeated the multiple temperature component analysis of
Section \ref{sect:multicomp} accounting for projection in six sectors.
The results from the two sets of analyses appear to be quite similar,
so we do not show the versions accounting for projection here. In the
0.5 keV plot there is a low level of gas over much of the core, but
more at the centre. In the 1 keV plot there is little evidence for gas
at that temperature, except right in the core and in the outer
regions. At 2 and 4 keV, the cool swirl we see in the projected map is
seen again.  At 8 keV the outer parts of the core are visible, and
there is some apparent presence of gas near the centre.  At 16 keV
there are some differences between the projected map and the map we
present here.  Accounting for projection, we do not see gas near 16
keV as extensively distributed to the N of core, but we do see it to
the S in the outer regions.  If we examine the abundances produced
using this analysis, the Ni and Fe abundances still appear to drop
towards the centre in several sectors.

Additionally, if we perform the same analysis as in Section
\ref{sect:proj_abun} but adding a second temperature component at half
of the fitted temperature, none of the profiles look substantially
different to the single temperature results.

\section{Radio lobes}
\label{sect:radiolobes}
We examined the X-ray emission in detail from the features associated
with the radio lobes, using the \textsc{projct} model in
\textsc{xspec} to account for projection effects.  Spectra were
extracted from sectors moving out from the radio lobes. We examined
the inner SW, NE, and outer S and NW radio lobes. Early work on the
filling factor of the holes using the 25~ks dataset was reported by
Schmidt et al (2002).

\begin{figure}
  \includegraphics[angle=-90,width=\columnwidth]{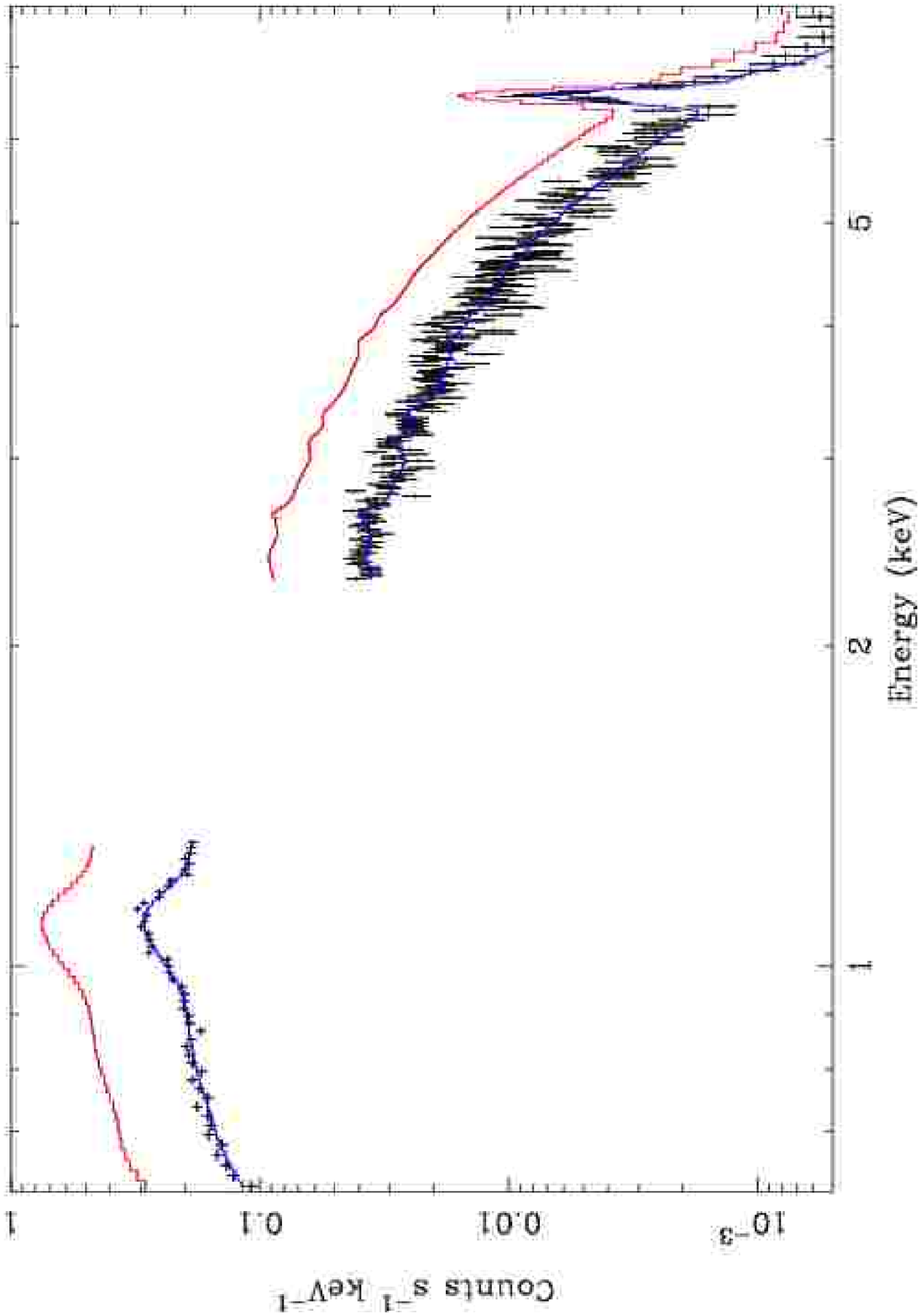}
  \caption{The points show a spectrum extracted from the inner SW
    radio lobe. The upper line shows a model for the spectrum assuming
    there is no emission from the radio lobe, and the only
    contribution is projected emission (assuming spherical symmetry in
    a sector). The lower line shows the same model, but reduced by a
    \textsc{mekal} component to account for the excess emission on the
    plane of the sky.}
  \label{fig:lobespec}
\end{figure}

We found that this simple deprojection analysis failed. There was too
little emission from the inner SW, NE and outer NE lobe regions than
would be expected from the projected emission of overlying gas,
assuming spherical symmetry in the sectors.  Fig.~\ref{fig:lobespec}
shows the projected spectrum of the inner SW radio lobe (points). The
upper line in the plot is the expected projected emission from the
region if the lobe itself were void. Since we found less emission than
we excepted from projection, there is an enhancement outside the lobes
on the plane of the sky which is not present along the line of sight.
This effect may simply be due to the radio lobes displacing the hot
gas outward, so leading to excess emission beyond the lobes. To
account for this difference, we fit the deficit in emission in the SW
lobe with a thermal spectrum, resulting as the reduced prediction
shown as the lower line in Fig.~\ref{fig:lobespec}. The temperature of
the thermal component is $\sim 3$ keV.  In contrast, there is more
soft emission in the outer S lobe than would be expected by
projection. Taking into account these soft components, we find little
evidence for any emission from the radio lobes themselves.

The radio spectral index in the region of the SW lobe is about unity
(Fabian et al 2002) and the 330~MHz radio luminosity about $3\times
10^{39}\ergps$ from the maps of Pedlar et al (1990). This is
considerably smaller than our limit for deprojected power-law X-ray
emission with the same spectral index of about $2\times 10^{43}\ergps$
(once the 3 keV soft component is taken into account).  The radio to
X-ray luminosities should be roughly in the ratio of the energy
density in the magnetic field to that in the cosmic microwave
background (assumed to be source of target photons for inverse Compton
scattering). This gives a lower limit on the magnetic field of
$B>10^{-7}$~G, much less than the likely field of $10^{-5}$~G (Fabian
et al 2002). Reworking this result using the formulae of Harris \&
Grindlay (1979) gives a limit about 1.5 times smaller.  This means
that our current X-ray observations are far from detecting the inverse
Compton emission, mainly due to the brightness of the thermal gas.

\section{Discussion}
\subsection{Temperature structure}
The high resolution images of the projected emission-weighted
temperature in Section \ref{sect:temperature}, illustrate that there
is a considerable wealth of temperature structure in the cluster.
There is a factor of 4.3 in projected temperature between the coolest
gas we observe (2.4 keV) to the hottest (10.3 keV). If we take the
coolest blob and use a neighbouring region as a background spectrum,
the best fitting temperature is $1.45^{+0.28}_{-0.06}$ keV. This is
comparable with the temperature of the X-ray emission associated with
the H$\alpha$ filaments (Fabian et al 2003b).

Accounting for projection effects (Fig.~\ref{fig:deproj20_plots}), we
observe the gas near the nucleus goes down to $\sim 1.8 \keV$
(although spherical symmetry assumptions fail very close to the
central object).  The shortest radiative cooling time we measure in
this analysis is around 0.18 Gyr (although if we underestimated the
abundance of material in the core, this period will be shorter).

Looking at the results from the multiple temperature component
analysis (Fig.~\ref{fig:multicomp}), then there may be some gas,
especially near the core, at temperatures near 0.5 keV with a low VFF.
There may be a significant amount of gas near 1 keV around the nucleus
too. Surprisingly, the gas at 0.5 keV appears to take up more volume
than that at 1 keV. These results, particularly the surprising 0.5 keV
map, depend on there not being significant systematics there.  Once
the response below 0.6 keV is more accurately determined we will be
able to confirm whether gas is at these temperatures.  If there is gas
at low temperatures then this may indicate there could be gas cooling,
but at rates reduced from earlier studies.

Another surprising result from fitting multiple temperature components
is the existence of a band of gas near 8 keV to the N of the nucleus.
Some of the band coincides with the position of the HVS, but it is
more extensive. It would be surprising if this were shocked gas from
the HVS given its VFF and extent. Although we do have limited spatial
resolution, some of the band appears coincident with the weak shock
identified to the NE (Fabian et al 2003a). The remainder of this gas
appears to be within this radius, so it is possible we are observing
post-shocked gas.

The 16~keV map is difficult to interpret. It is not uniform, so is
unlikely to be due to incomplete background subtraction.  In addition
there is no indication of emission where the gas is the hottest,
especially to the SE corner of the map.  It may still be a background
effect if there are other significant systematics, but reassuringly
the count rate in a 10 to 12 keV band agrees with the rate in the
blank-sky observation we have used for background spectra to 2
per~cent.  The most probable explanation is that there is diffuse
non-thermal emission over the cluster core.  The morphology of this
emission is not matched to the radio lobes, although at the location
of the radio lobes there appears to be emission. If this is
non-thermal emission, rather than some sort of unidentified systematic
uncertainty, then it could be due to a different population of
particles to those producing the radio lobes. It may, for instance, be
associated with the radio mini halo in the Perseus core (Pedlar 1990;
see also Fabian \& Kembhavi 1982; Gitti, Brunetti \& Setti 2002).  The
radio mini halo does have some morphological resemblance to the region
where the 16~keV component is strongest.  The presence of a hard
component over the inner region of the Perseus cluster and other
nearby clusters had also been indicated by previous \emph{ASCA}
observations (Allen et al 2001).

\begin{figure}
  \includegraphics[angle=-90,width=\columnwidth]{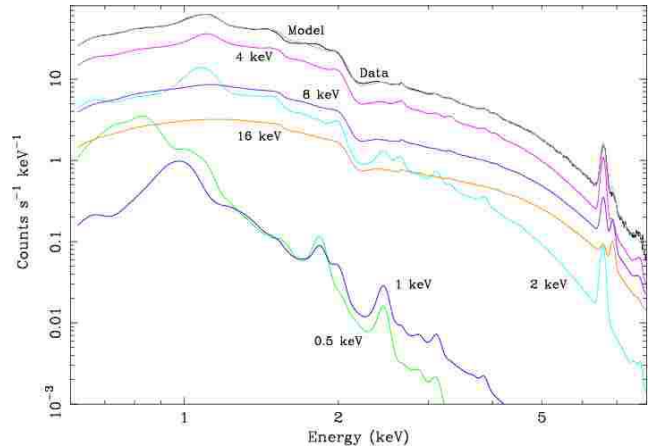}
  \caption{The contributions of the temperature components to the
    muli-temperature model, which is shown in grey. The model was
    fitted to the total ACIS-S3 spectrum, which is plotted in black
    beneath the total model. The spectrum contains $1.5 \times 10^7$
    counts.}
  \label{fig:multicomp_spec}
\end{figure}

In Fig.~\ref{fig:multicomp_spec} we show the individual temperature
component contributions to a multi-component fit of the entire ACIS-S3
spectrum.  If we replace the 16 keV component with a more physically
relevant power-law component (which incidentally gives a better fit to
the spectrum), then the best-fitting power-law index is 1.3, and its
luminosity is $3.3 \times 10^{43} \ergps$ in the 1 to 10~keV band.
This luminosity is around three times greater than the luminosity of
the nucleus as reported by Churazov et al (2003a; agreeing with the
luminosity calculated from our moderately piled-up spectrum of the
central source). The $\chi^2$ of the fit reduces by around 200 with a
hard component present.  Future X-ray missions, such as
\emph{ASTRO-E2}, will be able to easily detect this hard-component if
it is not due to background effects.

\subsection{Absorption structure}
A basic interpretation of Fig.~\ref{fig:NH_map} would indicate there
could be evidence of absorbing material associated with the coolest
gas. Unfortunately this may be a result of either instrument
calibration, and the low energy QE degradation.

\subsection{Abundance structure}
The metallicity of the cores of cooling flow clusters has long been
known to differ from the outer parts of the same clusters. An iron
metallicity gradient is often observed (Fukazawa et al 1994; Allen \&
Fabian 1998; Irwin \& Bregman 2001; De Grandi \& Molendi 2001).
\emph{Chandra} data sometimes reveals that it drops again in the
centre (Sanders \& Fabian 2002; Johnstone et al 2002; Schmidt et al
2002; Blanton, Sarazin \& McNamara 2003).  The behaviour of the
abundances of other elements is not straightforward (e.g. Tamura et al
2001; Finoguenov et al 2002; Ettori et al 2002), with oxygen showing
little change across the core and nickel showing a strong excess
(Dupke \& Arnaud 2001).

A general picture was established of Type II supernovae dominating the
bulk of the gas (especially the outer gas) and the central abundance
peak being dominated by Type Ia supernovae. The details of the
observations do not however fit completely with such a simple picture
and consideration has been made of an early population of hypernovae
contributing to all the gas (e.g. Loewenstein 2001) and of there being
different classes of Type Ia events (see Baumgartner et al 2003 and
Loewenstein 2003 for an overall discussion of all supernova types).
Other issues that could be relevant here are an `Fe bias' due to the
gas really being multiphase but treated as single-phase (Buote \&
Fabian 1998; Buote 1999), rapid cooling of metal-rich regions (Fabian
et al 2001; Morris \& Fabian 2003), sedimentation of metals (Fabian \&
Pringle 1977; Gilfanov \& Syunyaev 1984), resonance-line scattering
(Gilfanov et al 1987), gas flows (Fabian 2003) and photoionization by
the central active nucleus.

We find in Perseus that O and Mg (perhaps our most unreliable
measurements) show no central peak whereas Fe, Si, S, Ca and Ni show
an off-centre peak. Ne is the only element which peaks at the centre.
Ni reaches the highest metallicity which is plausibly an indication of
strong Type Ia enhancement from the central galaxy NGC\,1275. The
peaks in Si and S may indicate that the Type Ia events produce
significant quantities of Si and S, although a combination of Type Ia
and II may be more relevant. Given that the spectrum of the body of
NGC\,1275 is, despite being a giant elliptical, of spectral type A,
then a steady occurrence of supernovae are expected, of both Type Ia and II
(SN\,1968A in NGC\,1275 was of Type I; Barbon, Cappellaro \& Turatto
1989; Meusinger \& Brunzendorf 1996; Capetti 2002). The central Ne
peak unexpected, but it should be noted that the Ne-K
lines occur within the band containing the Fe-L lines so some spectral
confusion could occur if the temperature structure of the gas is not
simple. (Ettori et al 2002 find a broad central Ne peak in A\,1795.)

The abundances and radial trends we find, ignoring Mg, are very
similar to those found in M87 (Finoguenov et al 2002; Gastaldello \&
Molendi 2002; Matsushita, Finoguenov \& B\"ohringer 2003). Si however
is more abundant in M87, peaking at a value about twice that found
here in Perseus. The overall agreement is impressive given that the
temperatures in Perseus are 2.5 -- 3 times those at the same radius in
M87. The ratio S/Si is about 1.5 in Perseus but less than one in M87.
Matsushita et al (2003) argue for a significant Si contribution from
SN Ia.

The excess mass of iron in the inner 70~kpc of the Perseus cluster is
about $2\times 10^8\Msun$. Excess here means exceeding a metallicity
of 0.4. The similar mass for M87 is $7\times 10^7\Msun$. The 2MASS
total $J$ magnitudes of the two galaxies indicate that NGC\,1275 is
about 1.6 times the stellar mass of M87, whereas the excess iron
masses differ by a factor of three. The overall difference may be due
to continued supernova activity in NGC\,1275.

Resonance scattering (Gilfanov et al 1987) has been ruled out for the
bulk of the Fe-K line emission in the Perseus cluster by studies of
the Fe-K$\alpha$/K$\beta$ ratio with \emph{XMM-Newton} (Churazov et al
2003b; Gastaldello \& Molendi 2003).  Mild motions of the intracluster
medium can account for this.  Nevertheless, it is possible that the
central emission and some of the L-shell lines are still thick to
resonance scattering.

This could account for the central differences in Fe-K and Fe-L
abundance measurements and indeed for the general drop in central
abundances. The central Ne peak may be an indication that resonance
scattering is affecting the Fe-L emission at the centre, in the
following manner. The expected Ne\,X emission (12.2A) coincides with
non-resonant L-shell emission from FeXXIII and FeXXII. If the Fe-L
resonance lines from FeXXIV and FeXXIII (at 10.6, 11, 11.4 and 11.8A)
are significantly scattered near the centre then we fit an
artificially low iron abundance to that region. Emission at 12.2A is
then wrongly attributed to Ne. In other words, the present evidence
supports a resonance scattering interpretation of the apparent central
drop in metal abundances from many species. The apparent peak in Ne is
due to Fe-L not Ne. Spectral fitting of projected spectra in the
central regions show that this hypothesis is tenable. We shall pursue
these issues in detail elsewhere.

Finally, we note the similarity of our mean abundance-temperature
values (Fig.~\ref{fig:T_varplots}~[bottom]) of the gas in the core of
the Perseus cluster with those of the large ASCA sample studied by
Baumgartner et al (2003; their Fig.~2). They eliminated the brightest
clusters from this plot and also were usually measuring the whole
observed cluster. This suggests a universal $Z-kT$ relation which
would be very puzzling. However, the Centaurus cluster and M87 have
profiles which, although of similar shape, are both stronger and
shifted to lower temperatures. There is not therefore a fixed
universal profile, although the peaked shape seems to be universal. An
overall summary is that most cluster gas which is above 6 keV has a
metallicity below about 0.3, whereas most gas at 3 keV has a
metallicity (iron) of 0.4 or higher. At the lowest temperatures
observed the metallicity drops to 0.2 or less. We do not understand
this result{\footnote{Note that if the apparent metallicity of cluster
    gas is low at the lowest X-ray temperatures, rises to a peak
    between 1 and 4 keV, then drops at higher temperature for some
    reason not yet understood, then X-ray spectra of a cooling flow of
    such gas would appear to stop cooling just below that peak.}.

\section{Conclusions}
We have studied a 191~ks \emph{Chandra} image of the core of the
Perseus cluster for small-scale temperature and abundance structures.
The obvious X-ray surface brightness features are seen to be due to
temperature variations, with corresponding density variations ensuring
approximate azimuthal pressure balance. Such variations are seen on
scales down to above 2.5~arcsec ($\sim1$~kpc) in the brightest
regions.

We have made extensive searches for multi-temperature gas components.
There is little evidence in the deprojected spectra for any widespread
components at a temperature more than a factor of two away from the
local temperature. There is some evidence for a widespread hot
component (at 16~keV) in the projected data. This emission could
alternatively be due to an extended nonthermal component. The solution
to the cooling flow problem in the Perseus cluster must produce a
fairly uniform mass distribution of gas from 2.5 to 8~keV, with little
gas below~2 keV.

The abundances of Si, S, Ar, Ca, Fe and Ni rise inward from about
100~kpc, peaking at about 30--40~kpc. Most of these abundances level
out or drop again inward of the peak. The extent of the drop is
unclear, but it is plausibly explained by resonance scattering. O and
Mg are more uniform across the core and Ne shows a central peak. The
overall abundance pattern is similar to that found by others using
XMM-Newton data of M87 in the Virgo cluster, except that S/Si exceeds
unity in the Perseus cluster core. The abundance peaks are likely due
to Type Ia supernovae in the central galaxy, NGC\,1275.

\section*{Acknowledgements}
ACF and SWA thank the Royal Society for support. We are grateful for
the help of Mark Bautz and Maxim Markevitch on instrumentation issues.

% required because of bug in MN2e style file
% throws away figs otherwise
\clearpage

\end{document}